\documentclass{article}

    \usepackage[final]{neurips_2025}

\usepackage[utf8]{inputenc} 
\usepackage[T1]{fontenc}    
\usepackage{hyperref}       
\usepackage{url}            
\usepackage{booktabs}       
\usepackage{amsfonts}       
\usepackage{nicefrac}       
\usepackage{microtype}      
\usepackage{xcolor}         

\usepackage{graphicx}
\usepackage{subfigure}
\usepackage{subcaption}
\usepackage{lipsum}
\usepackage{wrapfig}
\usepackage{amsmath}
\usepackage{amssymb}
\usepackage{multirow,makecell}
\usepackage{mathtools}
\usepackage{amsthm}
\usepackage[textsize=tiny]{todonotes}

\usepackage{enumitem}
\usepackage{float}
\usepackage{graphicx}
\usepackage{subcaption}
\usepackage{tcolorbox}
\usepackage[ruled,vlined]{algorithm2e}

\title{TEMPO: Temporal Multi-scale Autoregressive Generation of Protein Conformational Ensembles}

\author{
\textbf{Yaoyao Xu\textsuperscript{$\S$ \P }},
Di Wang\textsuperscript{\P},
Zihan Zhou\textsuperscript{$\S$},
Tianshu Yu\textsuperscript{$\S$ \dag},
Mingchen Chen\textsuperscript{\P \dag}
\\
\textsuperscript{$\S$} School of Data Science, The Chinese University of Hong Kong, Shenzhen \\
\textsuperscript{\P} Changping Laboratory, Beijing\\
 \texttt{\{yaoyaoxu,zihanzhou1\}@link.cuhk.edu.cn},~~ \texttt{yutianshu@cuhk.edu.cn} \\
  \texttt{\{lotus,mingchenchen\}@cpl.ac.cn}\\
}

\begin{document}

\maketitle
\def\thefootnote{\dag}\footnotetext{\ Corresponding authors}\def\thefootnote{\arabic{footnote}}

\begin{abstract}
Understanding the dynamic behavior of proteins is critical to elucidating their functional mechanisms, yet generating realistic, temporally coherent trajectories of protein ensembles remains a significant challenge. In this work, we introduce a novel hierarchical autoregressive framework for modeling protein dynamics that leverages the intrinsic multi-scale organization of molecular motions. Unlike existing methods that focus on generating static conformational ensembles or treat dynamic sampling as an independent process, our approach characterizes protein dynamics as a Markovian process. The framework employs a two-scale architecture: a low-resolution model captures slow, collective motions driving major conformational transitions, while a high-resolution model generates detailed local fluctuations conditioned on these large-scale movements. This hierarchical design ensures that the causal dependencies inherent in protein dynamics are preserved, enabling the generation of temporally coherent and physically realistic trajectories. By bridging high-level biophysical principles with state-of-the-art generative modeling, our approach provides an efficient framework for simulating protein dynamics that balances computational efficiency with physical accuracy.
\end{abstract}
\section{Introduction}
The intersection of artificial intelligence and protein science has revolutionized our understanding of biological systems. Recent breakthroughs in AI for protein research have transformed structure and function prediction~\cite{potapenko2021highly,senior2020improved,xu2024demystify,kim2025easy}, protein design~\cite{anishchenko2021novo,dauparas2022robust,liu2025text}, and interaction modeling~\cite{bryant2022improved,evans2021protein,xu2024boosting,pacesa2025one}. However, while static structural understanding has advanced dramatically, accurately modeling protein dynamics remains an outstanding challenge at the frontier of computational biology.

Protein dynamics are characterized by two fundamental properties. First, they are inherently hierarchical and multi-scale, with motions naturally separating into slow collective movements (nanoseconds to microseconds) that typically correspond to functionally relevant conformational changes, and fast local fluctuations (picoseconds to nanoseconds) that reflect atomic-level interactions~\cite{frauenfelder1991energy,henzler2007dynamic}. This multi-scale organization forms the theoretical foundation for analytical methods like Principal Component Analysis and Normal Mode Analysis~\cite{brooks1983harmonic}, demonstrating that protein dynamics can be effectively decomposed into essential subspaces operating at different time scales. Second, protein motions exhibit strong temporal correlations, where the continuous evolution of conformational states follows specific pathways critical for biological functions. These temporally correlated dynamic ensembles have proven essential in understanding enzyme catalysis mechanisms~\cite{yabukarski2022ensemble}, characterizing drug-binding pathways~\cite{bowman2012equilibrium}, and elucidating allosteric regulation~\cite{popovych2006dynamically}. For instance, recent studies have demonstrated that analyzing dynamic ensembles can reveal cryptic binding sites that only become accessible through specific conformational transitions~\cite{boehr2009role}.

While molecular dynamics (MD) simulations can naturally capture both properties by solving Newton's equations of motion at atomic resolution, their computational demands make them impractical for large-scale applications. Even with specialized hardware and enhanced sampling techniques, MD simulations are typically limited to microsecond timescales and small protein systems, making it challenging to systematically study slow conformational changes or analyze large protein datasets~\cite{shaw2010atomic}.
Current generative approaches, particularly diffusion-based models~\cite{jing2024alphafold,lewis2024scalable,cheng20244d}, fundamentally fail to leverage these characteristics. These methods generate conformational ensembles by simultaneously producing and optimizing protein states independently, by learning an energy landscape rather than capturing the true sequential and multi-scale nature of protein dynamics. This approach cannot accurately represent the causal chain of events that governs protein conformational changes, limiting their ability to generate physically consistent trajectories.

Motivated by these challenges, we propose TEMPO -- a multi-scale autoregressive framework that models and generates protein dynamics across different temporal scales. Our approach combines a low-resolution model capturing essential conformational transitions with a high-resolution model generating detailed local fluctuations, directly translating biophysical principles into a computational framework for generating physically realistic trajectories. Extensive experiments demonstrate that our method achieves significant improvements over existing approaches.

Our work makes several key contributions:

\begin{itemize}
    \item \textbf{Algorithm Design}. TEMPO introduces a novel multi-scale framework that captures both protein collective motions and local fluctuations, enabling efficient trajectory generation orders of magnitude faster than MD simulations.
    \item \textbf{Performance and Metrics}. Our method achieves state-of-the-art performance in both structural accuracy and computational efficiency in various metrics, outperforming existing methods in matching MD ground truth while requiring fewer computational resources.
    \item \textbf{Extensive Analysis}. We demonstrate TEMPO's ability to capture biologically meaningful protein motions through comprehensive case studies and analyses.
\end{itemize}

\section{Related Work}
\textbf{Protein Ensemble Generation.}
Recent advances in deep learning have revolutionized the generation of protein conformational ensembles. Traditional approaches rely on MSA subsampling with AlphaFold2~\cite{potapenko2021highly}, which provides limited control over conformational diversity. Modern deep learning methods have introduced more sophisticated techniques. AlphaFlow~\cite{jing2024alphafold} fine-tunes single-state predictors under a flow matching framework to generate protein conformational ensembles. ESMFlow~\cite{jing2024alphafold} extends this approach by leveraging protein language models. BioEMU~\cite{lewis2024scalable} employs a diffusion-based framework to generate thermodynamically accurate ensembles. ConfDiff~\cite{wang2024protein} incorporates force-guided networks with diffusion models to enhance generation fidelity, while Str2Str~\cite{lustr2str} introduces a structure-to-structure translation framework with roto-translation equivariance. However, these methods primarily focus on generating conformational ensembles that match equilibrium distributions, without explicitly modeling the temporal evolution of protein structures.

\textbf{Learning Molecular Dynamics.}
Machine learning approaches have emerged as powerful tools for accelerating and enhancing molecular dynamics simulations. VAMPNet~\cite{mardt2022deep} pioneered the use of variational approaches for Markov processes in molecular kinetics. Recent works like DiffMD~\cite{wu2023diffmd} employ diffusion models to estimate conformational density gradients, while DFF~\cite{arts2023two} establishes connections between score-based generative models and molecular force fields. The Distributional Graphformer (DiG)~\cite{zheng2024predicting} predicts equilibrium distributions of molecular systems, enabling efficient conformational sampling. However, these methods often focus on general-purpose force field learning or small molecular systems, making them computationally intensive for large proteins.


\textbf{Multi-scale Dynamics Modeling.}
The inherent multi-scale nature of protein dynamics has long been recognized in computational biology. Traditional MD analysis methods decompose protein motions into slow collective changes, which are crucial for biological function, and fast local fluctuations contribute to overall stability. Recent deep learning approaches have begun to address this multi-scale characteristic. EigenFold~\cite{jing2023eigenfold} models protein structures as systems of harmonic oscillators, naturally inducing a cascading-resolution generative process along system eigenmodes. FoldFlow~\cite{bosese} proposes a family of flow-based generative models on SE(3), where the continuous-time dynamics naturally capture multi-scale structural variations - from global conformational changes to local refinements - through different time scales of the flow evolution. ITO~\cite{schreiner2023implicit} learns transition density operators that allow conditioning on arbitrary timesteps, focusing on coarse-grained C$\alpha$ representations with exponential distribution sampling. However, most existing approaches focus on either ensemble generation or short-timescale dynamics, without explicitly bridging the gap between different temporal scales in protein motion.

\textbf{Autoregressive Models in Structural Biology.}
While diffusion models have recently dominated protein structure generation, auto-regressive approaches are gaining traction for their ability to model temporally coherent and physically consistent dynamics. In the domain of bio-molecules, arDCA~\cite{trinquier2021efficient} applies a simple yet effective auto-regressive framework to model protein sequence distributions, capturing co-evolutionary couplings while enabling efficient sequence sampling and fitness prediction. For protein structure modeling, Structure Language Models~\cite{lu2024structure} employ latent-space auto-regression to efficiently generate diverse backbone conformations, while equivariant models such as EquiJump~\cite{costa2024equijump} build an $\mathrm{SO}(3)$-equivariant transport model that bridges long time intervals of all-atom protein MD by stochastically interpolating between snapshots. These advances demonstrate the promise of auto-regressive models in capturing complex bio-molecular dynamics across multiple timescales.

\section{Method}

\begin{figure}[t]
    \centering
    \includegraphics[width=\textwidth]{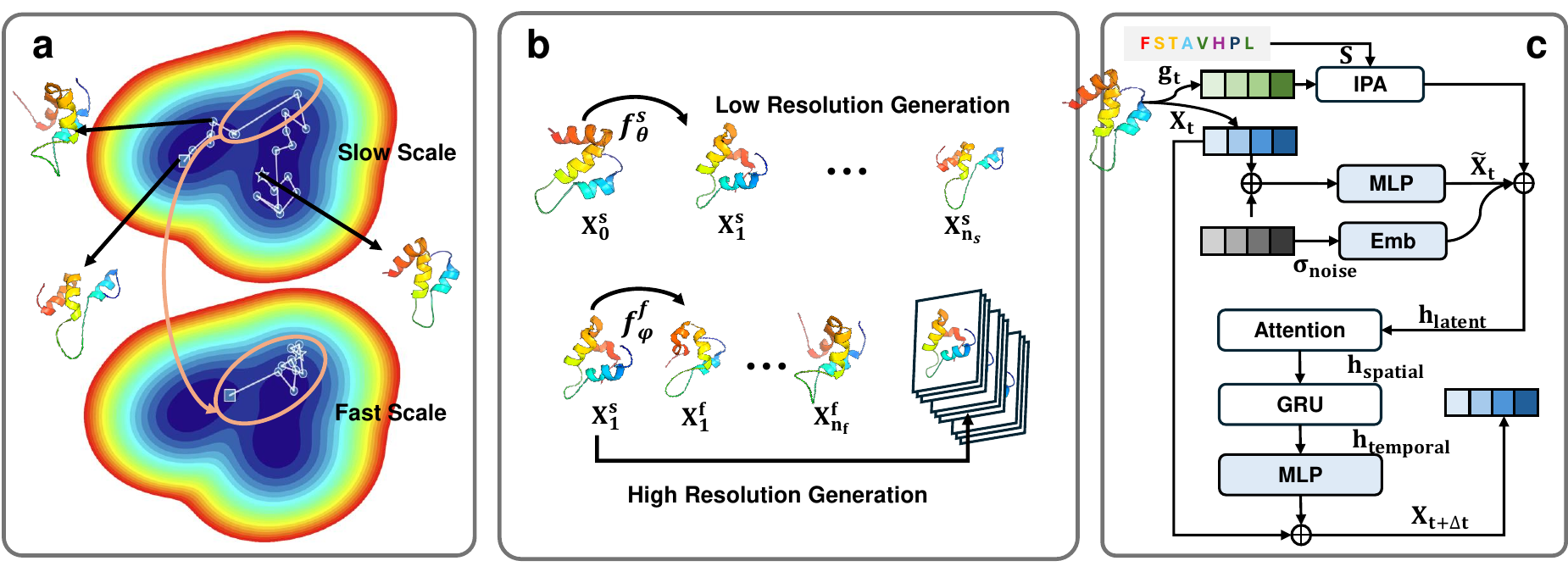}
    \caption{Overview of our multi-scale protein dynamics generation framework. \textbf{a)} The hierarchical free energy landscape of protein dynamics, where slow motions (upper) guide fast local fluctuations (lower). \textbf{b)} Our two-stage generation process: the low-resolution model $f^s_\theta$ captures slow collective motions, while the high-resolution model $f^f_\phi$ fills in detailed dynamics. \textbf{c)} The neural architecture that parameterizes both models features spatial-temporal encoding of protein conformations.}
    \label{fig:framework}
\end{figure}

\subsection{Preliminaries}

Given an initial protein structure $\mathbf{X}_0$ of sequence length $L$, our goal is to learn a generative model that produces protein backbone trajectories $\boldsymbol{\chi} = [\bf{X}_1, \ldots \bf{X}_T]$, where each $\bf{X}_i$ represents the backbone conformation at time step $i$. While full-atom protein structure representation has been widely adopted~\cite{potapenko2021highly,lewis2024scalable,jing2024alphafold}, we focus on backbone dynamics as they capture the essential conformational changes that determine protein function. This choice is motivated by several key observations: \textbf{(1)} many important biological processes, such as protein folding and large-scale conformational changes, are primarily determined by backbone movements~\cite{rose2006backbone}, \textbf{(2)} side-chain motions typically occur at faster timescales. They can be considered as local fluctuations around backbone configurations~\cite{ghosh2017watching}.

Consistent with standard representations in protein modeling~\cite{potapenko2021highly}, we describe each residue's backbone conformation using an $\mathrm{SE}(3)$ frame along with the torsion angles ($\phi$, $\psi$, $\omega$):
\begin{equation}
    \boldsymbol{\chi}_t^l = ((R, \mathbf{t}), (\phi, \psi, \omega)), \quad \boldsymbol{\chi}\in \left( \left[ \mathrm{SE}(3) \times \mathbb{T}^3 \right]^L\right)^T
\end{equation}
where $t$ denotes the time step and $l$ indicates the residue index. The $\mathrm{SE}(3)$ frame, representing the rigid body transformation, consists of a rotation component encoded as a unit quaternion from the positive real part $\hat{\mathbb{Q}}^+ \subset \mathbb{R}^4$ and a translation vector in $\mathbb{R}^3$, yielding a 7-dimensional representation. Each torsion angle is encoded as a 2D vector $[\sin(\theta), \cos(\theta)] \in S^2$ to avoid discontinuities at the periodic boundary. Combining these components, each residue at each time step is represented by a 13-dimensional vector:
\begin{equation}
    \boldsymbol{\chi}^j_t \in \left(\hat{ \mathbb{Q}}^+ \oplus \mathbb{R}^3 \right) \times (S^2)^3 \subset \mathbb{R}^{13}
\end{equation}
This representation captures the essential geometric features of protein backbone dynamics while maintaining computational efficiency, and the detailed data processing could be found in Appendix~\ref{appendix:protein_tokenization}.

\subsection{SDE-based Protein Dynamics Modeling}
Protein dynamics in solution naturally follows Langevin dynamics, which describes the motion of particles under both conservative forces and random collisions with solvent molecules~\cite{yang2006effective}. The Langevin equation captures two key aspects of protein motion: \textbf{(1)} Deterministic forces arising from inter-atomic interactions that drive conformational changes, \textbf{(2)} Random forces from thermal fluctuations that contribute to the stochastic nature of protein dynamics~\cite{copperman2015predicting}.

Motivated by this physical principle, we model protein dynamics as a stochastic differential equation (SDE) process, which can be viewed as a continuous-time generalization of the Langevin dynamics. In our framework, the evolution of protein conformations follows a combination of deterministic drift and stochastic diffusion. Specifically, for a protein conformation $X_t$ at time $t$, its temporal evolution can be described as:
\begin{equation}
dX_t = \mu(X_t)dt + \sigma dW_t
\end{equation}
where $\mu(X_t)$ represents the drift term that captures the deterministic dynamics, $\sigma$ is the diffusion coefficient, and $W_t$ denotes a standard Brownian motion. The drift term $\mu(X_t)$ is learned by our model, while the stochastic component is simulated through Gaussian noise injection.

In discrete time steps, our model approximates this continuous SDE process as:
\begin{equation}
X_{t+\Delta t} = X_t + f_\theta(X_t)\Delta t + \epsilon_t\sqrt{\Delta t}
\end{equation}
where $f_\theta$ is our neural network model parameterized by $\theta$ that learns the drift dynamics, $\Delta t$ is the time step, and $\epsilon_t \sim \mathcal{N}(0, \sigma^2I)$ represents the Gaussian noise. This formulation allows our model to capture both the deterministic conformational changes and the stochastic nature of protein dynamics.This approach is theoretically justified as we learn the conditional expectation $\mathbb{E}[X_{t+1}|X_t]$ in the finite time-step regime, mirroring how numerical MD integrators operate with deterministic updates plus controlled stochastic components for temperature regulation~\cite{yang2006effective,copperman2015predicting}.

However, protein dynamics typically exhibits non-Markovian behavior at short time scales~\cite{husic2018markov}, meaning that future states depend on multiple previous states rather than just the current one. To account for this memory effect, we extend our model to consider multiple timesteps:
\begin{equation}
\mathbf{X}_{t+1:t+2} = f_\theta(\mathbf{X}_{t-1:t})\Delta t + \epsilon_t\sqrt{\Delta t}
\end{equation}
where $\mathbf{X}_{t-1:t}$ represents two consecutive frames at times $t-1$ and $t$, and the model predicts the next two frames $\mathbf{X}_{t+1:t+2}$. This design choice is supported by prior research in physical system modeling~\cite{zwanzig1961memory} and latent ODEs~\cite{rubanova2019latent}, where incorporating temporal memory has been shown to significantly improve the accuracy of dynamical predictions.

\subsection{Multi-scale Dynamics Learning}
Building upon the SDE-based framework, we decompose protein motion into a bio-physically motivated two-timescale formulation to capture both slow collective motions and fast local fluctuations\cite{frauenfelder1991energy,henzler2007dynamic}. This decomposition is realized through coupled stochastic differential equations:
\begin{equation}
d\mathbf{X}^s_t = \mu_s(\mathbf{X}^s_t)dt + \sigma_s dW^s_t
\end{equation}
\begin{equation}
d\mathbf{X}^f_t = \mu_f(\mathbf{X}^f_t)dt + \sigma_f dW^f_t
\end{equation}
where $\mathbf{X}^s_t$ and $\mathbf{X}^f_t$ represent protein conformations at slow and fast timescales, respectively, with corresponding drift terms $\mu_s$ and $\mu_f$, and independent Brownian motions $W^s_t$, $W^f_t$.

To learn these multi-scale dynamics, we employ a unified neural architecture that operates at both timescales. Specifically, the drift terms $\mu_s$ and $\mu_f$ are parameterized by the same spatiotemporal encoder architecture $f$ but trained separately to capture scale-specific features. The discrete time evolution at each scale follows:
\begin{equation}
\mathbf{X}^s_{t+\Delta t_s} = \mathbf{X}^s_t + f^s_\theta(\mathbf{X}^s_t)\Delta t_s + \epsilon^s_t\sqrt{\Delta t_s}
\end{equation}
\begin{equation}
\mathbf{X}^f_{t+\Delta t_f} = \mathbf{X}^f_t + f^f_\phi(\mathbf{X}^f_t)\Delta t_f + \epsilon^f_t\sqrt{\Delta t_f}
\end{equation}
where $\Delta t_s$ and $\Delta t_f$ represent the time steps for slow and fast dynamics, with neural networks $f^s_\theta$ and $f^f_\phi$ learning the respective drift dynamics.

During inference, we employ a hierarchical sampling strategy that mirrors the natural organization of protein dynamics. Starting from an initial conformation $\mathbf{X}^s_0$, we first generate a sparse trajectory $\{\mathbf{X}^s_0, \mathbf{X}^s_{\Delta t_s}, ..., \mathbf{X}^s_{n_s\Delta t_s}\}$ using the slow-scale model $f^s_\theta$, which captures collective motions like domain reorientations. These slow-scale conformations then serve as anchoring states for the fast-scale model $f^f_\phi$, which generates the complete fine-grained trajectory $\{\mathbf{X}^f_t: t \in [0, n_f\Delta t_f]\}$, ensuring that fast local dynamics remain consistent with the broader conformational changes.

\subsection{Spatiotemporal Protein Encoder}
Here we detail the neural architecture that parameterizes the drift dynamics at both timescales. Our encoder design captures both spatial relationships between protein residues and temporal correlations in conformational dynamics. As illustrated in Figure~\ref{fig:framework}, the network comprises three functional components: input representation, spatial-temporal encoding, and conformational prediction.

The input representation module processes protein conformations $\mathbf{X}_t\in\mathbb{R}^{L\times13}$ with added noise terms that model the stochastic nature of protein dynamics. Specifically, we sample a noise scale $\sigma_{\text{noise}} \sim \mathcal{U}(a, b)$ and generate Gaussian noise $\epsilon \sim \mathcal{N}(0, \mathbf{I})$ to obtain noisy conformations $\tilde{\mathbf{X}}_t = \mathbf{X}_t + \sigma_{\text{noise}}\epsilon$. In parallel, protein sequences are embedded into feature vectors $\mathbf{s} \in \mathbb{R}^{L\times d_{\text{hidden}}}$ through a learnable embedding layer. These sequence features are combined with frame representations $\mathbf{g}_t \in \mathrm{SE}(3)^L$ and noise scale in an Invariant Point Attention (IPA)~\cite{potapenko2021highly} module to capture geometric relationships. The final latent representation is computed as:

\begin{equation}
    \mathbf{h}_{\text{latent}} = \mathrm{MLP}(\tilde{\mathbf{X}}_t) + \mathrm{IPA}(\mathbf{g}_t, \mathbf{s}, \sigma_{\text{noise}}) + \mathrm{Embed}(\sigma_{\text{noise}})
\end{equation}

where $\mathrm{MLP}: \mathbb{R}^{13} \rightarrow \mathbb{R}^{d_{\text{hidden}}}$ projects the input features to hidden dimension $d_{\text{hidden}}$, and $\mathrm{Embed}$ maps the scalar noise intensity to a $d_{\text{hidden}}$-dimensional vector. The latent representation then undergoes spatial-temporal processing through:
\begin{equation}
    \mathbf{h}_{\text{spatial}} = \mathrm{MultiHeadAttention}(\mathbf{h}_{\text{latent}})
\end{equation}
\begin{equation}
    \mathbf{h}_{\text{temporal}} = \mathrm{GRU}(\mathbf{h}_{\text{spatial}})
\end{equation}
The spatial module captures inter-residue interactions while the temporal module encodes frame-to-frame dependencies. The output module generates conformational updates:
\begin{equation}
    \mathbf{X}_{t+\Delta t} = \mathbf{X}_t + \mathrm{MLP}(\mathbf{h}_{\text{temporal}})
\end{equation}
where $\mathrm{MLP}: \mathbb{R}^{d_{\text{hidden}}} \rightarrow \mathbb{R}^{13}$ maps the latent features back to the conformational space.

The training objectives for both timescales follow the same formulation, consisting of two terms: a reconstruction loss measuring the mean squared error between predicted and ground truth conformations and a physical constraint loss penalizing steric clashes between backbone atoms:

\begin{equation}
    \mathcal{L}_{\text{total}} = \|\mathbf{X}_{t+\Delta t} - \hat{\mathbf{X}}_{t+\Delta t}\|^2 + \lambda \sum_{i\neq j} \mathrm{ReLU}(1.2\text{Å} - \|r_i - r_j\|)
\end{equation}

where $r_i$ and $r_j$ are positions of backbone atoms from different residues, and the minimum distance threshold of 1.2Å is chosen following the widely adopted steric criteria in Rosetta~\cite{alford2017rosetta}. This objective is applied independently to train the slow and fast dynamics models, with appropriate time intervals $\Delta t_s$ and $\Delta t_f$, respectively.

\section{Experiments}
\subsection{Experimental Settings}
\label{experiment}
\textbf{Datasets.}
We conduct experiments on two comprehensive molecular dynamics datasets: mdCATH~\cite{mirarchi2024mdcath} and ATLAS~\cite{vander2024atlas}. For mdCATH, we randomly sampled 1,000 proteins and their 320K temperature trajectories with three independent seeds for training. Each protein sequence was truncated to 240 residues, and trajectories were standardized to 400 frames at 1ns intervals through periodic extension or truncation. We randomly selected 50 proteins for validation and 64 proteins for testing, ensuring no overlap with the training set. For ATLAS, we follow the data split and processing protocol established by MDGen~\cite{jinggenerative}. We rigorously quantified sequence similarity using mmseqs2, finding an average of 18.93\% sequence similarity between training and test sets for mdCATH and 18.3\% for ATLAS, well below standard thresholds (40\%) for sequence relatedness. 

\textbf{Baselines.}
To evaluate our method's capability in capturing both protein dynamics and ensemble properties, we conduct comprehensive comparisons with four state-of-the-art baselines: BioEMU~\cite{lewis2024scalable}, AlphaFlow and ESMFlow~\cite{jing2024alphafold}, and MDGen~\cite{jinggenerative}. While the first three methods primarily focus on generating conformational ensembles, MDGen, though not explicitly modeling dynamics, captures temporal evolution through training on the ATLAS dataset~\cite{vander2024atlas}. 


\textbf{Implementation details.}
Our multi-scale modeling approach captures protein dynamics at two temporal resolutions. The low-resolution model generates trajectories at $20$ns intervals, characterizing major structural transitions, while the high-resolution model operates at $1$ns resolution to capture local fluctuations. We empirically chose the 20ns/1ns hierarchy based on established biophysical principles where the 20ns interval effectively captures major conformational transitions between different states in the free energy surface as visualized in Figure~\ref{fig:framework}a, while the 1ns resolution represents the dataset's finest temporal sampling. The generation process follows a hierarchical strategy. The low-resolution model first produces a sequence of conformational states $\{X_t^s\}_{t=1}^{20}$ at $\Delta t = 20\text{ns}$.  Specifically, each high-resolution segment is initialized by the corresponding low-resolution state ($X_{t+\Delta t}^f = f_\theta(X_t^s)$), followed by autoregressive sampling ($X_{t+k\Delta t}^f = f_\theta(X_{t+(k-1)\Delta t}^f)$ for $k = 2,...,20$ where $\Delta t = 1\text{ns}$). This hierarchical process generates a complete trajectory while maintaining consistency across different temporal scales.

During training, both scale-models simulate the forward process of protein dynamics SDE through autoregressive sampling with noise scales uniformly sampled from $[0.01, 0.05]$. At inference time, while the low-resolution model maintains similar noise levels, we increase the noise scale to $5.0$ for the high-resolution model. This elevated noise level enables diverse conformational sampling on the learned energy surface.

\textbf{Evaluation Framework.}
Our comprehensive evaluation framework encompasses both ensemble properties and trajectory-specific characteristics. Following AlphaFlow, we analyze conformational flexibility through several complementary measures: \textbf{Dynamic Range} (the average C$\alpha$-RMSD between pairs of conformations within each ensemble, quantifying the overall conformational space explored), \textbf{Local Flexibility} (assessed through root mean square fluctuation (RMSF) analysis of atomic positions), and \textbf{Distribution Accuracy} (quantified using the root mean Wasserstein distance (RMWD) between predicted and ground truth conformational distributions).

The trajectory accuracy is measured by the backbone RMSD error between generated and ground truth trajectories relative to the native structure: $\text{Error}_{\text{frame}} = |\text{RMSD}_{\text{pred}} - \text{RMSD}_{\text{gt}}|$. This metric, averaged across all frames and test proteins, quantifies the model's ability to capture conformational change magnitudes accurately~\cite{lindorff2011fast}. Further biological validation includes contact dynamics analysis, where contacts between C$\alpha$ atoms (8\AA{} threshold) are classified as weak (initially present but dissociate in $>10\%$ of the ensemble) or transient (initially absent but form in $>10\%$ of the ensemble), with accuracy evaluated using Jaccard similarity between predicted and ground truth contact sets. Computational efficiency is assessed through the average inference time per protein in generating 400 snapshots. Additionally, we calculate the clash ratio, defined as the proportion of conformations containing steric clashes among the 400 generated snapshots for each protein. Detailed definitions of all metrics are provided in Appendix~\ref{appendix:metrics}.

\begin{table}[t]
    \caption{\textbf{Evaluation on mdCATH}. Comparing predicted ensembles with MD ensembles across various metrics. For protein flexibility and RMSF, ground truth values are in parentheses. Median values across 64 test ensembles are reported. The rightmost column shows TEMPO's performance on the up-sampling task. $r$: Pearson correlation; $J$: Jaccard similarity; $\mathcal{W}_2$: 2-Wasserstein distance.}
    \label{tab:md_results}
    \centering
    \small
    \resizebox{\textwidth}{!}{%
    \begin{tabular}{lccccc|c}
    \toprule
    \textbf{Metrics} & \textbf{TEMPO} & \textbf{BioEMU} & \textbf{Alpha\textsc{Flow}-MD} & \textbf{ESM\textsc{Flow}-MD} & \textbf{MDGEN} & \textbf{TEMPO}(Up)\\ 
    \midrule
    Pairwise RMSD($=$ 3.26) &\textbf{2.78} &13.82  &2.00 &2.32 & 1.11 &3.06  \\
    Pairwise RMSD $r$ $\uparrow$ & \textbf{0.77} &-0.02 &0.41 &0.26 & 0.71 &0.99 \\
    All-atom RMSF($=$ 1.64)  & \textbf{1.60} &10.08 &0.99 &1.18 & 0.56 &1.64  \\
    Global RMSF $r$ & \textbf{0.67} &0.13 &0.41 &0.34 & 0.67 &0.99 \\
    \midrule
    Root mean $\mathcal{W}_2$ $\downarrow$ &4.21 &10.70 &5.62 &4.08 &\textbf{3.36} &1.06 \\
    MD PCA $\mathcal{W}_2$ $\downarrow$ &\textbf{2.33} &2.49 &2.38 &2.36 &2.62 &0.63  \\
    \% PC-sim $>0.5$ $\uparrow$ &7.81 &9.38  &21.88 &\textbf{25.00} &17.19 &95.31 \\
    \midrule
    Weak contacts $J$ $\uparrow$ &0.43 &0.38  &0.42 &\textbf{0.51} & 0.41 &0.83 \\
    Trans. contacts $J$ $\uparrow$ &0.20 &0.12  &0.27  &\textbf{0.28} &0.20 &0.60 \\   
    \% Clash ratio $\downarrow$ &4.75 &19.7 &15.5 &5.23 &\textbf{0.42} &4.12 \\
    \midrule
    RMSD Error $\downarrow$  &\textbf{1.78} & - & - & - & 3.76 &0.60  \\
    \midrule
    Inference time (hour) &\textbf{0.006} &0.25  &4.5  &4.7  &0.008 & -  \\
    \bottomrule
    \end{tabular}}
    \vspace{-0.1in}
\end{table}

\subsection{Results and Analysis}
\label{results}

We evaluate our framework on both mdCATH and ATLAS datasets. Table~\ref{tab:md_results} presents results on mdCATH, while Table~\ref{tab:atlas_results} shows results on ATLAS where all methods are trained and tested following MDGen's protocol. TEMPO demonstrates consistent strong performance across both datasets, achieving state-of-the-art results on key metrics. 

\begin{table}[h]
\centering
\caption{\textbf{Evaluation on ATLAS.} All methods trained and tested following MDGen's protocol.}
\label{tab:atlas_results}
\begin{tabular}{lcccc}
\toprule
\textbf{Metrics} & \textbf{BioEMU} & \textbf{AlphaFlow-MD} & \textbf{MDGen} & \textbf{TEMPO} \\
\midrule
Pairwise RMSD $r$ $\uparrow$ & -0.02 & 0.48 & 0.48 & \textbf{0.91} \\
Global RMSF $r$ $\uparrow$ & 0.09 & 0.60 & 0.50 & \textbf{0.89} \\
Root mean W2 $\downarrow$ & 19.23 & 2.61 & 2.69 & \textbf{1.49} \\
MD PCA W2 $\downarrow$ & 3.61 & 1.52 & 1.89 & \textbf{0.60} \\
\% PC-sim $>$ 0.5 $\uparrow$ & 14 & 44 & 10 & \textbf{76} \\
Weak contacts $J$ $\uparrow$ & 0.26 & 0.51 & 0.62 & \textbf{0.74} \\
Trans. contacts $J$ $\uparrow$ & 0.06 & 0.29 & \textbf{0.41} & 0.38 \\
RMSD Error $\downarrow$ & - & - & 3.20 & \textbf{1.83} \\
\bottomrule
\end{tabular}
\end{table}

Our multi-scale framework demonstrates distinct advantages across three critical dimensions of protein dynamics modeling. On ATLAS, TEMPO achieves substantial improvements over baselines, with Pearson correlation of 0.91 for pairwise RMSD and 0.89 for global RMSF, significantly outperforming other methods. The model captures 76\% of principal components with similarity greater than 0.5, compared to 44\% for AlphaFlow and 10\% for MDGen, demonstrating superior ability to preserve essential collective motions. On mdCATH, in structural flexibility metrics, TEMPO achieves the closest match to molecular dynamics (MD) ground truth, with pairwise C$\alpha$-RMSD and residue-level RMSF closely matching MD values. The strong Pearson correlation ($r=0.77$) between predicted and ground truth RMSD values reveals our hierarchical SDE formulation preserves the intrinsic roughness of protein energy landscapes.

In distribution matching metrics, TEMPO achieves the lowest MD PCA Wasserstein distance (0.60 on ATLAS, 2.33 on mdCATH) across both datasets, suggesting accurate preservation of principal motion patterns. While MDGen shows slight advantages in average conformational Wasserstein distance on mdCATH, our approach achieves better backbone RMSD error on both datasets (1.83 on ATLAS, 1.78 on mdCATH compared to MDGen's 3.20 and 3.76). Furthermore, our physically constrained learning reduces steric clashes compared to ESMFlow, BioEMU, and AlphaFlow, validating the biological plausibility of generated conformations.

In computational efficiency, TEMPO generates complete 400-frame trajectories in approximately 22 seconds, significantly faster than AlphaFlow and ESMFlow. Among all baselines, only MDGen is specifically designed for trajectory generation, yet it requires prohibitive computational resources. In contrast, our multi-scale decomposition enables training on a single NVIDIA A100 GPU, with both slow-scale and fast-scale models operating within memory constraints.

These results collectively demonstrate that TEMPO's physics-informed multi-scale design achieves superior performance in conformational accuracy, temporal coherence, and computational efficiency across different datasets. Our experimental validation incorporates four analyses to verify TEMPO's spatiotemporal modeling: (1) Collective motion analysis to evaluate the capture of functionally relevant slow motions, (2) Up-sampling evaluates high-resolution reconstruction of local fluctuations, (3) State transition tracking examines temporal pathway fidelity, and (4) Free energy surface analysis to compare conformational sampling strategies. We focus subsequent analysis and visualizations on mdCATH as its free energy landscapes contain more distinct energy basins with clearer conformational transition pathways, better suited for analyzing dynamics modeling. Visual examples of generated protein trajectories are provided in Appendix~\ref{appendix:visualization}, offering an intuitive demonstration of our model's capability to capture protein dynamics. Additional ablation studies and detailed analyses are provided in the Appendix. All analyzed cases are identified by their PDB IDs.

\begin{figure}[h!]
    \centering
    \subfigure[1bl0A02]{
        \includegraphics[width=0.23\textwidth]{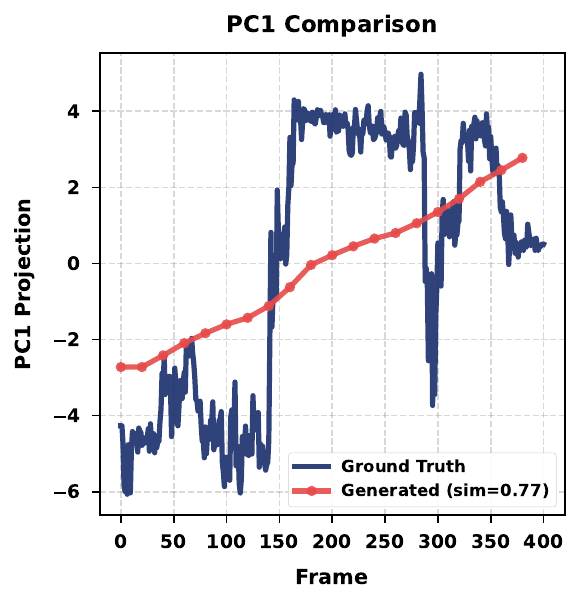}
    }
    \subfigure[5exeA02]{
        \includegraphics[width=0.24\textwidth]{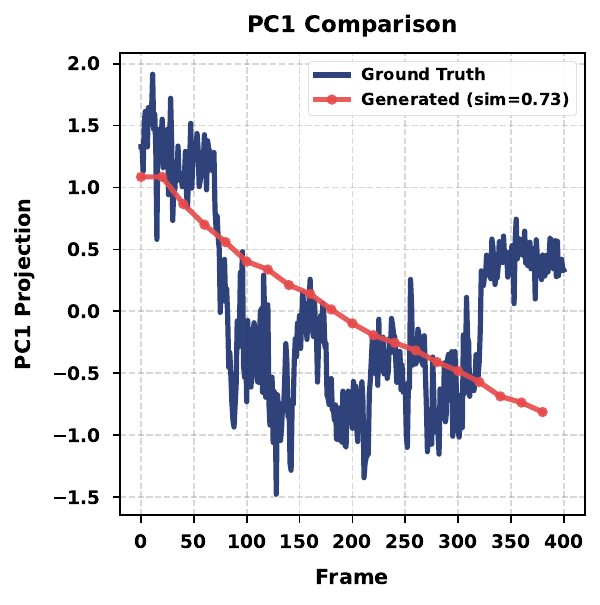}
    }
    \subfigure[3c0wA02]{
        \includegraphics[width=0.23\textwidth]{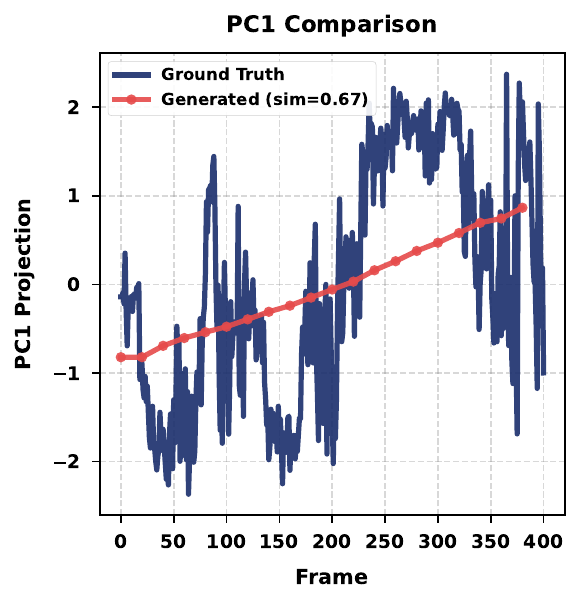}
    }
    \subfigure[]{
        \includegraphics[width=0.23\textwidth]{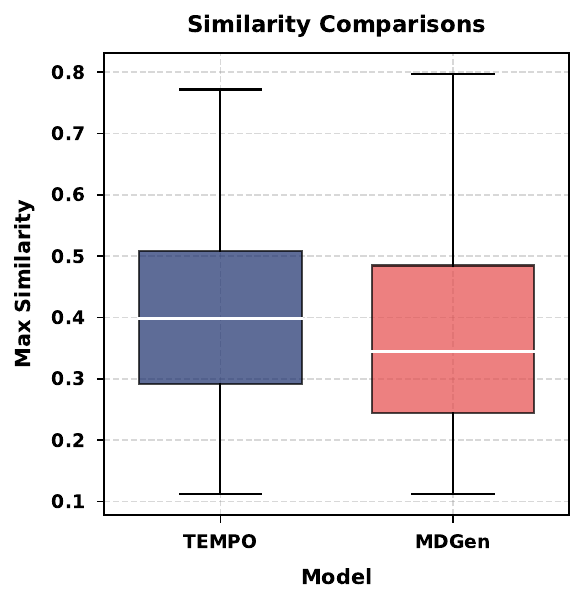}
    }
    \caption{Comparison of PC projections between MD and generated conformations. (a-c) PC1 trajectories from MD (blue) and our slow-scale trajectories (red) for three representative proteins. (d) Box plot of cosine similarity scores comparing TEMPO and MDGen across the test set.}
    \label{fig:pc1_comparison}
\end{figure}
\textbf{Slow Motion generation.}
Protein slow motions, occurring on microsecond to millisecond timescales, often correspond to functionally relevant conformational changes such as domain movements~\cite{henzler2007dynamic}. Capturing these collective motions is challenging due to their long timescales and coordinated nature. We evaluate our model's capability in capturing such motions through principal component analysis (PCA), as the first few PCs typically describe the dominant collective motions in protein dynamics~\cite{amadei1993essential}. For representative cases (Figure~\ref{fig:pc1_comparison}(a-c)), we projected both MD trajectories and our generated slow-scale conformations onto the first principal component (PC1) of MD trajectories. The cosine similarity between these projections ranges from 0.67 to 0.77, indicating strong alignment of collective motions. Extending this analysis across all test proteins (Figure~\ref{fig:pc1_comparison}(d)), we evaluated the maximum cosine similarity between the first two PCs of generated ensembles and MD trajectories to assess the capture of collective motions. TEMPO achieves better performance than MDGen, with a mean similarity of 0.41 compared to MDGen's 0.36, suggesting potential for capturing slow motions while highlighting the challenging nature of this task.

\begin{figure}[tb]
    \centering
    \subfigure[2eyzA03 md]{
        \includegraphics[width=0.23\textwidth]{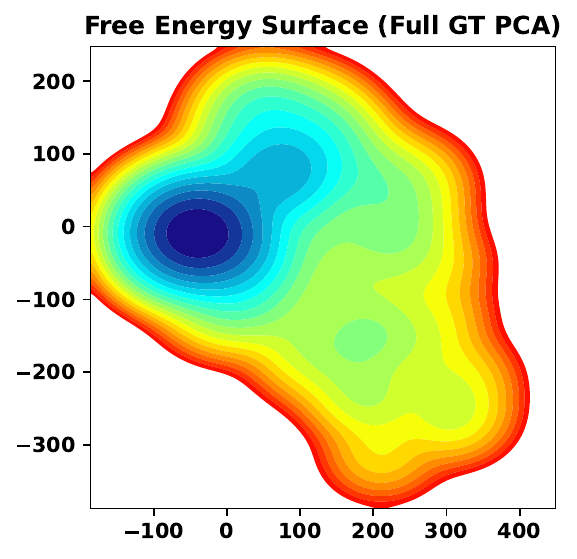}
    }
    \subfigure[2eyzA03 generated]{
        \includegraphics[width=0.23\textwidth]{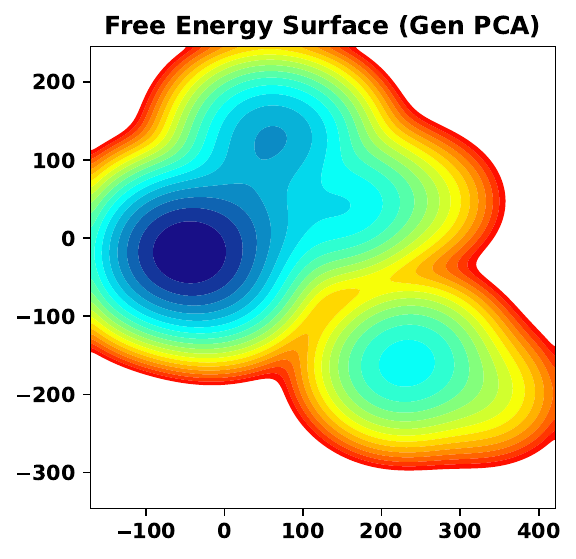}
    }
    \subfigure[3gyxA02 md]{
        \includegraphics[width=0.225\textwidth]{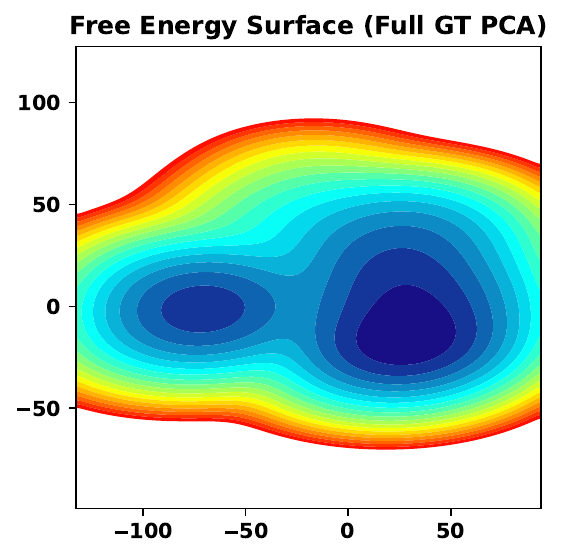}
    }
    \subfigure[3gyxA02 generated]{
        \includegraphics[width=0.22\textwidth]{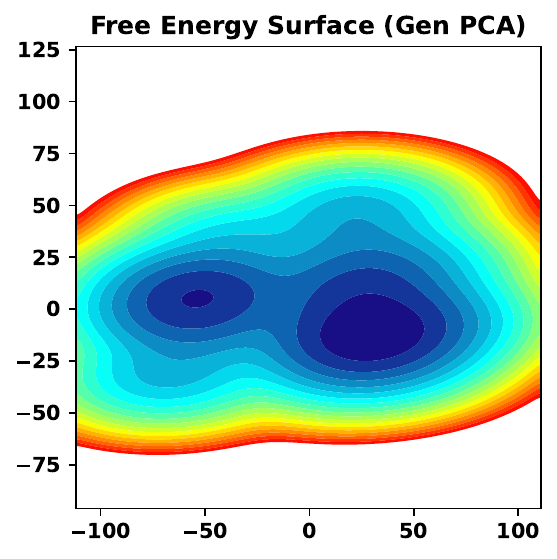}
    }
    \caption{FES comparison between MD trajectories (a,c) and generated ensembles (b,d) for proteins 2eyzA03 (a,b) and 3gyxA02 (c,d). Colors represent free energy values from low (blue) to high (red).}
    \label{fig:upsampling}
\end{figure}

\textbf{Up-sampling.} 
We evaluated our high-resolution model's capacity to capture detailed protein motions through up-sampling experiments, using ground truth low-resolution protein conformation as input. 
The conformational sampling quality was quantitatively assessed via free energy surface (FES) analysis in a reduced dimensionality space, obtained through PCA of protein backbone coordinates using Prody. 
The free energy landscapes were constructed by projecting conformations onto the first two principal components, followed by kernel density estimation and Boltzmann inversion at $300$K. 
Figure~\ref{fig:upsampling} demonstrates the FES contour plots for two representative test proteins, comparing the distributions between MD ensembles and our generated trajectories. 
The close correspondence in FES characteristics, particularly in the location and depth of energy minima, validates our model's ability.
More evaluation metrics across our test set (Table~\ref{tab:md_results}) further confirm that our high-resolution model effectively captures local protein fluctuations consistent with MD simulations. Additional case studies with extended protein sets are presented in Appendix~\ref{appendix:upsampling}.

\begin{figure}[t!]
    \centering
    \includegraphics[width=0.245\textwidth]{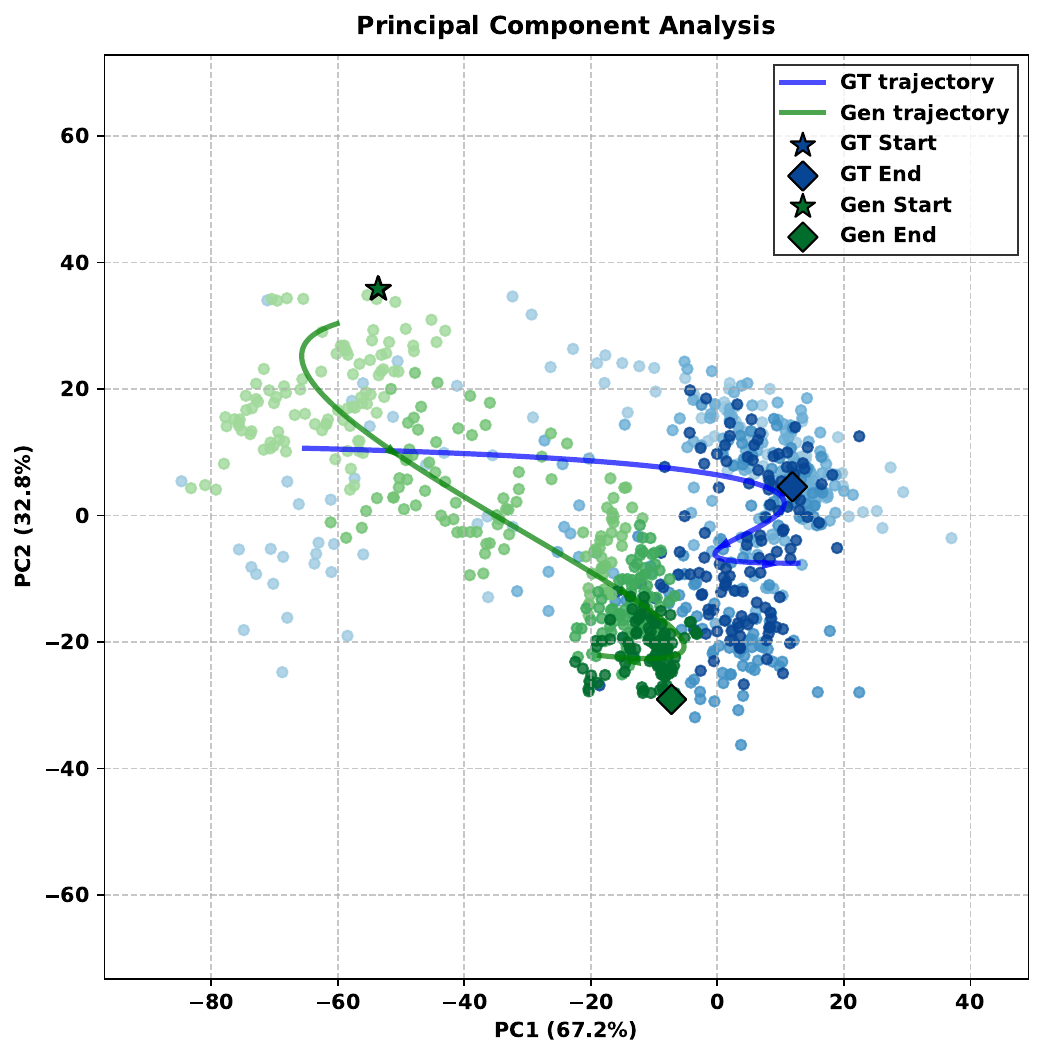}\includegraphics[width=0.245\textwidth]{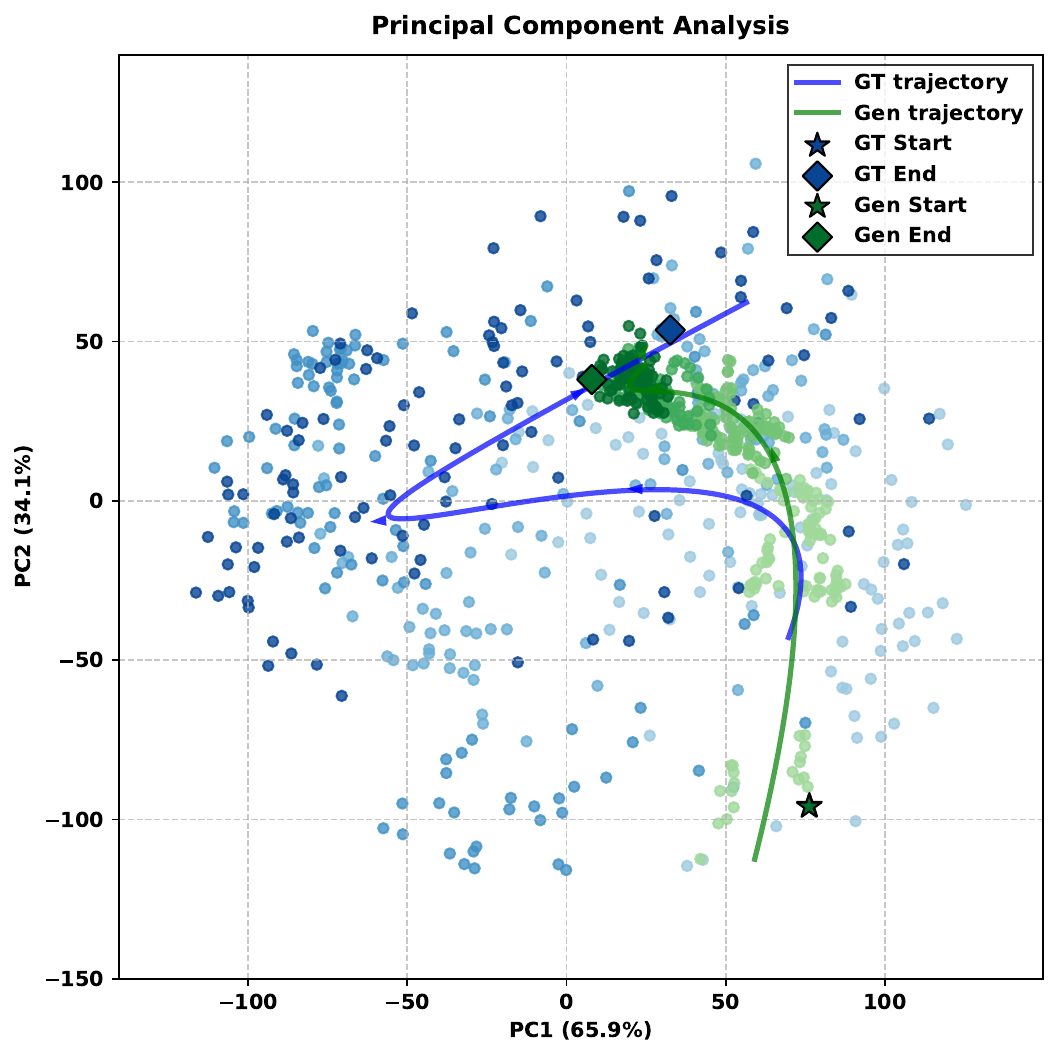}
    \includegraphics[width=0.245\textwidth]{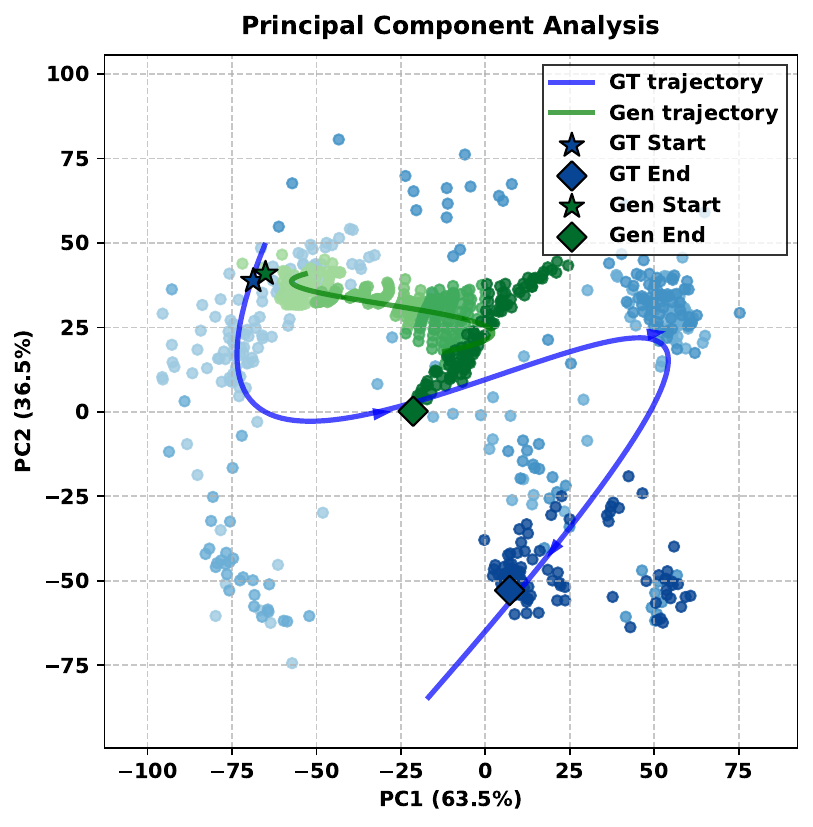}
    \includegraphics[width=0.245\textwidth]{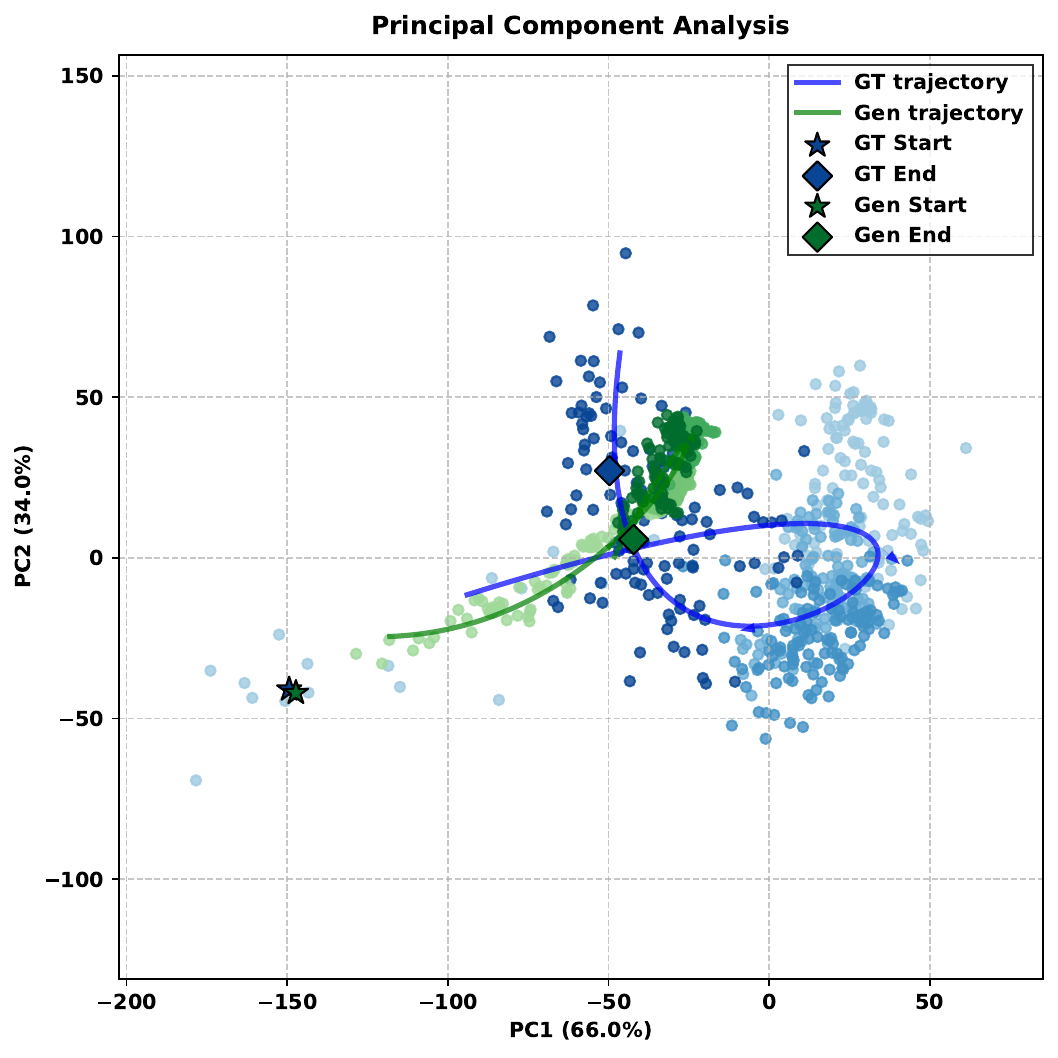}
    \includegraphics[width=0.245\textwidth]{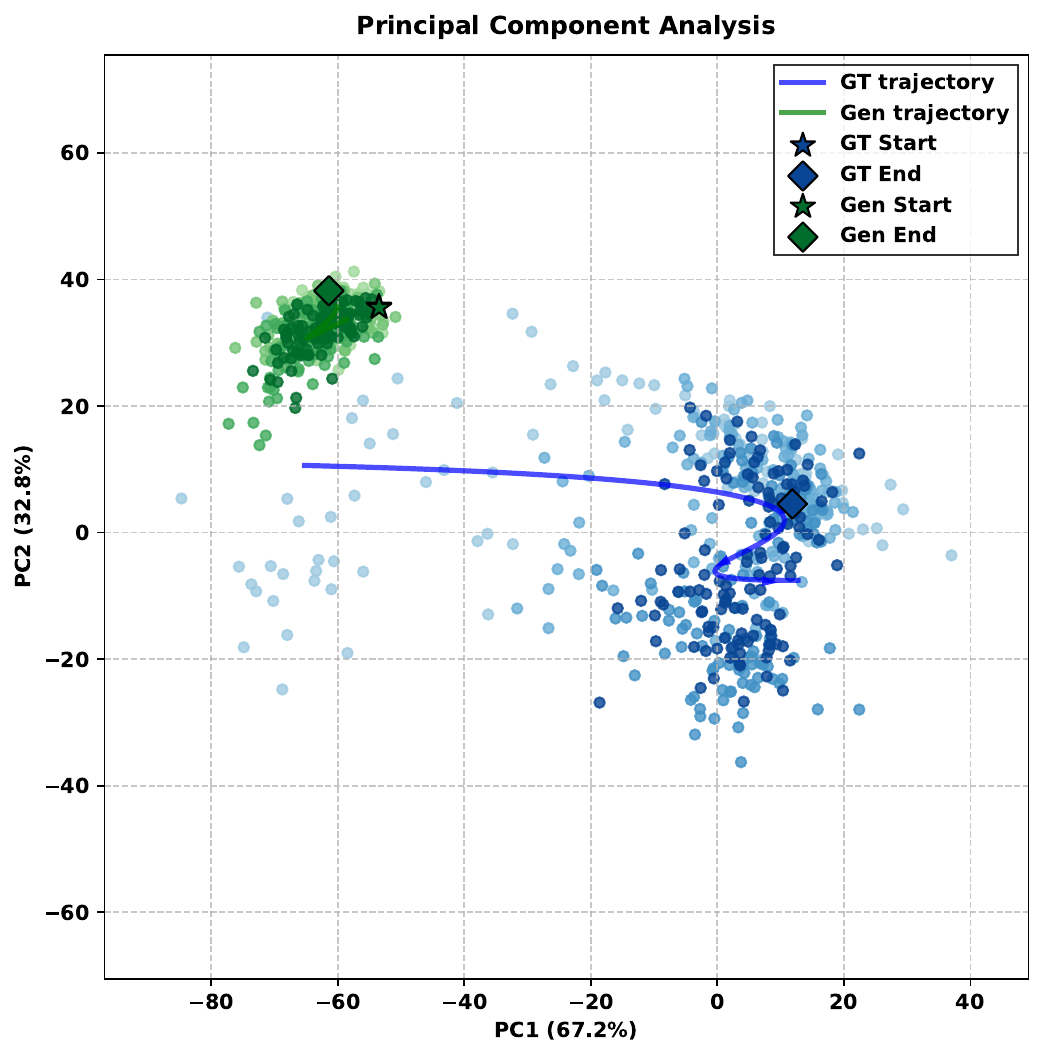}\includegraphics[width=0.245\textwidth]{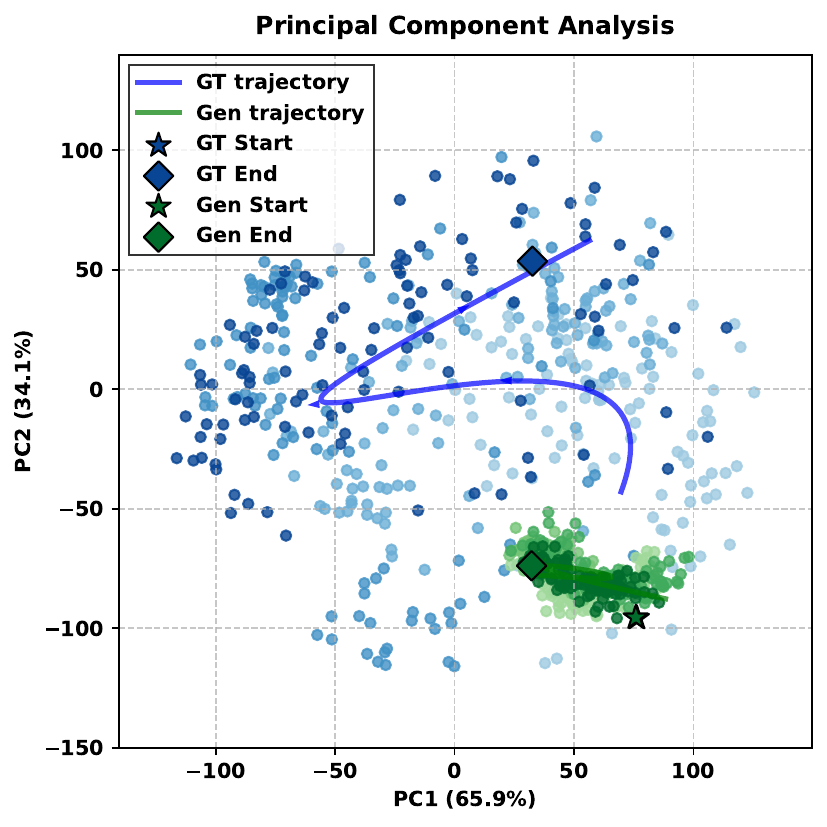}
    \includegraphics[width=0.245\textwidth]{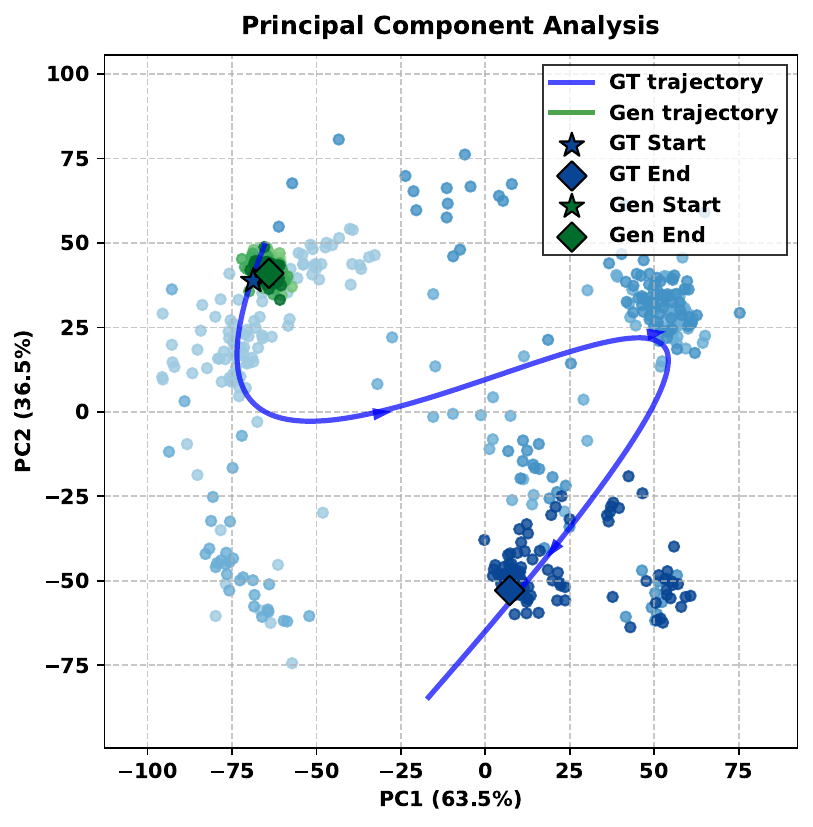}
    \includegraphics[width=0.245\textwidth]{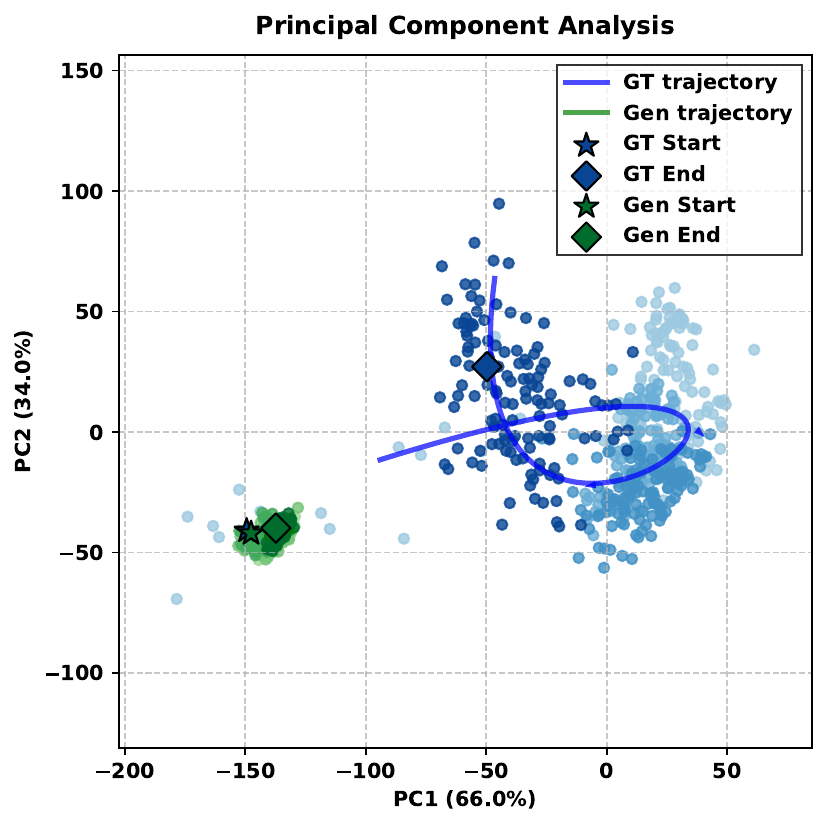}
    \caption{Comparison of conformational transitions in PC space between TEMPO and MDGen baseline (bottom). Ground truth MD trajectories are shown in blue, while generated trajectories are in green. The polynomial fitting curves highlight the temporal evolution of conformational changes (Protein from left to right: 2e9xB01, 1s79A00, 1bl0A02, 3cx5E01).}\label{fig:state_transition}\vspace{-3.9mm}
\end{figure}

\textbf{State Transition.}
After evaluating the two scale models separately, we further investigated TEMPO's ability to capture complete conformational transition pathways. Using four representative proteins from our test set, we analyzed how well our integrated framework reproduces the sequential nature of conformational changes. We first constructed the PCA space using MD trajectories as reference, then projected both our generated trajectories and MD trajectories onto the first two principal components to visualize the conformational landscape. As shown in Figure~\ref{fig:state_transition}, the upper row demonstrates TEMPO's ability to generate trajectories (green) that follow similar transition pathways as MD simulations (blue). In contrast, the MDGen tends to generate clustered conformations in limited regions of the PC space, failing to capture the full range of transitions. This comparison highlights the advantage of our temporal modeling approach over the frame-independent generation strategy, particularly in reproducing the sequential nature of conformational changes. The polynomial fitting curves further illustrate how our model better tracks the temporal evolution of these state transitions. Additional transition pathway analyses for an extended set of test proteins are provided in Appendix~\ref{appendix:transition}.

\begin{figure}[h]
    \centering
     \includegraphics[width=0.245\textwidth]{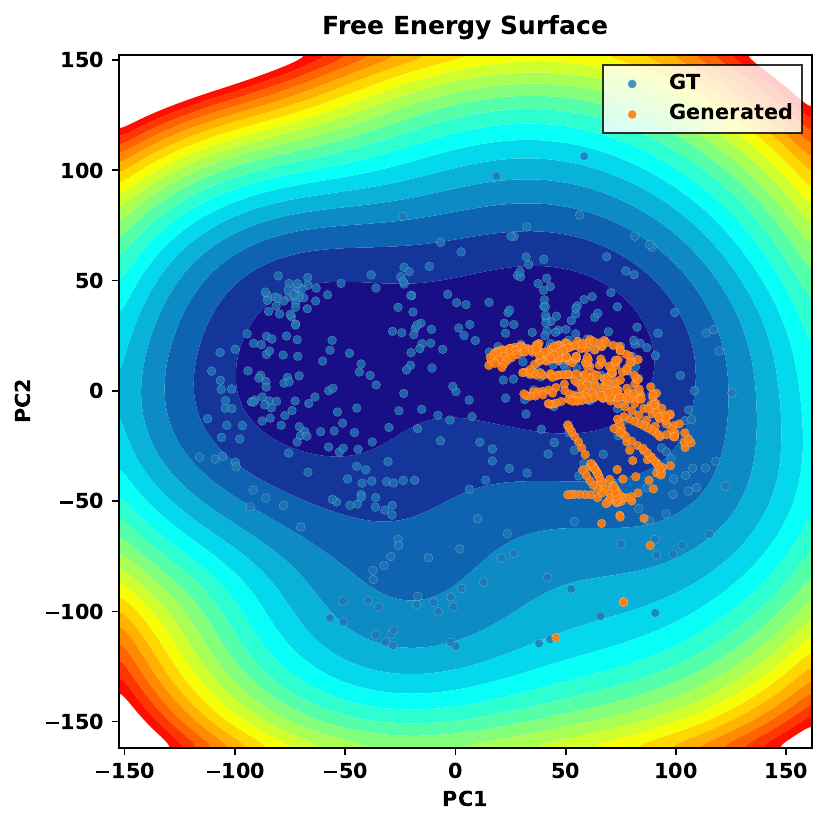}
    \includegraphics[width=0.245\textwidth]{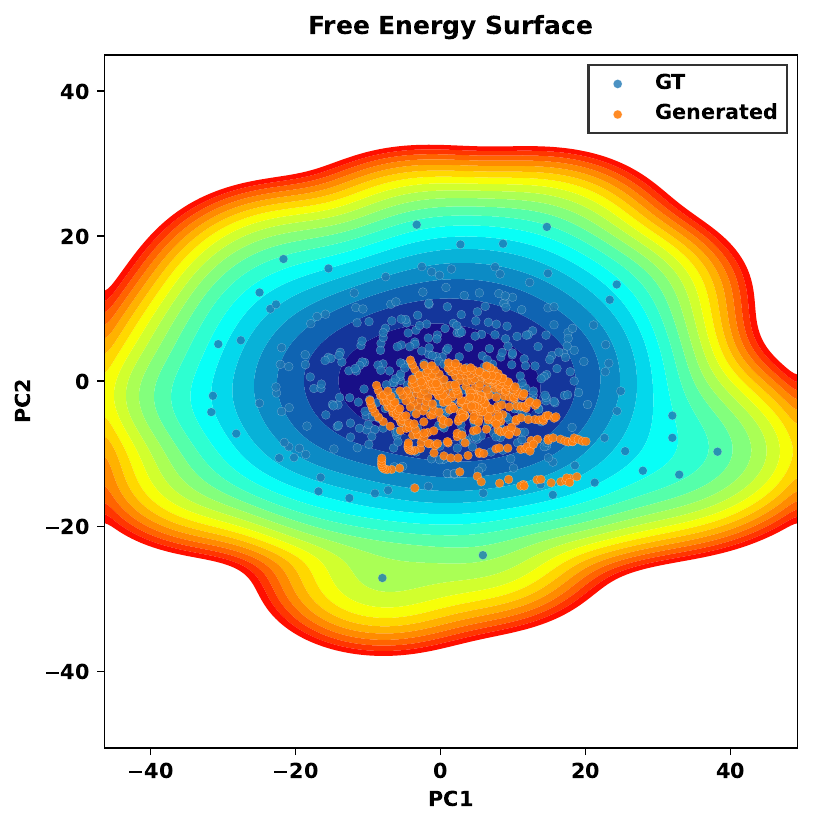}
    \includegraphics[width=0.245\textwidth]{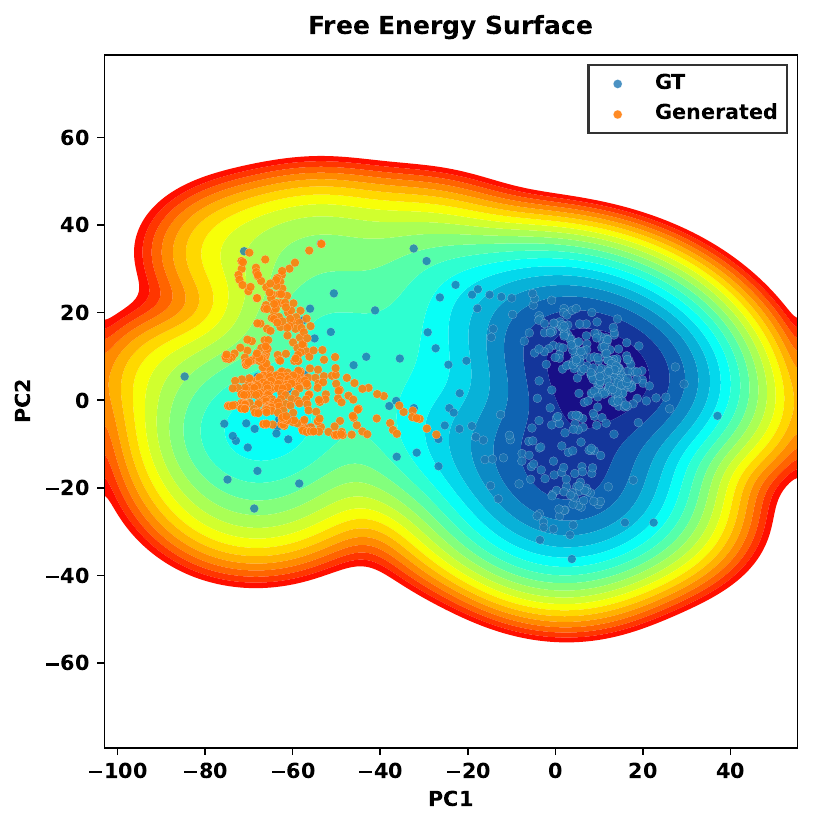}
    \includegraphics[width=0.245\textwidth]{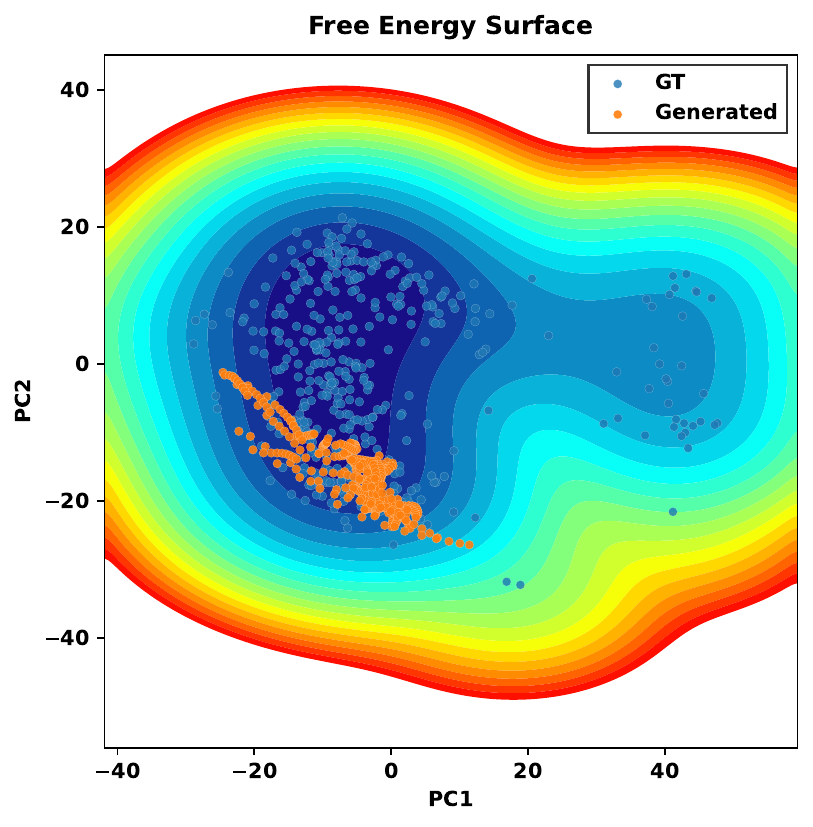}
    \includegraphics[width=0.245\textwidth]{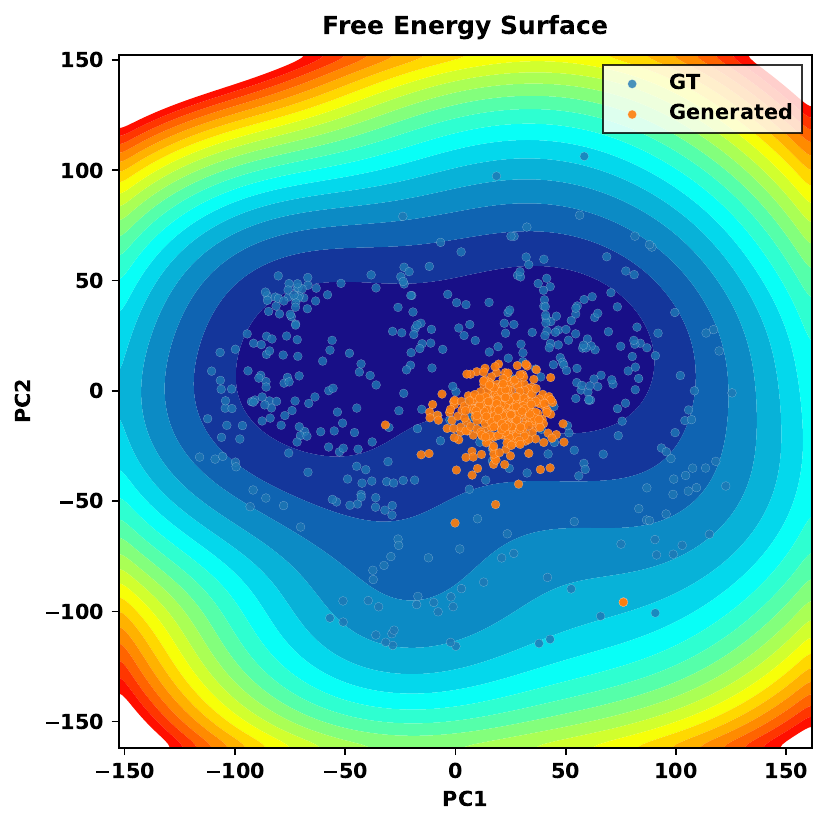}
    \includegraphics[width=0.245\textwidth]{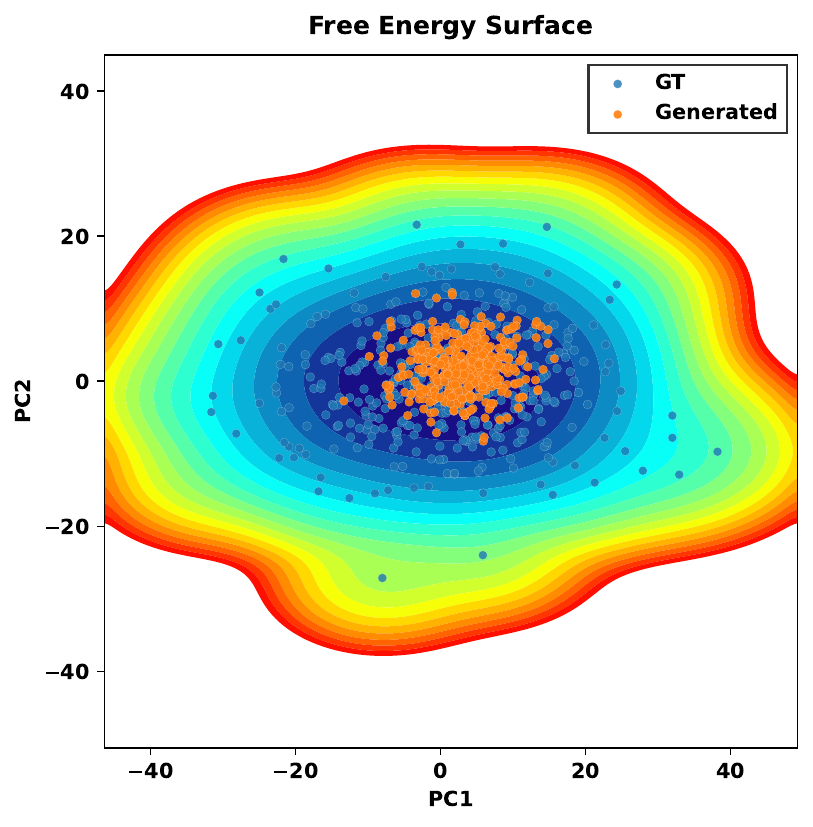}
    \includegraphics[width=0.245\textwidth]{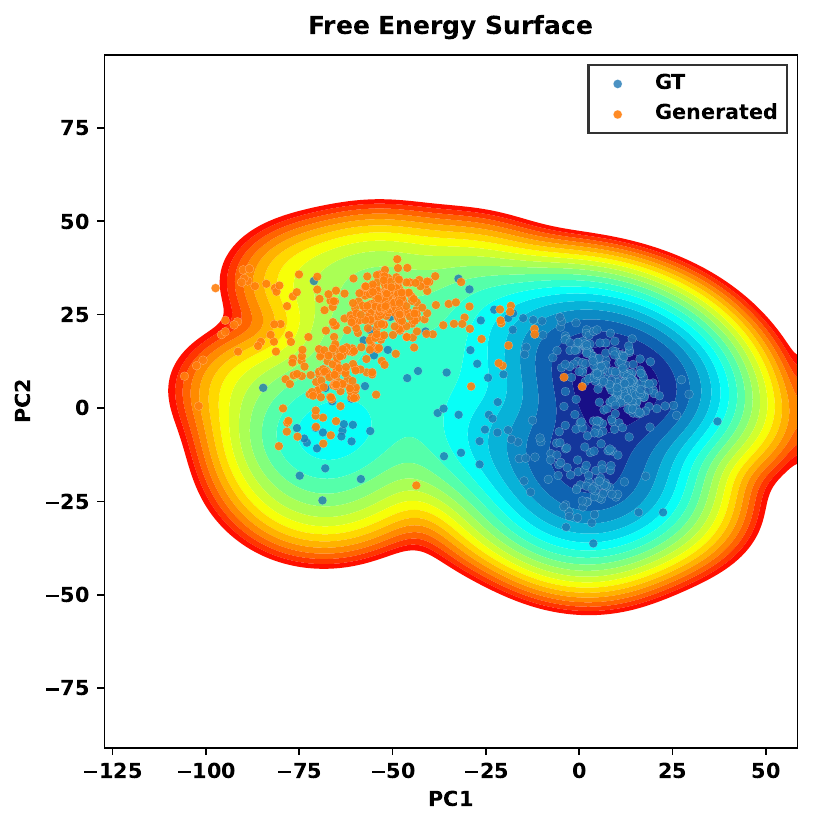}
    \includegraphics[width=0.245\textwidth]{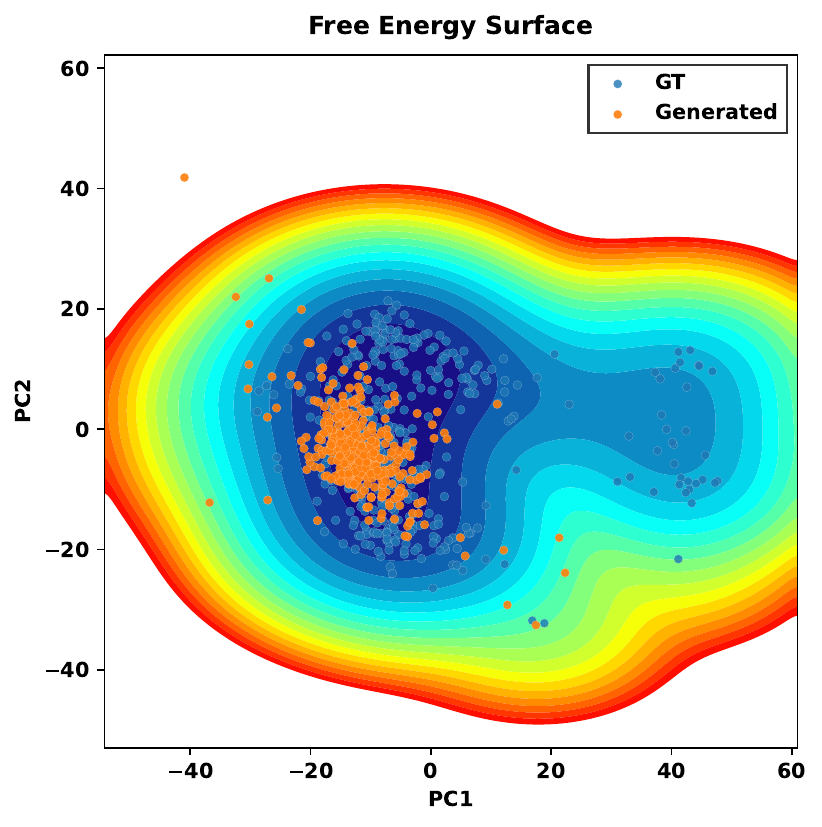}
    \caption{FES comparison between TEMPO (top row) and ESMFlow (bottom row) on four randomly selected test proteins. TEMPO's dynamic modeling shows more focused exploration of conformational space, whereas ESMFlow's independent sampling achieves broader coverage across the PCA-derived free energy surface (Protein from left to right: 1s79A00, 1x4tA01, 2e9xB01, 5b58T02).}
    \label{fig:FES}
\end{figure}

\textbf{Free Energy Surface Coverage.}
Our quantitative metrics effectively assess conformational stability and structural accuracy, but provide limited insight into the comprehensive exploration of conformational space. To address this, we established a reference framework using PCA derived from MD ensembles of four randomly selected proteins, subsequently constructing the corresponding free energy surface (FES). Projection of generated conformations onto this FES revealed that ESMFlow achieves broader coverage of the conformational landscape (Figure~\ref{fig:FES}), consistent with its design for independent sampling across the energy surface. 

This divergence reflects fundamentally different modeling objectives. ESMFlow optimizes conformational diversity through independent sampling, valuable for exploring thermodynamically accessible states. In contrast, TEMPO's focused sampling is a design feature that preserves temporal correlations and physical constraints governing real protein motion. Real proteins must respect energy barriers and cannot instantaneously jump between distant conformations; biological processes depend on sequential, pathway-dependent conformational changes where kinetic accessibility differs from thermodynamic accessibility. Such trajectory-aware modeling ultimately serves our primary objective of elucidating the kinetic mechanisms underlying biological function, rather than maximizing configurational sampling (additional analyses in Appendix~\ref{appendix:fes}).

\section{Conclusion}
In conclusion, our proposed TEMPO framework represents a significant advancement in the modeling and generation of protein dynamics by effectively addressing the inherent complexities of hierarchical and multi-scale behavior. Through the integration of a multi-scale autoregressive approach with stochastic differential equations, TEMPO successfully captures both slow collective motions and fast local fluctuations that characterize protein dynamics. Our comprehensive experimental validation demonstrates that TEMPO achieves superior performance across multiple metrics, from structural flexibility to distribution matching, while maintaining computational efficiency compared to existing methods. As we move forward, TEMPO's innovative design and proven capabilities hold the potential to facilitate further research in protein dynamics, ultimately contributing to a more comprehensive understanding of biological systems and their underlying mechanisms.

\section{Acknowledgment}
We express our sincere gratitude to Changping Laboratory for their research funding and support. This work is also supported in part by the Guangdong Provincial Key Laboratory of Mathematical Foundations for Artificial Intelligence (2023B1212010001). The resources and collaborative environment provided have been instrumental in enabling us to achieve the objectives of this project.

\newpage
\bibliographystyle{plain}
\bibliography{ref}

\begin{thebibliography}{10}

\bibitem{alford2017rosetta}
Rebecca~F Alford, Andrew Leaver-Fay, Jeliazko~R Jeliazkov, Matthew~J O’Meara, Frank~P DiMaio, Hahnbeom Park, Maxim~V Shapovalov, P~Douglas Renfrew, Vikram~K Mulligan, Kalli Kappel, et~al.
\newblock The rosetta all-atom energy function for macromolecular modeling and design.
\newblock {\em Journal of chemical theory and computation}, 13(6):3031--3048, 2017.

\bibitem{amadei1993essential}
Andrea Amadei, Antonius~BM Linssen, and Herman~JC Berendsen.
\newblock Essential dynamics of proteins.
\newblock {\em Proteins: Structure, Function, and Bioinformatics}, 17(4):412--425, 1993.

\bibitem{anishchenko2021novo}
Ivan Anishchenko, Samuel~J Pellock, Tamuka~M Chidyausiku, Theresa~A Ramelot, Sergey Ovchinnikov, Jingzhou Hao, Khushboo Bafna, Christoffer Norn, Alex Kang, Asim~K Bera, et~al.
\newblock De novo protein design by deep network hallucination.
\newblock {\em Nature}, 600(7889):547--552, 2021.

\bibitem{arts2023two}
Marloes Arts, Victor Garcia~Satorras, Chin-Wei Huang, Daniel Zugner, Marco Federici, Cecilia Clementi, Frank No{\'e}, Robert Pinsler, and Rianne van~den Berg.
\newblock Two for one: Diffusion models and force fields for coarse-grained molecular dynamics.
\newblock {\em Journal of Chemical Theory and Computation}, 19(18):6151--6159, 2023.

\bibitem{boehr2009role}
David~D Boehr, Ruth Nussinov, and Peter~E Wright.
\newblock The role of dynamic conformational ensembles in biomolecular recognition.
\newblock {\em Nature chemical biology}, 5(11):789--796, 2009.

\bibitem{bosese}
Joey Bose, Tara Akhound-Sadegh, Guillaume Huguet, Kilian FATRAS, Jarrid Rector-Brooks, Cheng-Hao Liu, Andrei~Cristian Nica, Maksym Korablyov, Michael~M Bronstein, and Alexander Tong.
\newblock Se (3)-stochastic flow matching for protein backbone generation.
\newblock In {\em The Twelfth International Conference on Learning Representations}.

\bibitem{bowman2012equilibrium}
Gregory~R Bowman and Phillip~L Geissler.
\newblock Equilibrium fluctuations of a single folded protein reveal a multitude of potential cryptic allosteric sites.
\newblock {\em Proceedings of the National Academy of Sciences}, 109(29):11681--11686, 2012.

\bibitem{brooks1983harmonic}
Bernard Brooks and Martin Karplus.
\newblock Harmonic dynamics of proteins: normal modes and fluctuations in bovine pancreatic trypsin inhibitor.
\newblock {\em Proceedings of the National Academy of Sciences}, 80(21):6571--6575, 1983.

\bibitem{bryant2022improved}
Patrick Bryant, Gabriele Pozzati, and Arne Elofsson.
\newblock Improved prediction of protein-protein interactions using alphafold2.
\newblock {\em Nature communications}, 13(1):1265, 2022.

\bibitem{cheng20244d}
Kaihui Cheng, Ce~Liu, Qingkun Su, Jun Wang, Liwei Zhang, Yining Tang, Yao Yao, Siyu Zhu, and Yuan Qi.
\newblock 4d diffusion for dynamic protein structure prediction with reference guided motion alignment.
\newblock {\em arXiv preprint arXiv:2408.12419}, 2024.

\bibitem{copperman2015predicting}
J~Copperman and MG~Guenza.
\newblock Predicting protein dynamics from structural ensembles.
\newblock {\em The Journal of chemical physics}, 143(24), 2015.

\bibitem{costa2024equijump}
Allan dos~Santos Costa, Ilan Mitnikov, Franco Pellegrini, Ameya Daigavane, Mario Geiger, Zhonglin Cao, Karsten Kreis, Tess Smidt, Emine Kucukbenli, and Joseph Jacobson.
\newblock Equijump: Protein dynamics simulation via so (3)-equivariant stochastic interpolants.
\newblock {\em arXiv preprint arXiv:2410.09667}, 2024.

\bibitem{dauparas2022robust}
Justas Dauparas, Ivan Anishchenko, Nathaniel Bennett, Hua Bai, Robert~J Ragotte, Lukas~F Milles, Basile~IM Wicky, Alexis Courbet, Rob~J de~Haas, Neville Bethel, et~al.
\newblock Robust deep learning--based protein sequence design using proteinmpnn.
\newblock {\em Science}, 378(6615):49--56, 2022.

\bibitem{evans2021protein}
Richard Evans, Michael O’Neill, Alexander Pritzel, Natasha Antropova, Andrew Senior, Tim Green, Augustin {\v{Z}}{\'\i}dek, Russ Bates, Sam Blackwell, Jason Yim, et~al.
\newblock Protein complex prediction with alphafold-multimer.
\newblock {\em biorxiv}, pages 2021--10, 2021.

\bibitem{frauenfelder1991energy}
Hans Frauenfelder, Stephen~G Sligar, and Peter~G Wolynes.
\newblock The energy landscapes and motions of proteins.
\newblock {\em Science}, 254(5038):1598--1603, 1991.

\bibitem{ghosh2017watching}
Ayanjeet Ghosh, Joshua~S Ostrander, and Martin~T Zanni.
\newblock Watching proteins wiggle: Mapping structures with two-dimensional infrared spectroscopy.
\newblock {\em Chemical reviews}, 117(16):10726--10759, 2017.

\bibitem{henzler2007dynamic}
Katherine Henzler-Wildman and Dorothee Kern.
\newblock Dynamic personalities of proteins.
\newblock {\em Nature}, 450(7172):964--972, 2007.

\bibitem{husic2018markov}
Brooke~E Husic and Vijay~S Pande.
\newblock Markov state models: From an art to a science.
\newblock {\em Journal of the American Chemical Society}, 140(7):2386--2396, 2018.

\bibitem{jing2024alphafold}
Bowen Jing, Bonnie Berger, and Tommi Jaakkola.
\newblock Alphafold meets flow matching for generating protein ensembles.
\newblock In {\em Proceedings of the 41st International Conference on Machine Learning}, pages 22277--22303, 2024.

\bibitem{jing2023eigenfold}
Bowen Jing, Ezra Erives, Peter Pao-Huang, Gabriele Corso, Bonnie Berger, and Tommi~S Jaakkola.
\newblock Eigenfold: Generative protein structure prediction with diffusion models.
\newblock In {\em ICLR 2023-Machine Learning for Drug Discovery workshop}.

\bibitem{jinggenerative}
Bowen Jing, Hannes Stark, Tommi Jaakkola, and Bonnie Berger.
\newblock Generative modeling of molecular dynamics trajectories.
\newblock In {\em The Thirty-eighth Annual Conference on Neural Information Processing Systems}.

\bibitem{kim2025easy}
Gyuri Kim, Sewon Lee, Eli Levy~Karin, Hyunbin Kim, Yoshitaka Moriwaki, Sergey Ovchinnikov, Martin Steinegger, and Milot Mirdita.
\newblock Easy and accurate protein structure prediction using colabfold.
\newblock {\em Nature Protocols}, 20(3):620--642, 2025.

\bibitem{lewis2024scalable}
Sarah Lewis, Tim Hempel, Jos{\'e} Jim{\'e}nez-Luna, Michael Gastegger, Yu~Xie, Andrew~YK Foong, Victor~Garc{\'\i}a Satorras, Osama Abdin, Bastiaan~S Veeling, Iryna Zaporozhets, et~al.
\newblock Scalable emulation of protein equilibrium ensembles with generative deep learning.
\newblock {\em bioRxiv}, pages 2024--12, 2024.

\bibitem{lindorff2011fast}
Kresten Lindorff-Larsen, Stefano Piana, Ron~O Dror, and David~E Shaw.
\newblock How fast-folding proteins fold.
\newblock {\em Science}, 334(6055):517--520, 2011.

\bibitem{liu2025text}
Shengchao Liu, Yanjing Li, Zhuoxinran Li, Anthony Gitter, Yutao Zhu, Jiarui Lu, Zhao Xu, Weili Nie, Arvind Ramanathan, Chaowei Xiao, et~al.
\newblock A text-guided protein design framework.
\newblock {\em Nature Machine Intelligence}, pages 1--12, 2025.

\bibitem{lu2024structure}
Jiarui Lu, Xiaoyin Chen, Stephen~Zhewen Lu, Chence Shi, Hongyu Guo, Yoshua Bengio, and Jian Tang.
\newblock Structure language models for protein conformation generation.
\newblock In {\em NeurIPS 2024 Workshop on AI for New Drug Modalities}.

\bibitem{lustr2str}
Jiarui Lu, Bozitao Zhong, Zuobai Zhang, and Jian Tang.
\newblock Str2str: A score-based framework for zero-shot protein conformation sampling.
\newblock In {\em The Twelfth International Conference on Learning Representations}.

\bibitem{mardt2022deep}
Andreas Mardt, Tim Hempel, Cecilia Clementi, and Frank No{\'e}.
\newblock Deep learning to decompose macromolecules into independent markovian domains.
\newblock {\em Nature Communications}, 13(1):7101, 2022.

\bibitem{mirarchi2024mdcath}
Antonio Mirarchi, Toni Giorgino, and Gianni De~Fabritiis.
\newblock mdcath: A large-scale md dataset for data-driven computational biophysics.
\newblock {\em Scientific Data}, 11(1):1299, 2024.

\bibitem{pacesa2025one}
Martin Pacesa, Lennart Nickel, Christian Schellhaas, Joseph Schmidt, Ekaterina Pyatova, Lucas Kissling, Patrick Barendse, Jagrity Choudhury, Srajan Kapoor, Ana Alcaraz-Serna, et~al.
\newblock One-shot design of functional protein binders with bindcraft.
\newblock {\em Nature}, pages 1--10, 2025.

\bibitem{popovych2006dynamically}
Nataliya Popovych, Shangjin Sun, Richard~H Ebright, and Charalampos~G Kalodimos.
\newblock Dynamically driven protein allostery.
\newblock {\em Nature structural \& molecular biology}, 13(9):831--838, 2006.

\bibitem{potapenko2021highly}
Alex~Bridgland Potapenko, Clemens Meyer, Simon~AA Kohl, Andrew~J Ballard, Andrew Cowie, Bernardino Romera-Paredes, Stanislav Nikolov, Rishub Jain, and Demis Hassabis.
\newblock Highly accurate protein structure prediction with alphafold.
\newblock {\em Nature}, 596:583, 2021.

\bibitem{rose2006backbone}
George~D Rose, Patrick~J Fleming, Jayanth~R Banavar, and Amos Maritan.
\newblock A backbone-based theory of protein folding.
\newblock {\em Proceedings of the National Academy of Sciences}, 103(45):16623--16633, 2006.

\bibitem{rubanova2019latent}
Yulia Rubanova, Ricky~TQ Chen, and David~K Duvenaud.
\newblock Latent ordinary differential equations for irregularly-sampled time series.
\newblock {\em Advances in neural information processing systems}, 32, 2019.

\bibitem{satorras2021n}
V{\i}ctor~Garcia Satorras, Emiel Hoogeboom, and Max Welling.
\newblock E (n) equivariant graph neural networks.
\newblock In {\em International conference on machine learning}, pages 9323--9332. PMLR, 2021.

\bibitem{schreiner2023implicit}
Mathias Schreiner, Ole Winther, and Simon Olsson.
\newblock Implicit transfer operator learning: Multiple time-resolution models for molecular dynamics.
\newblock {\em Advances in Neural Information Processing Systems}, 36:36449--36462, 2023.

\bibitem{senior2020improved}
Andrew~W Senior, Richard Evans, John Jumper, James Kirkpatrick, Laurent Sifre, Tim Green, Chongli Qin, Augustin {\v{Z}}{\'\i}dek, Alexander~WR Nelson, Alex Bridgland, et~al.
\newblock Improved protein structure prediction using potentials from deep learning.
\newblock {\em Nature}, 577(7792):706--710, 2020.

\bibitem{shaw2010atomic}
David~E Shaw, Paul Maragakis, Kresten Lindorff-Larsen, Stefano Piana, Ron~O Dror, Michael~P Eastwood, Joseph~A Bank, John~M Jumper, John~K Salmon, Yibing Shan, et~al.
\newblock Atomic-level characterization of the structural dynamics of proteins.
\newblock {\em Science}, 330(6002):341--346, 2010.

\bibitem{shoemake1985animating}
Ken Shoemake.
\newblock Animating rotation with quaternion curves.
\newblock In {\em Proceedings of the 12th annual conference on Computer graphics and interactive techniques}, pages 245--254, 1985.

\bibitem{tian2024visual}
Keyu Tian, Yi~Jiang, Zehuan Yuan, Bingyue Peng, and Liwei Wang.
\newblock Visual autoregressive modeling: Scalable image generation via next-scale prediction.
\newblock {\em Advances in neural information processing systems}, 37:84839--84865, 2024.

\bibitem{trinquier2021efficient}
Jeanne Trinquier, Guido Uguzzoni, Andrea Pagnani, Francesco Zamponi, and Martin Weigt.
\newblock Efficient generative modeling of protein sequences using simple autoregressive models.
\newblock {\em Nature communications}, 12(1):5800, 2021.

\bibitem{vander2024atlas}
Yann Vander~Meersche, Gabriel Cretin, Aria Gheeraert, Jean-Christophe Gelly, and Tatiana Galochkina.
\newblock Atlas: protein flexibility description from atomistic molecular dynamics simulations.
\newblock {\em Nucleic acids research}, 52(D1):D384--D392, 2024.

\bibitem{wang2024protein}
Yan Wang, Lihao Wang, Yuning Shen, Yiqun Wang, Huizhuo Yuan, Yue Wu, and Quanquan Gu.
\newblock Protein conformation generation via force-guided se (3) diffusion models.
\newblock In {\em International Conference on Machine Learning}, pages 56835--56859. PMLR, 2024.

\bibitem{watson2023novo}
Joseph~L Watson, David Juergens, Nathaniel~R Bennett, Brian~L Trippe, Jason Yim, Helen~E Eisenach, Woody Ahern, Andrew~J Borst, Robert~J Ragotte, Lukas~F Milles, et~al.
\newblock De novo design of protein structure and function with rfdiffusion.
\newblock {\em Nature}, 620(7976):1089--1100, 2023.

\bibitem{wu2023diffmd}
Fang Wu and Stan~Z Li.
\newblock Diffmd: a geometric diffusion model for molecular dynamics simulations.
\newblock In {\em Proceedings of the AAAI conference on artificial intelligence}, volume~37, pages 5321--5329, 2023.

\bibitem{xu2024demystify}
Yaoyao Xu, Xuxi Chen, Tong Wang, Huan He, Tianlong Chen, and Manolis Kellis.
\newblock Demystify the secret function in protein sequence via conditional diffusion models.
\newblock In {\em ICLR 2024 Workshop on Generative and Experimental Perspectives for Biomolecular Design}, 2024.

\bibitem{xu2024boosting}
Yaoyao Xu, Xinjian Zhao, Xiaozhuang Song, Benyou Wang, and Tianshu Yu.
\newblock Boosting protein language models with negative sample mining.
\newblock In {\em Joint European Conference on Machine Learning and Knowledge Discovery in Databases}, pages 199--214. Springer, 2024.

\bibitem{yabukarski2022ensemble}
Filip Yabukarski, Tzanko Doukov, Margaux~M Pinney, Justin~T Biel, James~S Fraser, and Daniel Herschlag.
\newblock Ensemble-function relationships to dissect mechanisms of enzyme catalysis.
\newblock {\em Science Advances}, 8(41):eabn7738, 2022.

\bibitem{yang2006effective}
Sichun Yang, Jos{\'e}~N Onuchic, and Herbert Levine.
\newblock Effective stochastic dynamics on a protein folding energy landscape.
\newblock {\em The Journal of chemical physics}, 125(5), 2006.

\bibitem{zheng2024predicting}
Shuxin Zheng, Jiyan He, Chang Liu, Yu~Shi, Ziheng Lu, Weitao Feng, Fusong Ju, Jiaxi Wang, Jianwei Zhu, Yaosen Min, et~al.
\newblock Predicting equilibrium distributions for molecular systems with deep learning.
\newblock {\em Nature Machine Intelligence}, 6(5):558--567, 2024.

\bibitem{zwanzig1961memory}
Robert Zwanzig.
\newblock Memory effects in irreversible thermodynamics.
\newblock {\em Physical Review}, 124(4):983, 1961.

\end{thebibliography}

\clearpage
\appendix


\section{Limitation}
\label{limitaion}
While TEMPO demonstrates promising capabilities in protein dynamics generation, several aspects warrant further exploration. The current model shows limited generalization capability to unseen proteins, especially for cases involving large conformational changes. This could be addressed through expanded training data incorporating diverse protein architectures and enhanced model designs for capturing large-scale motions. Our two-scale temporal decomposition, though effective for the demonstrated trajectory lengths, might benefit from auto-regressive resolution mechanisms similar to those successful in image generation~\cite{tian2024visual} when modeling longer timescale dynamics.

The framework's current limitation to single-protein backbone dynamics could be extended in several directions. Incorporating side-chain reconstruction and multi-molecular interactions could enable modeling of protein-ligand binding dynamics and DNA recognition processes, critical downstream applications in drug discovery and biological mechanism studies. Additionally, while existing metrics provide useful validation, developing more biologically-grounded evaluation protocols could better assess trajectory quality through direct correlation with experimental observables and functional outcomes. These directions collectively suggest rich potential for expanding both the scope and practical utility of deep learning-based protein dynamics modeling.

\section{Protein Structure Tokenization}
\label{appendix:protein_tokenization}

We represent protein structures using a combination of local reference frames and torsion angles, resulting in a rotation and translation equivariant representation. For a protein with $L$ amino acids, our tokenization procedure is as follows:

\paragraph{\textrm{SE(3)} Frame Representation.} For each residue $i$, we construct a local reference frame using the backbone atoms (N, C$\alpha$, C) following the approach similar to AlphaFold~\cite{potapenko2021highly}. Specifically:
\begin{itemize}
    \item The origin $O_i$ is placed at the C$\alpha$ atom position
    \item The x-axis $\hat{x}_i$ is aligned with the normalized C$\alpha$-N bond vector
    \item The temporary vector $\vec{v}_i$ is the normalized C$\alpha$-C bond vector
    \item The z-axis $\hat{z}_i$ is computed as $\hat{z}_i = \frac{\hat{x}_i \times \vec{v}_i}{|\hat{x}_i \times \vec{v}_i|}$
    \item The y-axis $\hat{y}_i$ completes the right-handed coordinate system: $\hat{y}_i = \hat{z}_i \times \hat{x}_i$
\end{itemize}

Each \textrm{SE(3)} frame consists of a rotation matrix $R_i \in \mathrm{SO}(3)$ and a translation vector $t_i \in \mathbb{R}^3$:
\begin{equation}
    R_i = [\hat{x}_i, \hat{y}_i, \hat{z}_i] \in \mathbb{R}^{3 \times 3}, \quad t_i = O_i \in \mathbb{R}^3
\end{equation}

To obtain a compact representation, we convert the rotation matrix to a unit quaternion $q_i \in \mathbb{S}^3$ following~\cite{shoemake1985animating}:
\begin{equation}
    q_i = \begin{bmatrix}
        q_{w,i} \\
        q_{x,i} \\
        q_{y,i} \\
        q_{z,i}
    \end{bmatrix} = \begin{bmatrix}
        \frac{1}{2}\sqrt{1 + R_i[0,0] + R_i[1,1] + R_i[2,2]} \\
        \frac{R_i[2,1] - R_i[1,2]}{4q_{w,i}} \\
        \frac{R_i[0,2] - R_i[2,0]}{4q_{w,i}} \\
        \frac{R_i[1,0] - R_i[0,1]}{4q_{w,i}}
    \end{bmatrix}
\end{equation}

Combined with the translation vector, this yields a 7-dimensional vector $[q_{w,i}, q_{x,i}, q_{y,i}, q_{z,i}, t_{x,i}, t_{y,i}, t_{z,i}]$ for each residue, resulting in a tensor of shape $[L, 7]$ for the entire protein.

\paragraph{Torsion Angle Representation.} We complement the \textrm{SE(3)} frames with the backbone torsion angles ($\phi$, $\psi$, $\omega$), which define the protein's conformation. For residue $i$, these angles are defined as:
\begin{align}
    \phi_i &= \mathrm{dihedral}(C_{i-1}, N_i, C\alpha_i, C_i) \\
    \psi_i &= \mathrm{dihedral}(N_i, C\alpha_i, C_i, N_{i+1}) \\
    \omega_i &= \mathrm{dihedral}(C\alpha_i, C_i, N_{i+1}, C\alpha_{i+1})
\end{align}

Rather than using the raw angles, we represent each angle as a 2D vector $[\sin(\theta), \cos(\theta)]$ to avoid discontinuities at $\pm\pi$, as commonly done in protein structure prediction models~\cite{watson2023novo}. This results in a 6-dimensional vector per residue:
\begin{equation}
    v_{\text{torsion},i} = [\sin(\phi_i), \cos(\phi_i), \sin(\psi_i), \cos(\psi_i), \sin(\omega_i), \cos(\omega_i)]
\end{equation}

This yields a tensor of shape $[L, 6]$ for the protein's torsion information.

\paragraph{Final Representation.} We concatenate the \textrm{SE(3)} frame and torsion angle representations to obtain a comprehensive protein structure encoding of shape $[L, 13]$, where a 13-dimensional vector represents each residue:
\begin{equation}
    v_i = [q_{w,i}, q_{x,i}, q_{y,i}, q_{z,i}, t_{x,i}, t_{y,i}, t_{z,i}, \sin(\phi_i), \cos(\phi_i), \sin(\psi_i), \cos(\psi_i), \sin(\omega_i), \cos(\omega_i)]
\end{equation}

\paragraph{Rotation and Translation Equivariance.}  
Our final protein representation—comprising unit quaternions and translation vectors for \textrm{SE(3)} frames and backbone torsion angles is equivariant to global rigid-body motions. Specifically, a global rotation $R_g \in \mathrm{SO}(3)$ and translation $t_g \in \mathbb{R}^3$ transform each local frame $(R_i, t_i)$ as $(R_g R_i, R_g t_i + t_g)$. When using unit quaternions $q_i$ to represent $R_i$, this corresponds to a left quaternion multiplication:
\[
q_i' = q_g \otimes q_i, \quad t_i' = R_g t_i + t_g
\]
where $q_g$ is the unit quaternion corresponding to $R_g$, and $\otimes$ denotes quaternion multiplication.

The torsion angles ($\phi$, $\psi$, $\omega$), represented as $[\sin(\theta), \cos(\theta)]$ pairs, are internal degrees of freedom and remain invariant under global \textrm{SE(3)} transformations. Therefore, our 13-dimensional representation $v_i$ is globally equivariant and captures both spatial orientation and internal conformation. This formulation ensures that the learned model respects 3D geometric symmetries, consistent with the principles of equivariant neural networks~\cite{satorras2021n}.

\section{Algorithm}
\label{appendix:algorithm}
Algorithm 1 describes our multi-scale training procedure, which is applied to both low-resolution and high-resolution models. The algorithm implementation differs in how we prepare the training trajectories for each scale:

\textbf{Low-resolution Training.}
For the slow-scale model ($\Delta t_s = 20$ns), we sample frames from the full trajectory at 20ns intervals. Starting from the native structure, the model learns to predict conformational changes over longer time scales, capturing major conformational transitions.

\textbf{High-resolution Training.}
For the high-resolution model ($\Delta t_f = 1$ns), we randomly sample continuous trajectory segments of $20$ns length with $1$ns intervals. Each segment is an independent training sequence, allowing the model to learn local fluctuations and fast conformational changes.
\paragraph{Notations:}
\begin{itemize}
    \item $g_t \in \mathrm{SE}(3)^L$: The frame sequence at time $t$, where each frame consists of rotation and translation components $(R, \mathbf{t})$ in \textrm{SE(3)} space
    \item $X_t \in \mathbb{R}^{L \times 13}$: Protein conformation at time $t$, 
where each residue 
includes 13 features as described in Section~\ref{appendix:protein_tokenization}.
    \item $T$: Number of timesteps used for both input and prediction windows
    \item $K$: Total number of frames in the training trajectory (differs between scales)
    \item $\sigma_{noise}$: Noise scale in our SDE formulation
\end{itemize}

 Algorithm 2 details our hierarchical inference strategy for generating complete protein dynamics trajectories. During inference, we first use the low-resolution model $f^s_\theta$ to predict conformations at coarse timesteps ($\Delta t_s = 20$ns), which captures slow collective motions. These coarse predictions then serve as anchoring points for the high-resolution model $f^f_\phi$, which fills in the intermediate frames at fine timesteps ($\Delta t_f = 1$ns) to capture local fluctuations. This hierarchical approach ensures that the generated trajectories maintain consistency between collective motions and fast local dynamics.

\begin{algorithm}[h]
\caption{Autoregressive Training of Protein Dynamics Model with Multiple Timesteps}
\LinesNumbered
\KwIn{ground truth frame sequences $\{g_t\}_{t=0}^{K-1} \in (\mathrm{SE}(3)^L)^K$, \\ 
      ground truth trajectories $\{X_t\}_{t=0}^{K-1} \in (\mathbb{R}^{L \times 13})^K$, \\
      amino acid identities $A \in \{1, \ldots, 20\}^L$, \\
      teaching force probability $p_{tf}$, \\
      number of timesteps $T$ (for both input and prediction)}
\KwOut{trained model parameters $\theta$}
\For{each training iteration}{
    \tcp{Initialize input buffers with first T frames}
    Initialize $X_{buffer} \gets \{X_i\}_{i=0}^{T-1}$ \;
    Initialize $g_{buffer} \gets \{g_i\}_{i=0}^{T-1}$ \;
    $\mathcal{L}_{total} \gets 0$ \;
    
    \For{$t = 0$ \KwTo $\lfloor K/T \rfloor - 2$}{
        $\sigma_{noise} \sim \mathcal{U}(0.01, 0.05)$ \;
        $\epsilon \sim \mathcal{N}(0, I)$ \;
        $h_{latent} \gets W_{proj}(X_{buffer} + \sigma_{noise}\epsilon)$ \;
        $s \gets \mathrm{Embed}(A)$ \;
        \For{$i = 1$ \KwTo $n_{ipa}$}{
            $h_{frame} \gets \mathrm{InvariantPointAttention}(s, g_{buffer}, \sigma_{noise})$ \;
        }
        
        \tcp{spatial processing}
        $h_{spatial} \gets h_{latent} + h_{frame} + \mathrm{Embed}(\sigma_{noise})$ \;
        \For{$j = 1$ \KwTo $n_{att}$}{
            $h_{spatial} \gets \mathrm{MultiHeadAttention}(h_{spatial})$ \;
        }
        
        \tcp{temporal processing}
        $h_{temporal} \gets h_{spatial}$ \;
        \For{$k = 1$ \KwTo $n_{GRU}$}{
            $h_{temporal} \gets \mathrm{GRU}(h_{temporal})$ \;
        }
        
        $X_{pred} \gets W_{proj}(h_{temporal})$ \;
        $g_{pred} \gets \mathrm{RigidTransformDecode}(X_{pred})$ \;

        $\mathcal{L}_{mse} \gets \|X_{pred} - X_{(t+1)T:(t+2)T}\|^2$ \;
        $\mathcal{L}_{clash} \gets \mathrm{ComputeClashScore}(X_{pred})$ \;
        $\mathcal{L}_{total} \gets \mathcal{L}_{total} + \mathcal{L}_{mse} + \lambda\mathcal{L}_{clash}$ \;
        
        \tcp{Update buffers}
        $r \sim \mathcal{U}(0,1)$ \;
        \eIf{$r > p_{tf}$}{
            $X_{buffer} \gets X_{pred}$ \;
            $g_{buffer} \gets g_{pred}$ \;
        }{
            $X_{buffer} \gets X_{(t+1)T:(t+2)T}$ \;
            $g_{buffer} \gets g_{(t+1)T:(t+2)T}$ \;
        }
    }
    Update model parameters $\theta$ using $\nabla_\theta\mathcal{L}_{total}$ \;
}
\label{appendix:algorith_1}
\end{algorithm}

\begin{algorithm}[h]
\caption{Multi-scale Inference for Protein Dynamics Generation}
\LinesNumbered
\KwIn{initial frames $\{X_i\}_{i=0}^{T-1}, \{g_i\}_{i=0}^{T-1}$, \tcp*{T frames as initial buffer} \\
      coarse timestep $\Delta t_s = 20$ns, \\
      fine timestep $\Delta t_f = 1$ns, \\
      total simulation time $T_{total}$, \\
      trained coarse-resolution model $f^s_\theta$, \\
      trained fine-resolution model $f^f_\phi$
}
\KwOut{complete trajectory $\mathcal{X} = \{X_t\}_{t=0}^{T_{total}-1}$}
$X^s_{buffer} \gets \{X_i\}_{i=0}^{T-1}$\;
$g^s_{buffer} \gets \{g_i\}_{i=0}^{T-1}$\;
$\mathcal{X} \gets \{X_i\}_{i=0}^{T-1}$\;
\For{$t = 0$ \KwTo $n_s$}{
    $X^s_{pred}, g^s_{pred} \gets f^s_\theta(X^s_{buffer}, g^s_{buffer})$\;
    $X^f_{buffer} \gets X^s_{pred}$\; 
    $g^f_{buffer} \gets  g^s_{pred}$\;
    Append $X^s_{pred}$ to $\mathcal{X}$\;
        \For{$k = 0$ \KwTo $n_f$}{
        $X^f_{pred}, g^f_{pred} \gets f^f_\phi(X^f_{buffer}, g^f_{buffer})$ \tcp{Model forward pass}
        Append $X^f_{pred}$ to $\mathcal{X}$\;
        Update $X^f_{buffer}, g^f_{buffer}$ with prediction\;
    }

    Update $X^s_{buffer}, g^s_{buffer}$ with prediction\;
}
\Return{$\mathcal{X}$}
\end{algorithm}

\section{Evaluation Metrics}
\label{appendix:metrics}

We employ a comprehensive set of metrics to evaluate both structural accuracy and dynamic properties of generated protein ensembles. Our evaluation framework follows established protocols in protein ensemble generation~\cite{jing2024alphafold}, adapted for trajectory-based assessment.

\subsection{Structural Flexibility Metrics}

\textbf{Pairwise RMSD.} For each ensemble, we quantify overall conformational diversity as the average C$\alpha$-RMSD between all pairs of conformations. This metric captures the range of conformational space explored by the ensemble. We report both the absolute values and Pearson correlation coefficient between predicted and ground truth pairwise RMSD distributions to assess whether our model captures the relative flexibility patterns across different proteins.

\textbf{Root Mean Square Fluctuation (RMSF).} To assess local flexibility, we compute the RMSF for each residue, measuring the standard deviation of atomic positions across the ensemble after optimal alignment. The Pearson correlation between predicted and ground truth RMSF profiles indicates how well the model captures residue-level flexibility patterns.

\subsection{Distribution Accuracy Metrics}

\textbf{Root Mean Wasserstein Distance (RMWD).} To generalize all-atom RMSD to ensemble comparison, we define the root mean Wasserstein distance between ensembles $\mathcal{X}$ and $\mathcal{Y}$ as:
\begin{equation}
\text{RMWD}(\mathcal{X}, \mathcal{Y}) = \sqrt{\frac{1}{N} \sum_{i=1}^N \mathcal{W}_2^2\left(\mathcal{N}[\mathcal{X}_i], \mathcal{N}[\mathcal{Y}_i]\right)}
\end{equation}
where $\mathcal{N}[\mathcal{X}_i]$ denotes a 3D Gaussian fitted to the positional distribution of the $i$th atom in ensemble $\mathcal{X}$. This metric reduces to standard RMSD for single structures and provides a distributional measure of positional accuracy.

\textbf{Principal Component Analysis.} To evaluate collective motions, we project the joint distribution of C$\alpha$ positions onto principal components computed from the MD ensemble. We measure: (1) the 2-Wasserstein distance between predicted and ground truth ensembles in the PC space (MD PCA W2), and (2) the cosine similarity between the dominant principal components. A similarity $>0.5$ indicates successful capture of the dominant collective motion. We report the percentage of test proteins achieving this threshold (\% PC-sim $>$ 0.5).

\subsection{Dynamic Property Metrics}

\textbf{Contact Dynamics.} We analyze intermittent contacts to assess if the model captures thermal fluctuations. For each ensemble, we identify: (1) \emph{weak contacts} - C$\alpha$ pairs ($<$8Å) in the native structure that dissociate in $>10\%$ of ensemble structures, and (2) \emph{transient contacts} - pairs not in contact in the native structure but associate in $>10\%$ of ensemble structures. We compute Jaccard similarity between predicted and ground truth contact sets.

\textbf{Trajectory Accuracy.} For trajectory generation methods, we measure the backbone RMSD error: $\text{Error}_{\text{frame}} = |\text{RMSD}_{\text{pred}} - \text{RMSD}_{\text{gt}}|$, where RMSD is computed relative to the native structure. This metric quantifies the model's ability to accurately capture the magnitude of conformational changes over time.

\textbf{Clash Ratio.} We compute the proportion of generated conformations containing steric clashes, defined as backbone atom pairs from different residues with distance $<1.2$Å. This metric validates the physical plausibility of generated structures.

All metrics are computed on backbone atoms (N, C$\alpha$, C, O) after optimal rigid-body alignment to the native structure, unless otherwise specified. We report median values across the test set to ensure robustness to outliers.

\section{Trajectory Visualization}
\label{appendix:visualization}
\begin{figure}[t]
    \centering
    \includegraphics[width=\textwidth]{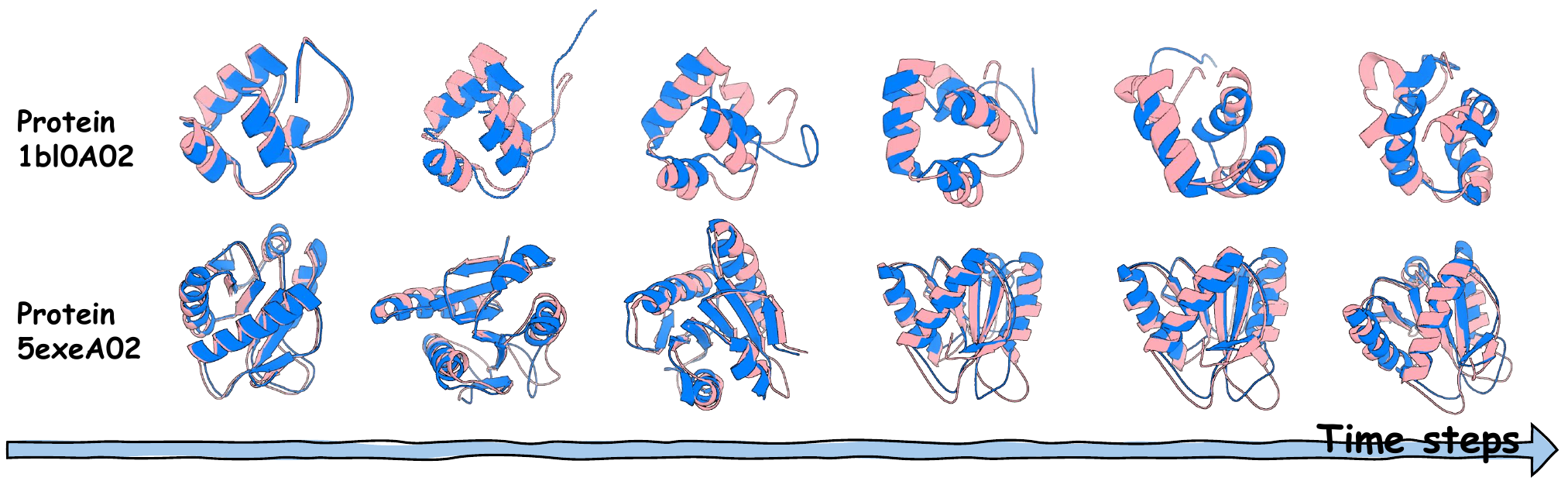}
    \caption{Visualization of protein conformational changes over time for two representative proteins (1bl0A02 and 5exeA02). Each row shows the temporal evolution of one protein. MD ground truth trajectories are shown in blue, while TEMPO-generated trajectories are shown in pink, demonstrating the close structural alignment between generated and reference structures throughout the trajectory.}
    \label{fig:appendix_viz}
\end{figure}

To provide an intuitive visualization of our generation results, we present trajectory comparisons for two representative test proteins. Figure~\ref{fig:appendix_viz} shows snapshots of the trajectories at $80$ns intervals, where the ground truth MD conformations are shown in blue and TEMPO-generated conformations are shown in pink. These visualizations demonstrate the close structural alignment between our generated conformations and the MD reference states.

\section{Thermodynamic Accuracy Analysis}
\label{appendix:free_energy}

Following BioEMU's established evaluation methodology~\cite{lewis2024scalable}, we provide quantitative assessment of thermodynamic accuracy through free energy difference analysis. This analysis validates that our generated trajectories not only capture structural features but also preserve the underlying thermodynamic properties of protein dynamics.

\textbf{Free Energy Difference Computation.} We compute free energy differences by extracting reaction coordinates, specifically the fraction of native contacts, to calculate folding probabilities ($p_{\text{fold}}$). The free energy differences are then calculated as:
\begin{equation}
\Delta G = -kT \cdot \log\left(\frac{p_{\text{fold}}}{1-p_{\text{fold}}}\right)
\end{equation}
where $k$ is the Boltzmann constant and $T$ is the temperature. We measure the free energy error as $\Delta\Delta G = \Delta G_{\text{ground truth}} - \Delta G_{\text{predicted}}$.

\textbf{Results.} TEMPO achieves an average $\Delta\Delta G$ of 0.67 kcal/mol on the mdCATH test set, demonstrating good agreement with reference MD simulations within the acceptable range for biological applications (typically $<$1-2 kcal/mol). This result indicates that our method not only generates structurally accurate conformations but also preserves the thermodynamic properties of protein folding and unfolding processes.

\textbf{Free Energy Profiles.} Following BioEMU's evaluation framework, we analyze our generated trajectories through multiple perspectives:
\begin{itemize}
    \item \textbf{1D Free Energy Profiles:} We construct free energy profiles using three key reaction coordinates: RMSD from native structure, radius of gyration, and fraction of native contacts. Our generated trajectories exhibit high similarity to ground truth MD simulations across all three coordinates.
    \item \textbf{2D Free Energy Surfaces:} Two-dimensional free energy surface plots constructed from combinations of reaction coordinates show that TEMPO captures the essential features of the conformational landscape, including energy minima locations and barrier heights.
    \item \textbf{Time Series Analysis:} The fraction of native contacts time series from our generated trajectories closely matches the temporal evolution patterns observed in ground truth simulations, validating our model's ability to capture dynamic processes.
\end{itemize}

\begin{figure}[h]
\centering
\includegraphics[width=0.9\textwidth]{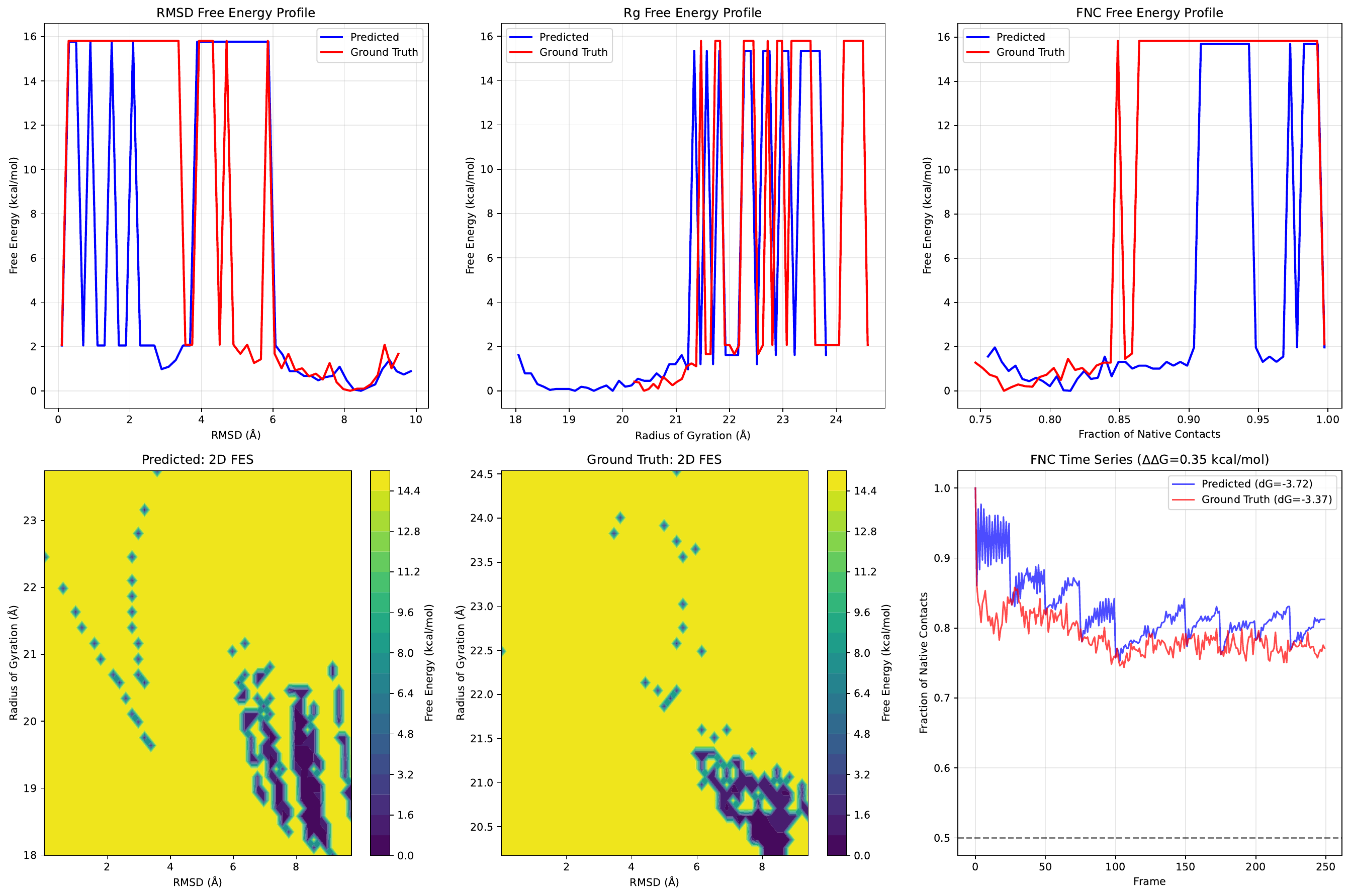}
\vspace{0.5em}
\includegraphics[width=0.9\textwidth]{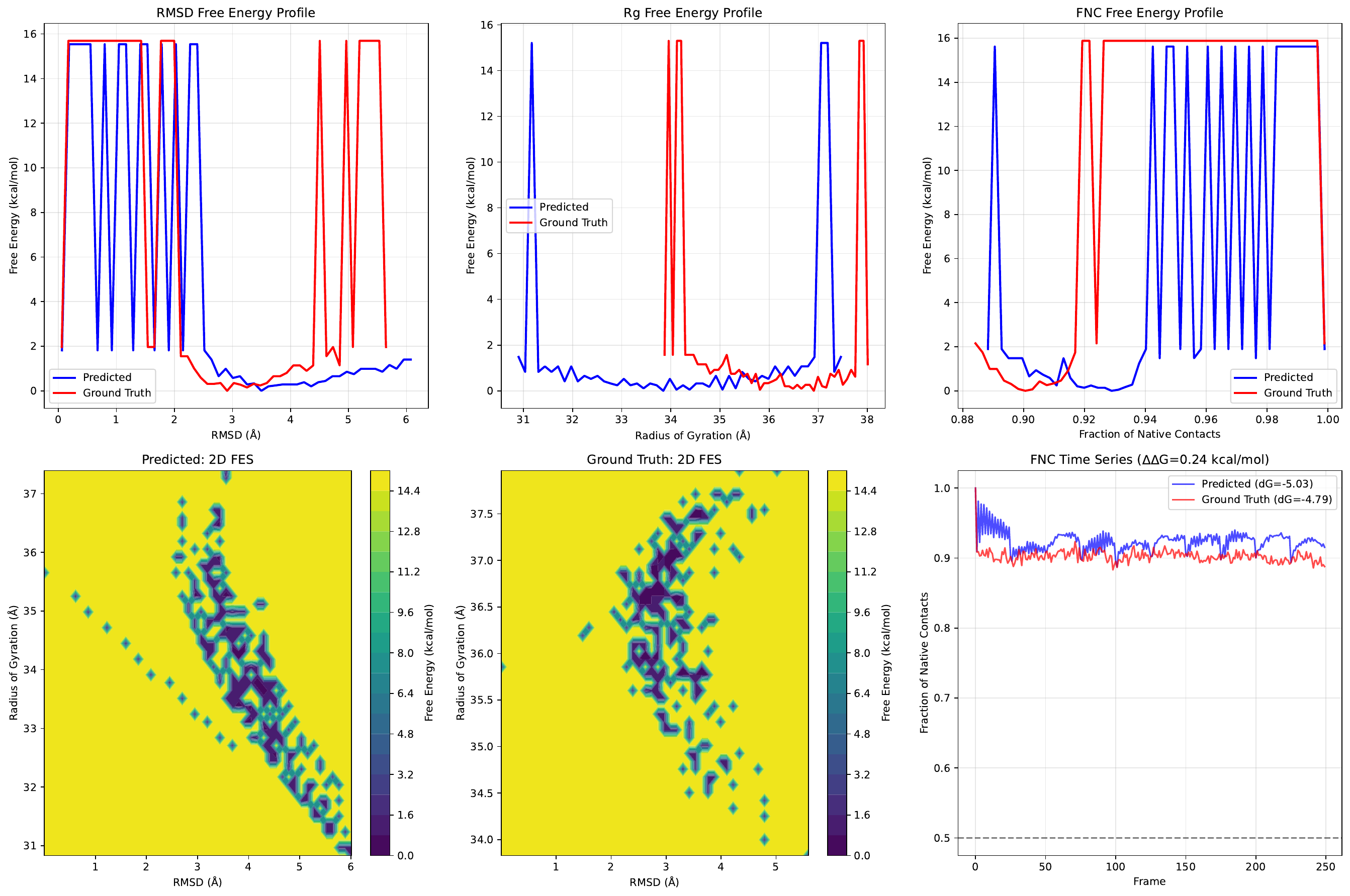}
\vspace{0.5em}
\caption{Thermodynamic accuracy analysis for two representative proteins (7p46\_A and 7asg\_A) selected from the ATLAS test set. For each protein, we show: 1D free energy profiles along three reaction coordinates (RMSD, radius of gyration, and fraction of native contacts), 2D free energy surfaces, and time series of fraction of native contacts. }
\label{fig:thermodynamic_analysis_atlas}
\end{figure}

\begin{figure}[h]
\centering

\includegraphics[width=0.9\textwidth]{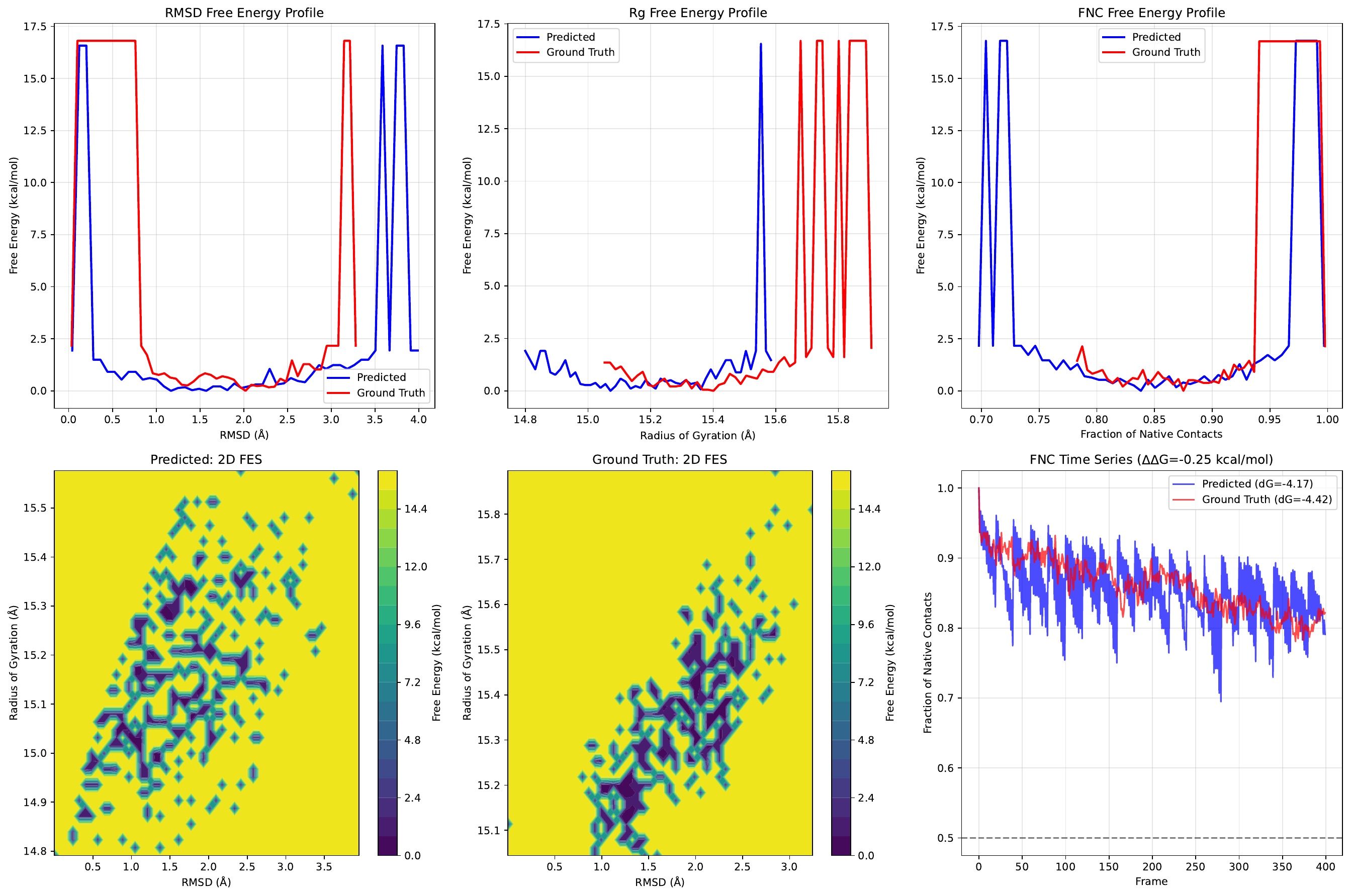}
\vspace{0.5em}
\includegraphics[width=0.9\textwidth]{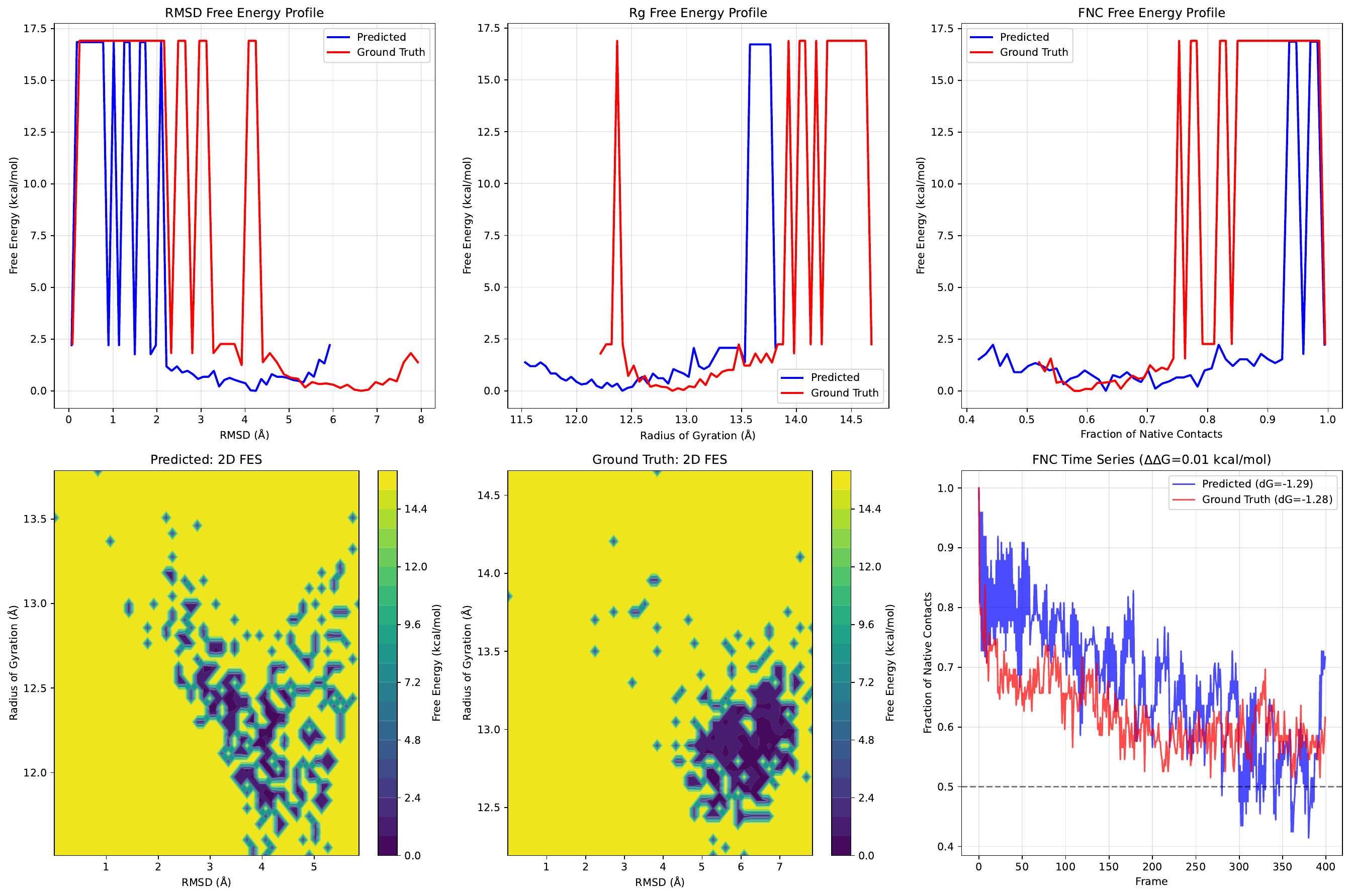}
\vspace{0.5em}
\caption{Thermodynamic accuracy analysis for two representative proteins (2ijd101 and 1rl0A02) selected from the mdCATH test set.}
\label{fig:thermodynamic_analysis_mdcath}
\end{figure}

These quantitative thermodynamic evaluations complement our structural metrics and demonstrate that TEMPO generates trajectories that are not only geometrically accurate but also thermodynamically consistent with reference MD simulations (see Figure~\ref{fig:thermodynamic_analysis_atlas} and Figure~\ref{fig:thermodynamic_analysis_mdcath} for visualizations).

\section{Ablation Study: Multi-scale vs Single-scale Generation}
\label{appendix:ablation}

While autoregressive methods typically suffer from error accumulation during sequential generation, our hierarchical multi-scale design mitigates this issue by anchoring fine-scale dynamics to coarse-scale predictions. To demonstrate the effectiveness of our multi-scale architecture, we compare TEMPO's full hierarchical framework against a single-scale baseline that directly generates 400 frames without multi-scale guidance on mdCATH. 

Table~\ref{tab:ablation} shows that the multi-scale approach significantly outperforms single-scale generation across all metrics. The multi-scale model achieves substantially lower RMSD error (1.78Å vs 8.62Å) after 400 frames, demonstrating controlled error propagation that maintains trajectory stability over extended generation periods. Furthermore, the multi-scale design preserves conformational diversity (pairwise RMSD of 2.78Å vs 7.46Å compared to ground truth 3.26Å) and structural flexibility patterns (Pearson correlation of 0.77 vs 0.14 for RMSD, 0.67 vs 0.15 for RMSF), while the single-scale approach shows severe degradation in capturing protein dynamics. These results validate that our hierarchical decomposition is crucial for generating physically realistic long protein trajectories.

\begin{table}[ht]
\centering
\caption{Ablation study comparing multi-scale TEMPO with single-scale baseline on mdCATH. Ground truth values are shown in parentheses where applicable. The single-scale model directly generates 400 frames without hierarchical guidance.}
\label{tab:ablation}
\begin{tabular}{lcc}
\toprule
\textbf{Metrics} & \textbf{TEMPO (Multi-scale)} & \textbf{TEMPO (Single-scale)} \\
\midrule
Pairwise RMSD (= 3.26) & \textbf{2.78} & 7.46 \\
Pairwise RMSD $r$ $\uparrow$ & \textbf{0.77} & 0.14 \\
All-atom RMSF (= 1.64) & \textbf{1.60} & 4.27 \\
Global RMSF $r$ $\uparrow$ & \textbf{0.67} & 0.15 \\
Root mean W2 $\downarrow$ & \textbf{4.21} & 8.27 \\
MD PCA W2 $\downarrow$ & \textbf{2.33} & 2.53 \\
\% PC-sim $>$ 0.5 $\uparrow$ & \textbf{7.81} & 3.12 \\
Weak contacts $J$ $\uparrow$ & \textbf{0.43} & 0.23 \\
Trans. contacts $J$ $\uparrow$ & \textbf{0.20} & 0.09 \\
RMSD Error $\downarrow$ & \textbf{1.78} & 8.62 \\
\bottomrule
\end{tabular}
\end{table}

\section{Training vs Test Performance Analysis}
\label{appendix:train_test}

We acknowledge that capturing all distribution modes represents a fundamental challenge in protein dynamics modeling. While our primary objective focuses on generating realistic conformational transitions rather than perfect mode coverage, we recognize the importance of understanding the performance gap between training and test scenarios.

\textbf{Performance Gap Analysis.} Table~\ref{tab:train_test} compares TEMPO's performance on training and test sets of mdCATH, where the training set results are evaluated on 60 randomly selected proteins from the training data. The training set demonstrates significantly better mode coverage accuracy, with 46.7\% of principal components achieving similarity greater than 0.5 compared to 7.81\% on the test set. Similarly, the training set shows improved performance across most metrics, including lower MD PCA Wasserstein distance (1.31 vs 2.33) and better contact prediction accuracy (0.53 vs 0.43 for weak contacts, 0.37 vs 0.20 for transient contacts).

This performance gap reflects the inherent complexity of generalizing to novel protein architectures in mdCATH, where most proteins contain distinct energy basins and unique conformational landscapes. The challenge of mode coverage is not unique to our method - even extensively pretrained methods like BioEMU show significant free energy surface (FES) deviations as shown in Appendix Figure~\ref{fig:appendix_fes}. The difficulty stems from the fundamental nature of protein dynamics: each protein has its own characteristic energy landscape shaped by its unique sequence and structure, making it challenging for models to perfectly capture the full conformational distribution of unseen proteins.

Despite the performance gap, our test set results still demonstrate strong capability in capturing essential protein dynamics, as evidenced by high structural flexibility correlations (Pearson $r = 0.77$ for pairwise RMSD, $r = 0.67$ for RMSF) and accurate trajectory generation (RMSD error of 1.78Å). The stronger training set performance validates our model's capacity to learn complex protein dynamics when sufficient data is available, suggesting that expanded training data incorporating diverse protein architectures could further improve generalization. 

\begin{table}[h]
\centering
\caption{Comparison of TEMPO performance on training set versus test set. Ground truth values are shown in parentheses. The training set demonstrates significantly better mode coverage.}
\label{tab:train_test}
\begin{tabular}{lcc}
\toprule
\textbf{Metrics} & \textbf{Train Sample} & \textbf{Test} \\
\midrule
Pairwise RMSD & 3.06 (2.51) & 2.78 (3.26) \\
Pairwise RMSD $r$ $\uparrow$ & \textbf{0.79} & 0.77 \\
All-atom RMSF & 1.16 (1.31) & 1.60 (1.64) \\
Global RMSF $r$ $\uparrow$ & \textbf{0.78} & 0.67 \\
Root mean W2 $\downarrow$ & \textbf{3.42} & 4.21 \\
MD PCA W2 $\downarrow$ & \textbf{1.31} & 2.33 \\
\% PC-sim $>$ 0.5 $\uparrow$ & \textbf{46.7} & 7.81 \\
Weak contacts $J$ $\uparrow$ & \textbf{0.53} & 0.43 \\
Trans. contacts $J$ $\uparrow$ & \textbf{0.37} & 0.20 \\
RMSD Error $\downarrow$ & \textbf{1.51} & 1.78 \\
\bottomrule
\end{tabular}
\end{table}

\section{Additional Up-sampling Analysis}
\label{appendix:upsampling}
\begin{figure}[h]
    \begin{minipage}{0.24\textwidth}
        \includegraphics[width=\linewidth]{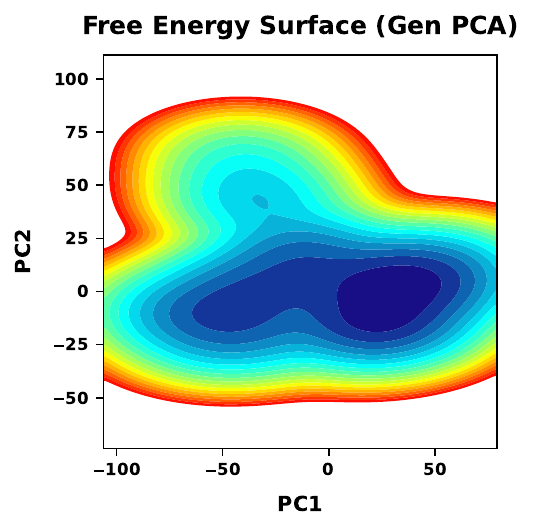}
    \end{minipage}
    \begin{minipage}{0.24\textwidth}
        \includegraphics[width=\linewidth]{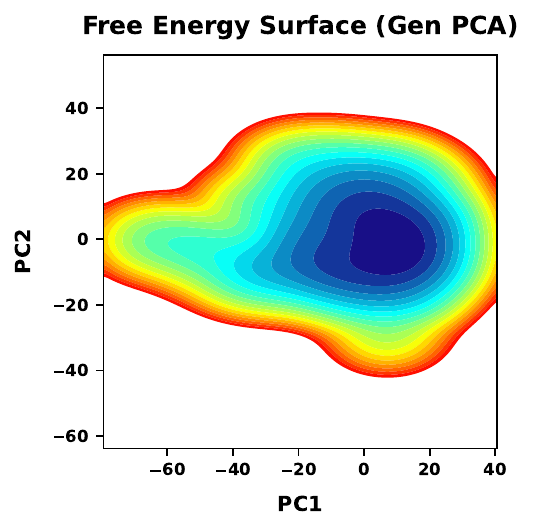}
    \end{minipage}
    \begin{minipage}{0.24\textwidth}
        \includegraphics[width=\linewidth]{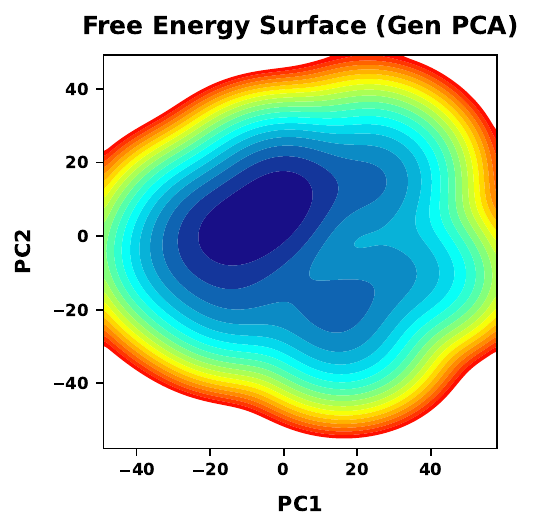}
    \end{minipage}
    \begin{minipage}{0.24\textwidth}
        \includegraphics[width=\linewidth]{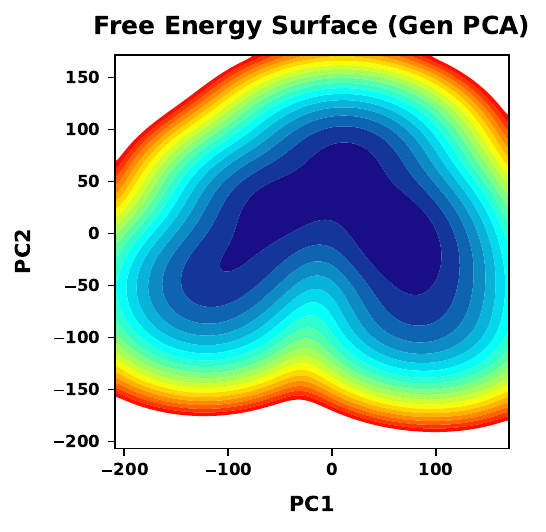}
    \end{minipage}
    
    \vspace{2mm}
    \begin{minipage}{0.24\textwidth}
        \includegraphics[width=\linewidth]{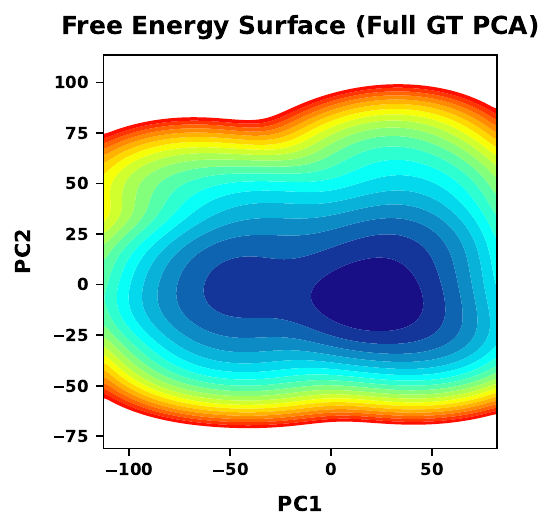}
    \end{minipage}
    \begin{minipage}{0.24\textwidth}
        \includegraphics[width=\linewidth]{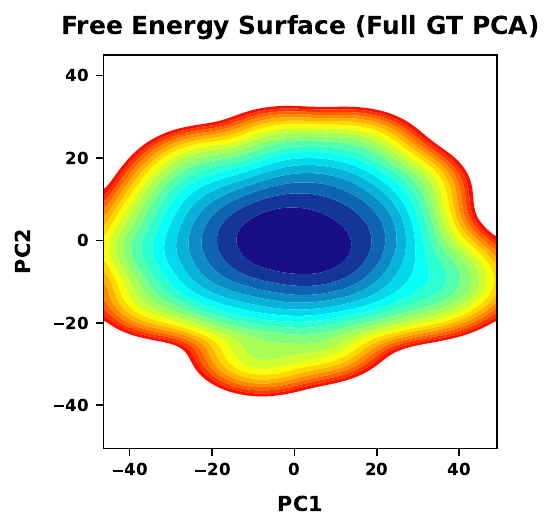}
    \end{minipage}
    \begin{minipage}{0.24\textwidth}
        \includegraphics[width=\linewidth]{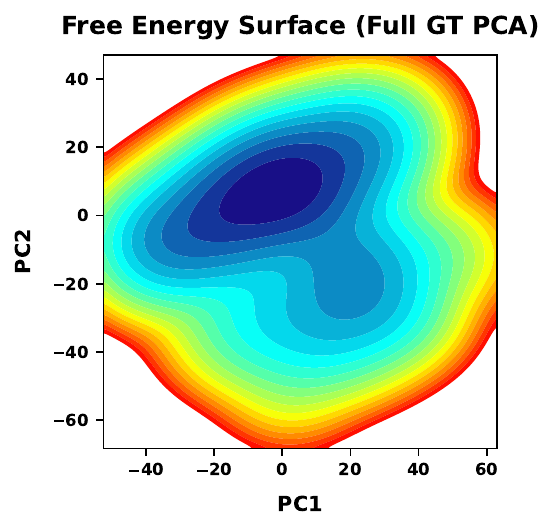}
    \end{minipage}
    \begin{minipage}{0.24\textwidth}
        \includegraphics[width=\linewidth]{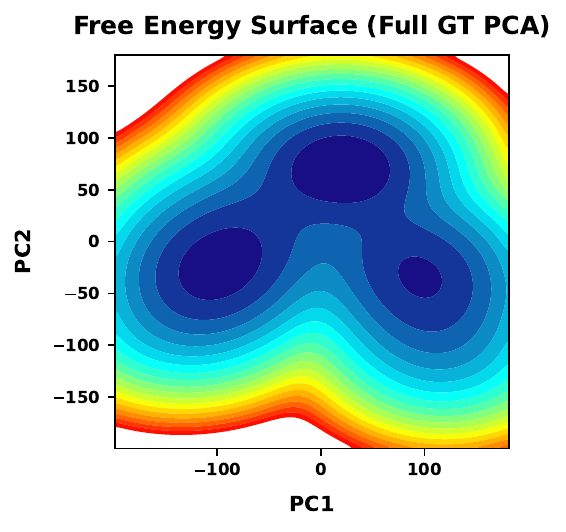}
    \end{minipage}
    
    \vspace{2mm}
    \begin{minipage}{0.24\textwidth}
        \includegraphics[width=\linewidth]{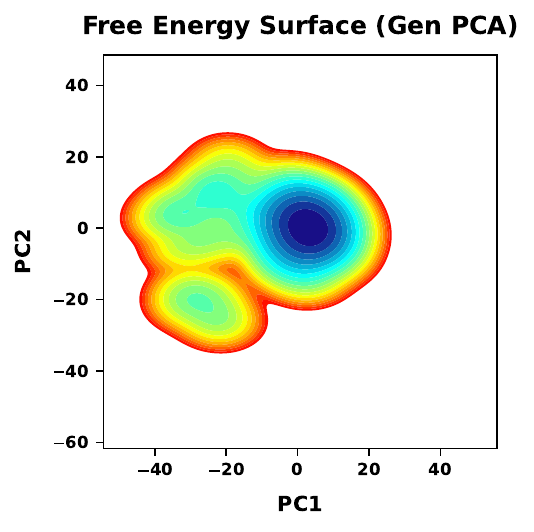}
    \end{minipage}
    \begin{minipage}{0.24\textwidth}
        \includegraphics[width=\linewidth]{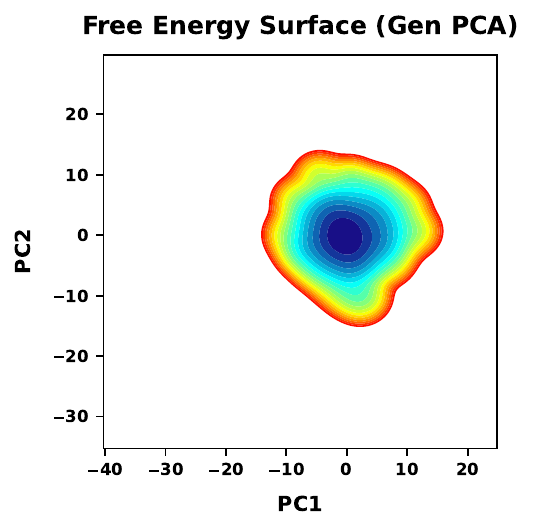}
    \end{minipage}
    \begin{minipage}{0.24\textwidth}
        \includegraphics[width=\linewidth]{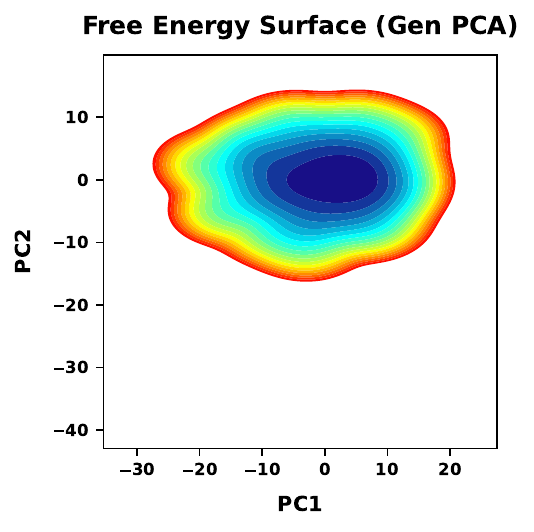}
    \end{minipage}
    \begin{minipage}{0.24\textwidth}
        \includegraphics[width=\linewidth]{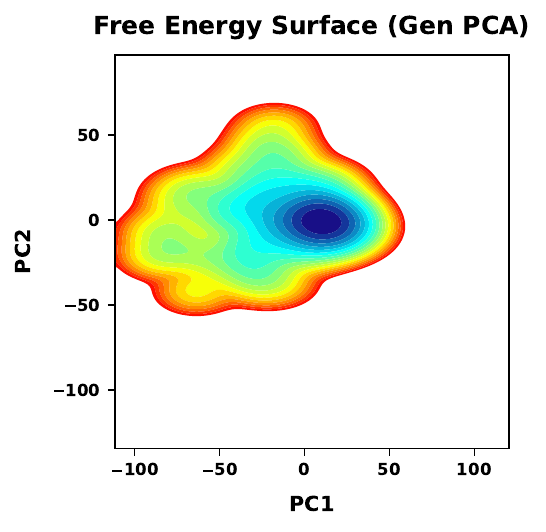}
    \end{minipage}
    
    \caption{FES plots of four randomly selected test proteins. The FES is computed from: (top) trajectories generated by TEMPO performing up-sampling tasks, (middle) ground truth molecular dynamics simulations, and (bottom) trajectories produced by the baseline model MDGen, performing simulation tasks. Each column represents a distinct test protein (from left to right: 1rl0A02, 1x4tA01, 1zpdA02, and 2ndpA00).}
    \label{app:upsampling_comparison}
\end{figure}

Additional up-sampling experiments across multiple test proteins further validate our high-resolution model's ability to generate full protein dynamics from ground truth low-resolution protein conformations. 
Figure~\ref{app:upsampling_comparison} shows the FES contour plots for four randomly selected proteins (1rl0A02, 1x4tA01, 1zpdA02, and 2ndpA00) computed from their backbone conformations.


\section{Additional State Transition Analysis}
\label{appendix:transition}
\begin{figure}[t]
    \centering
    \includegraphics[width=0.245\textwidth]{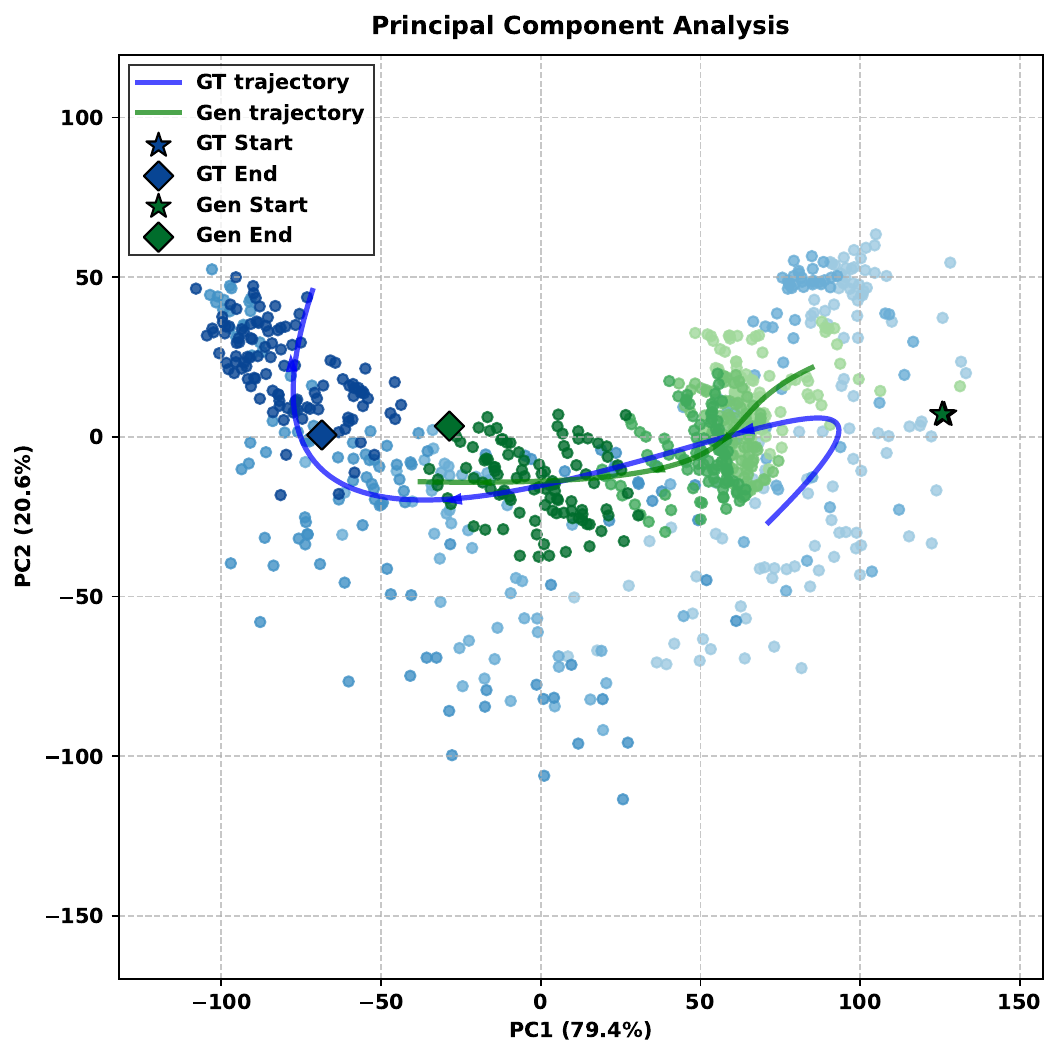}\includegraphics[width=0.245\textwidth]{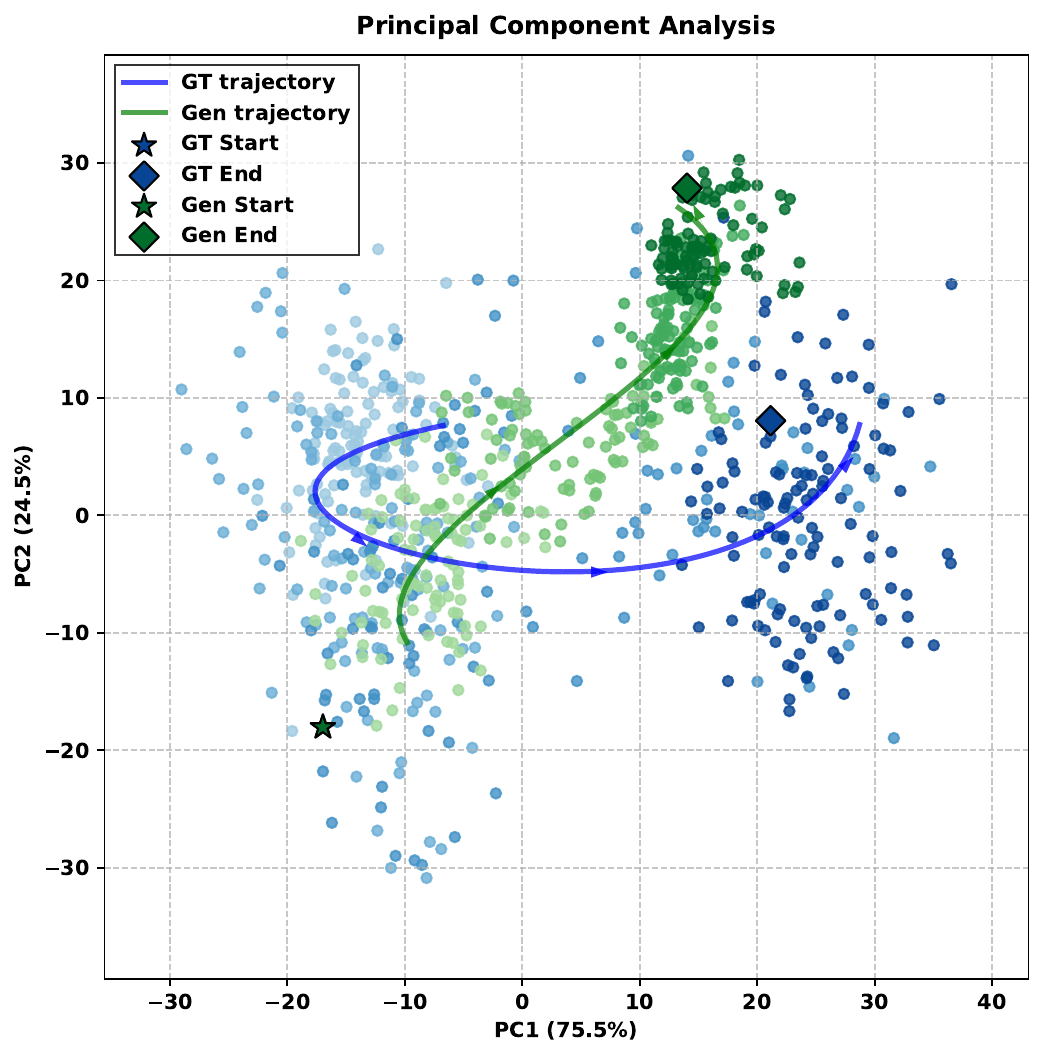}
    \includegraphics[width=0.245\textwidth]{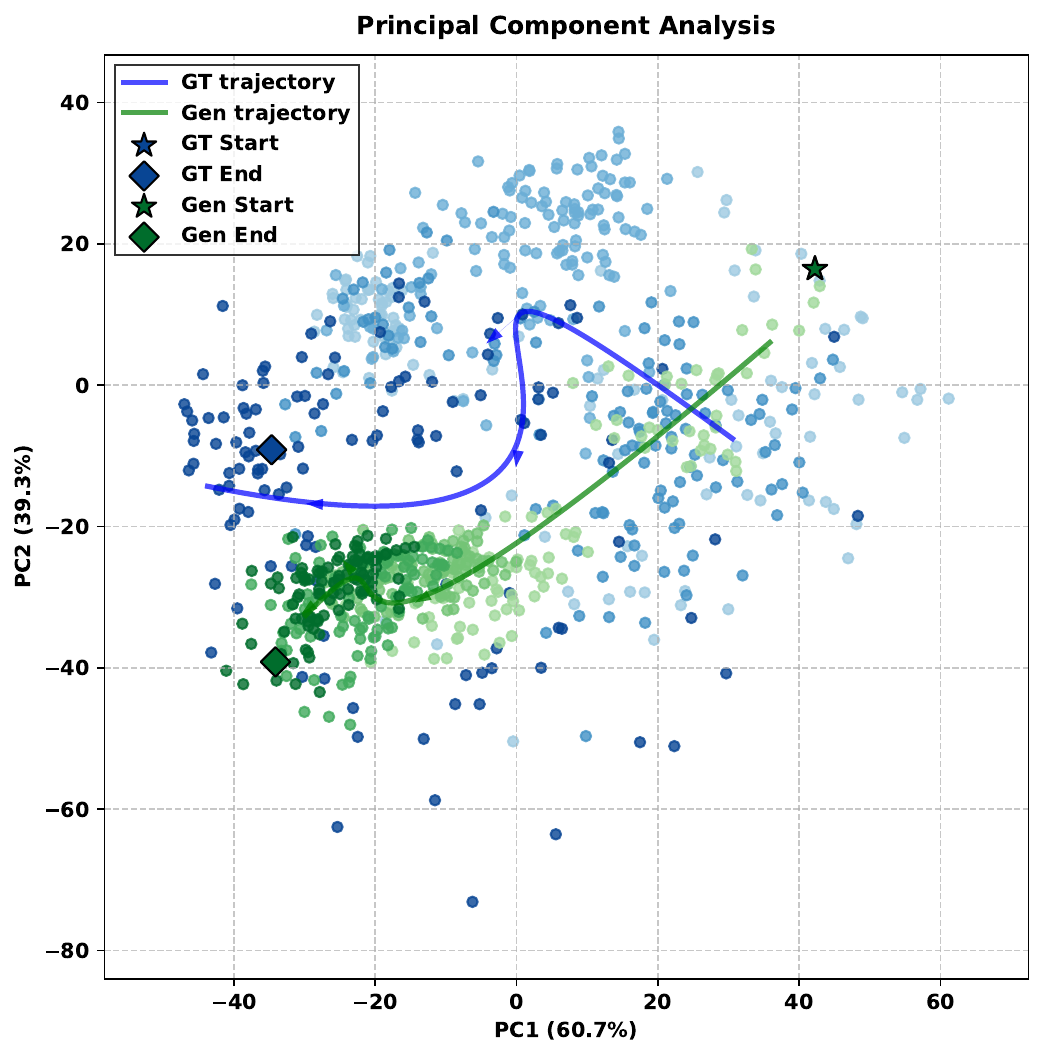}
    \includegraphics[width=0.245\textwidth]{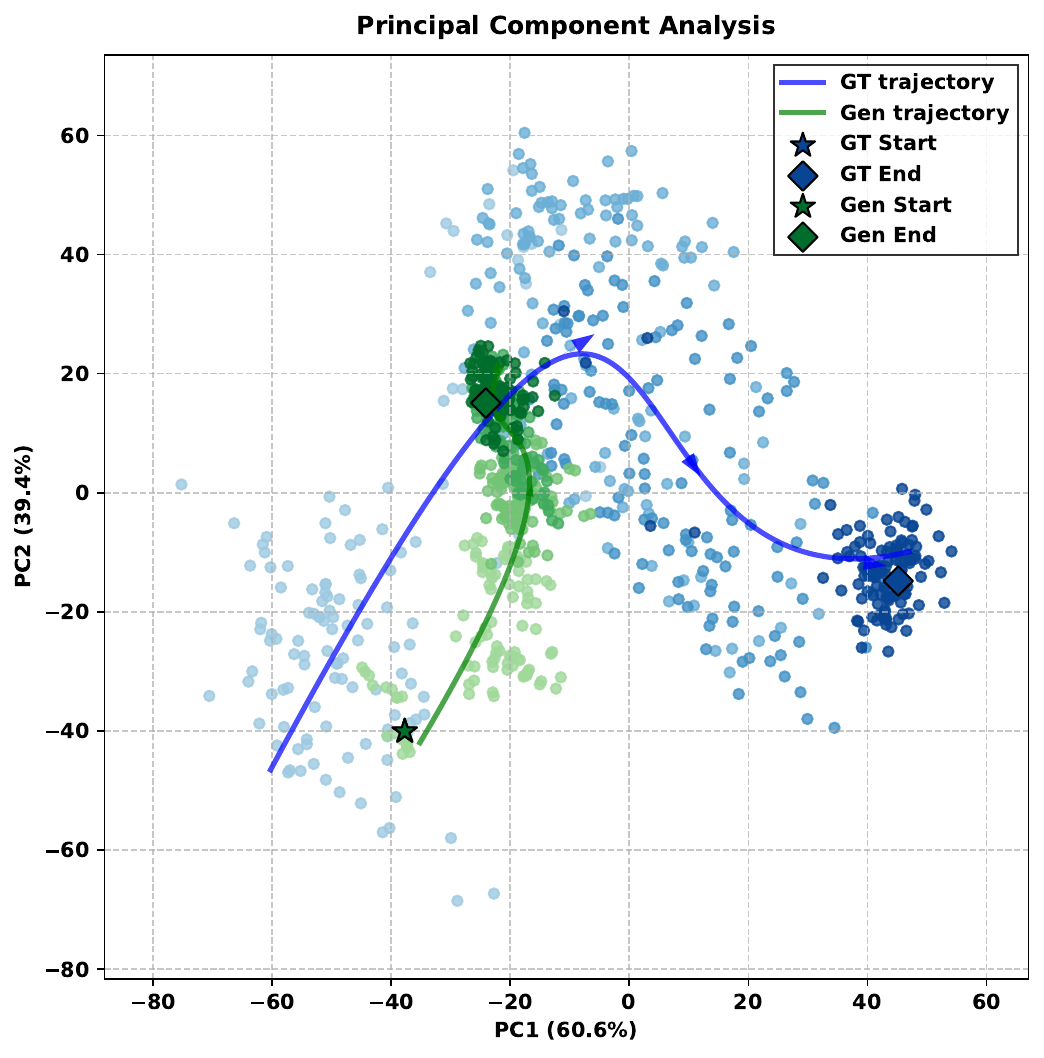}
    \includegraphics[width=0.245\textwidth]{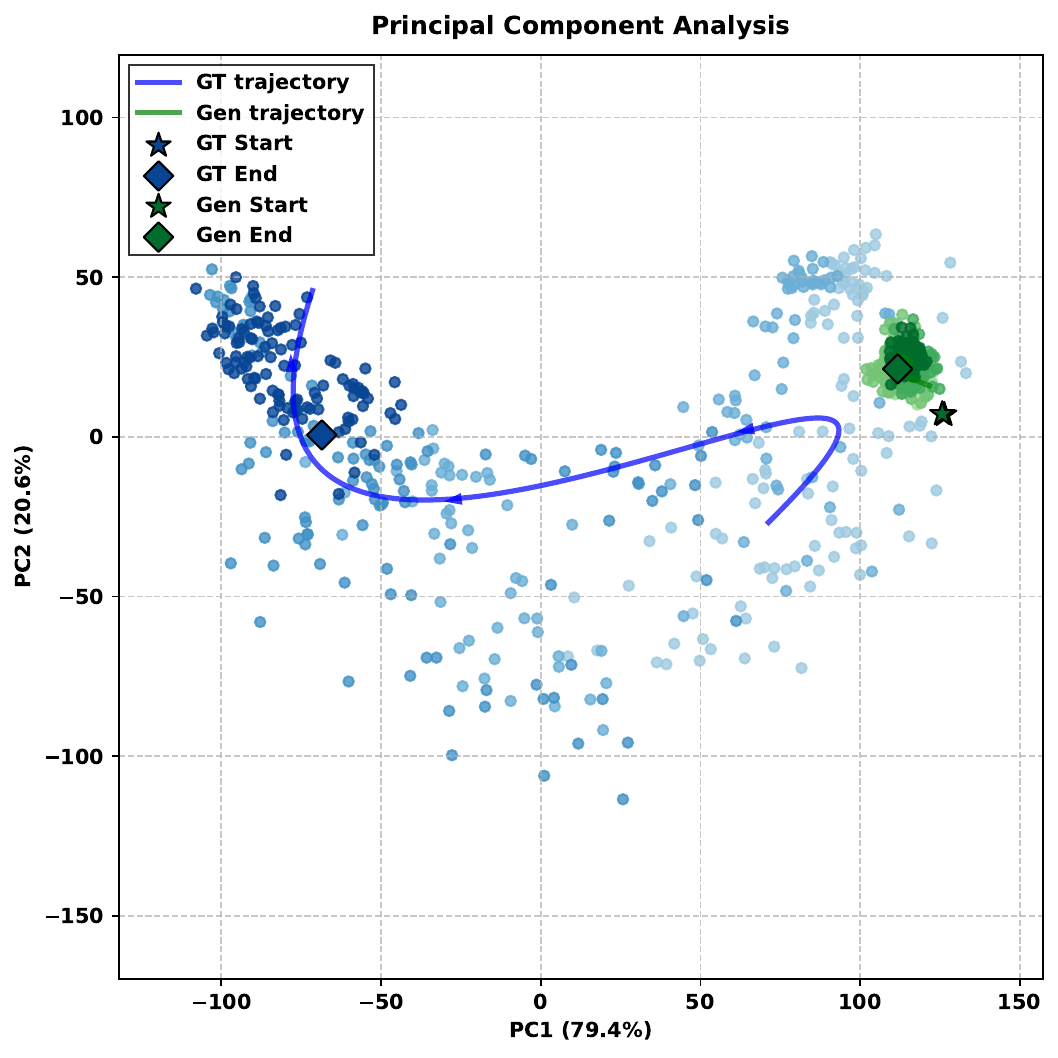}\includegraphics[width=0.245\textwidth]{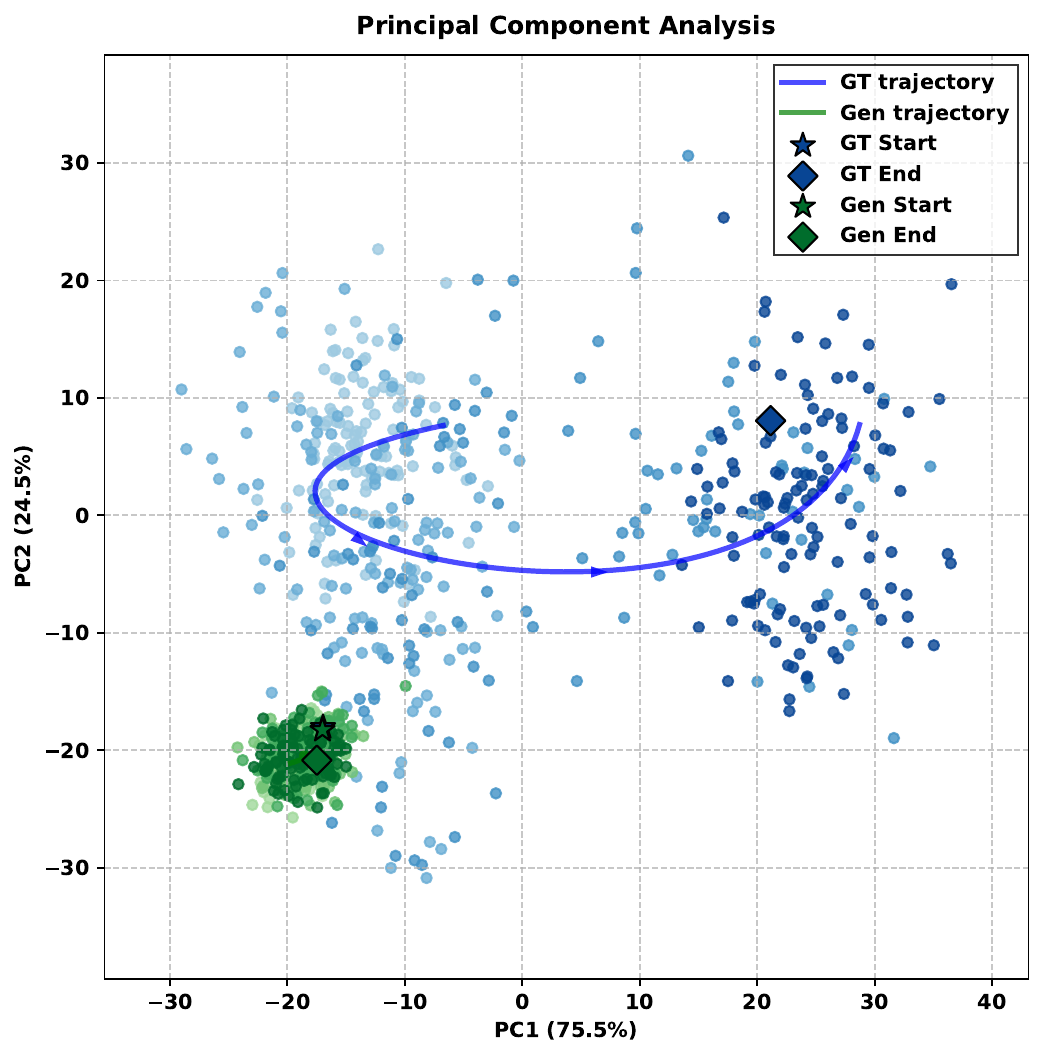}
    \includegraphics[width=0.245\textwidth]{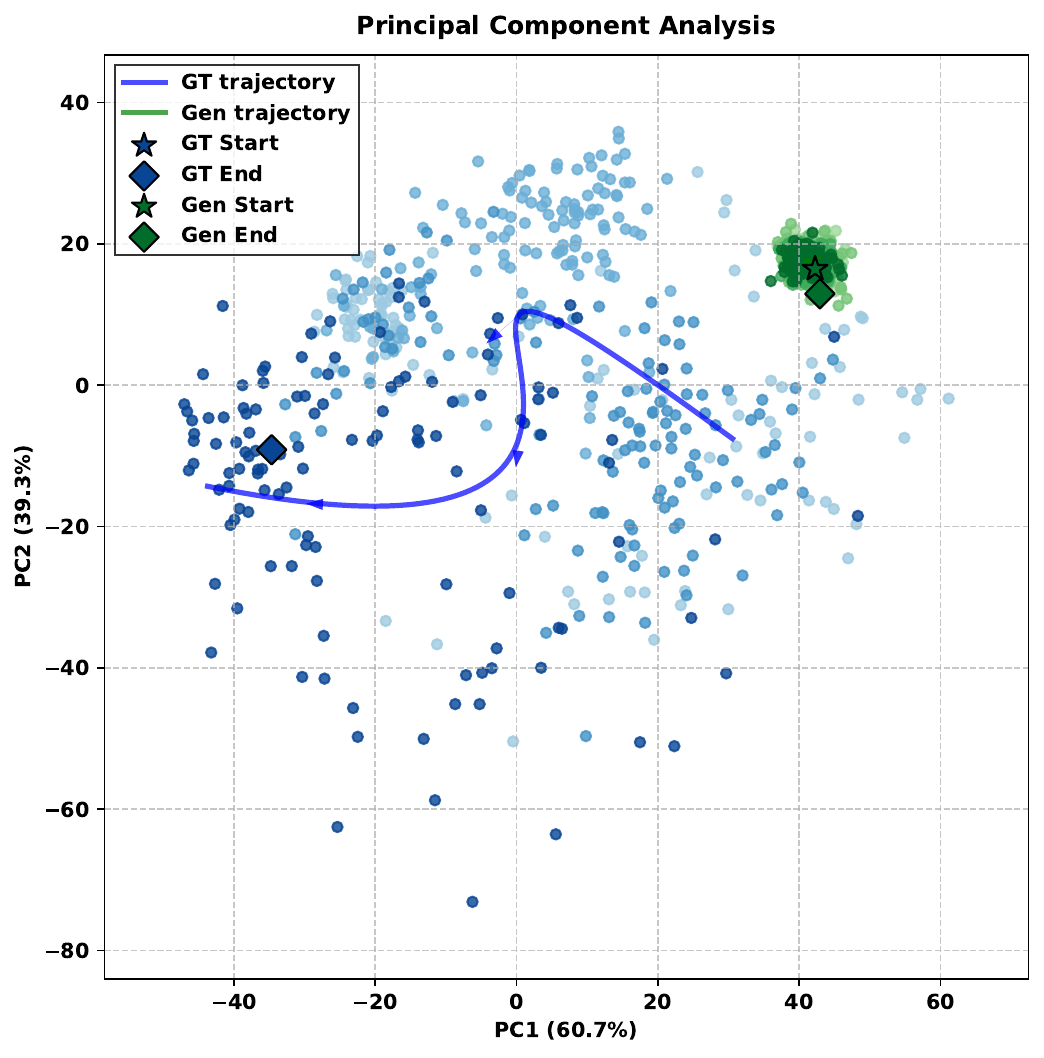}
    \includegraphics[width=0.245\textwidth]{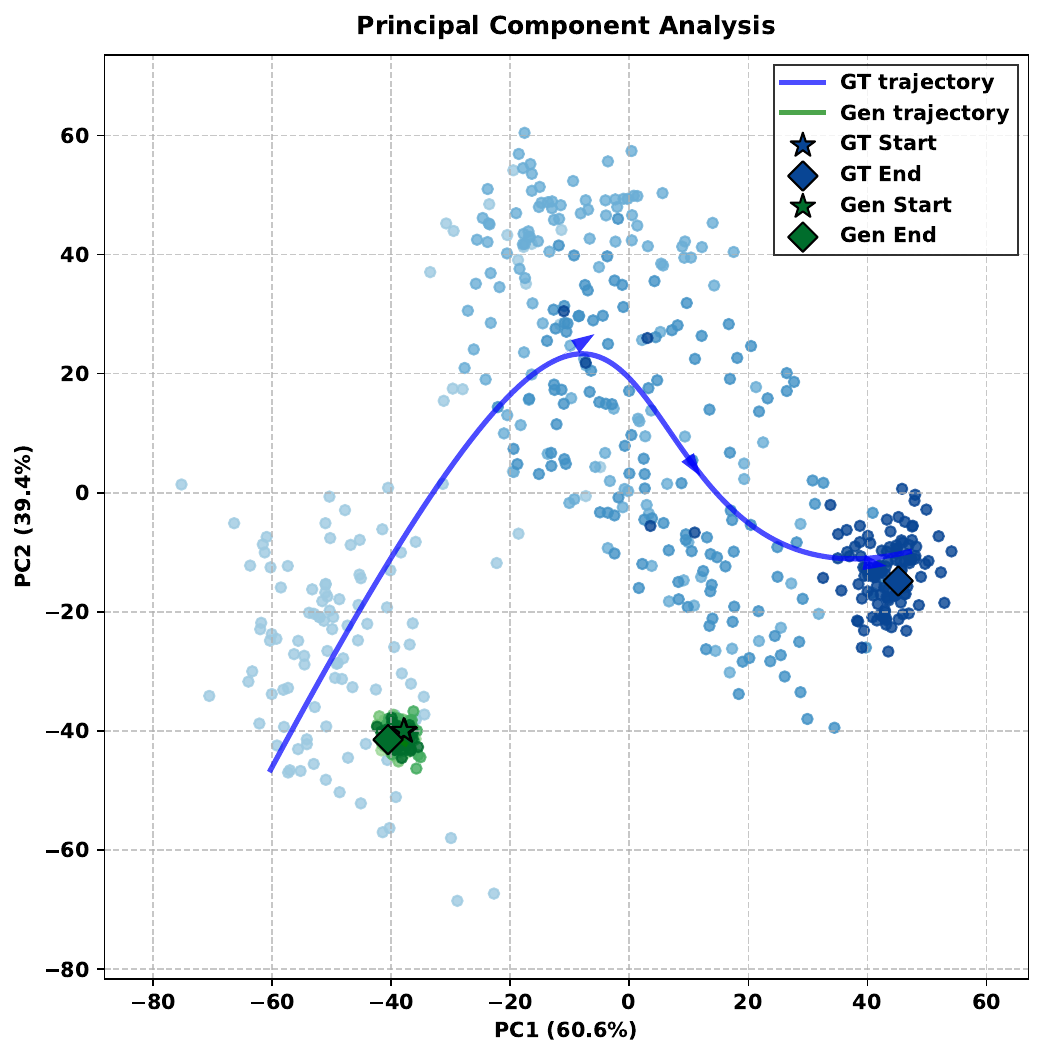}
    \caption{Comparison of conformational transitions in PC space between TEMPO and MDGen baseline (bottom). Ground truth MD trajectories are shown in blue, while generated trajectories are in green. The polynomial fitting curves highlight the temporal evolution of conformational changes.}\label{appdix:state_transition}
\end{figure}

To further validate our model's capability in capturing protein conformational transitions, we present additional state transition analyses on representative proteins from our test set. Figure~\ref{appdix:state_transition}, each subplot compares the conformational trajectories generated by TEMPO and MDGen with MD simulations in the space of the first two principal components. Consistent with our observations in the main text, these additional examples demonstrate TEMPO's robust ability to generate physically meaningful transition pathways. MDGen typically exhibits more clustered sampling patterns, fails to capture the continuous nature of conformational transitions. These results further support our framework's advantage in modeling the temporal dependencies inherent in protein dynamics.

\section{Additional Free Energy Surface Analysis}
\label{appendix:fes}
\begin{figure}[htbp]
    \centering
     \includegraphics[width=0.245\textwidth]{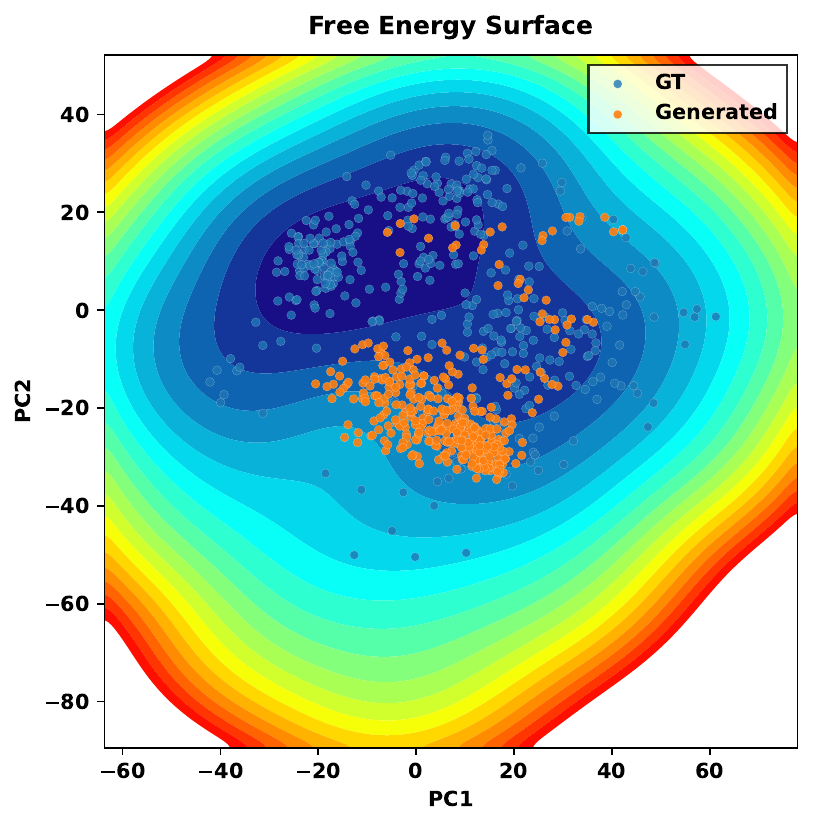}
    \includegraphics[width=0.245\textwidth]{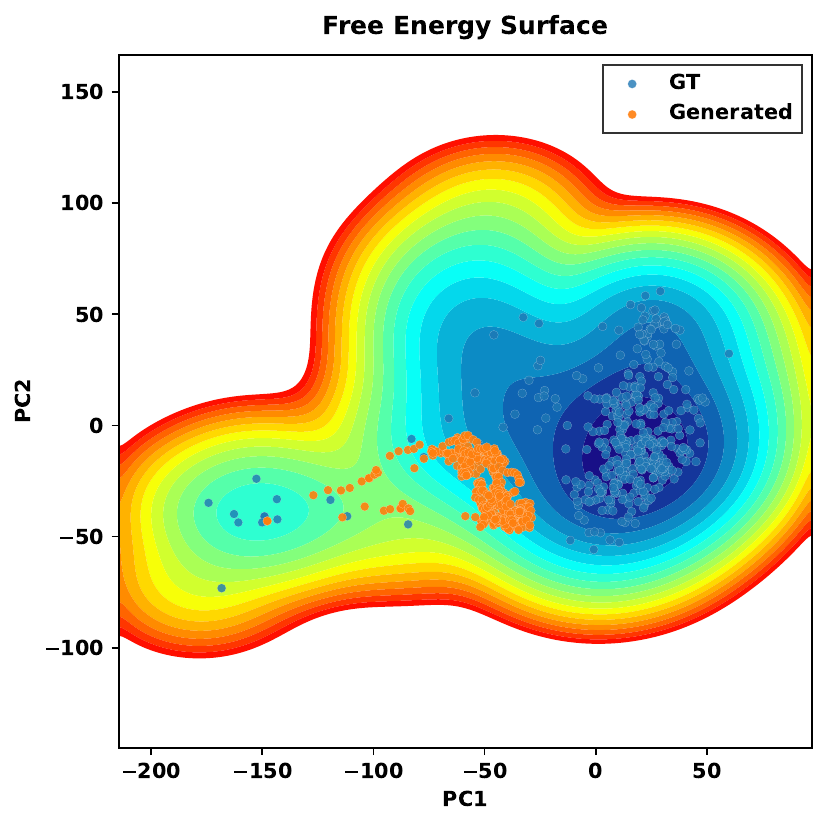}
    \includegraphics[width=0.245\textwidth]{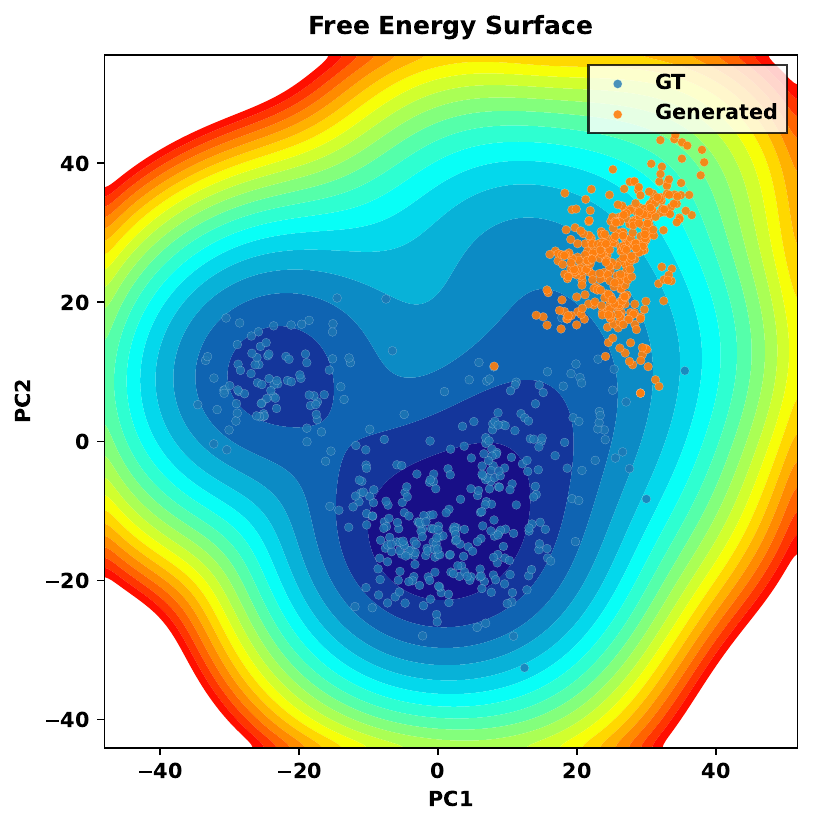}
    \includegraphics[width=0.245\textwidth]{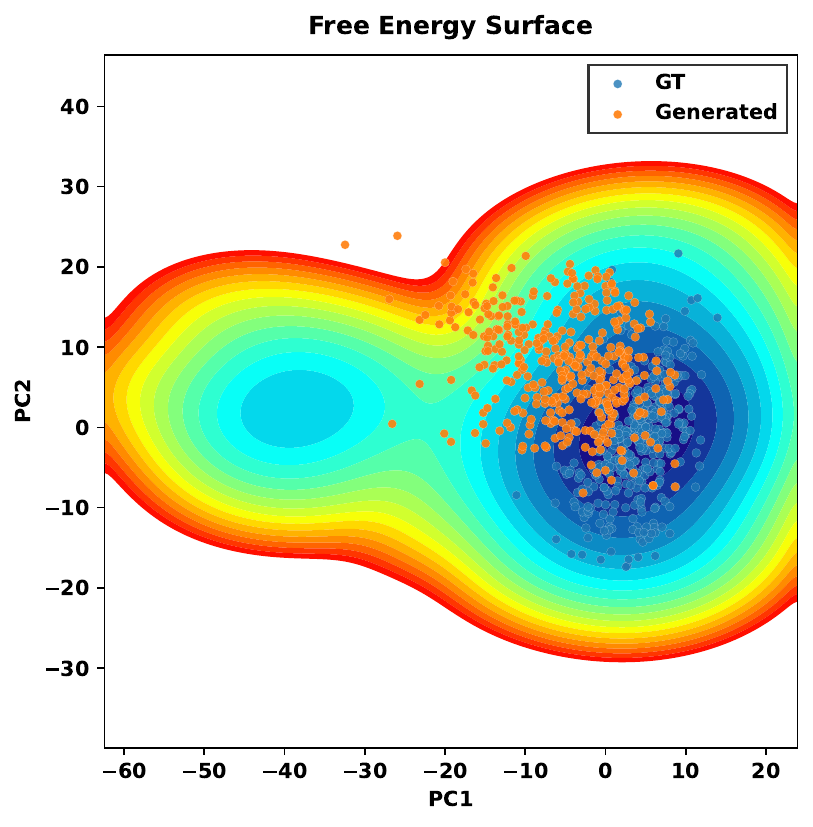}

    \includegraphics[width=0.245\textwidth]{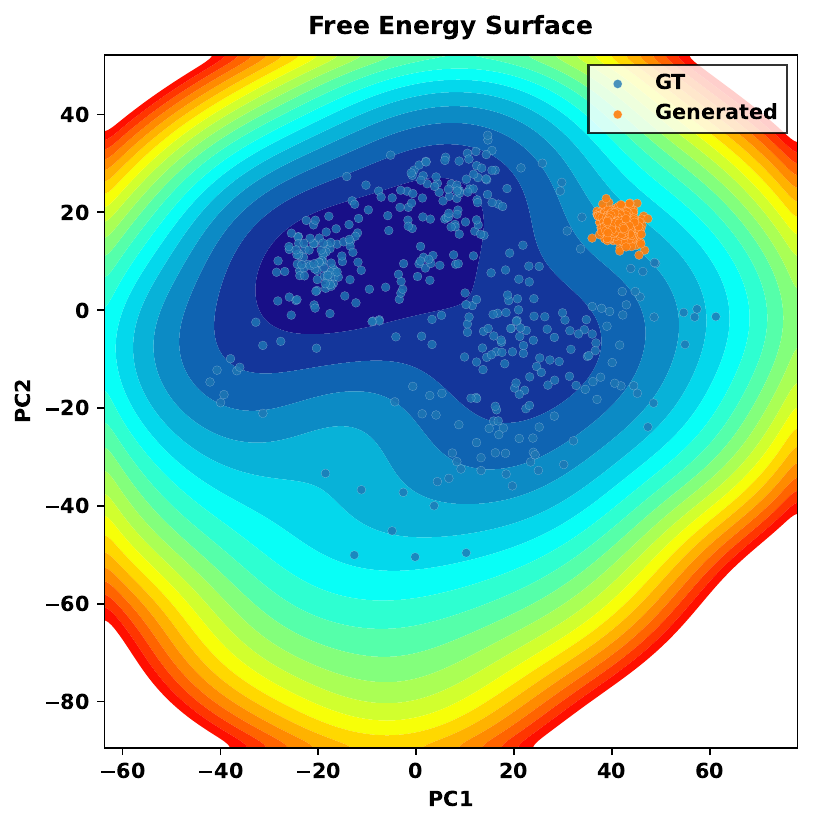}
    \includegraphics[width=0.245\textwidth]{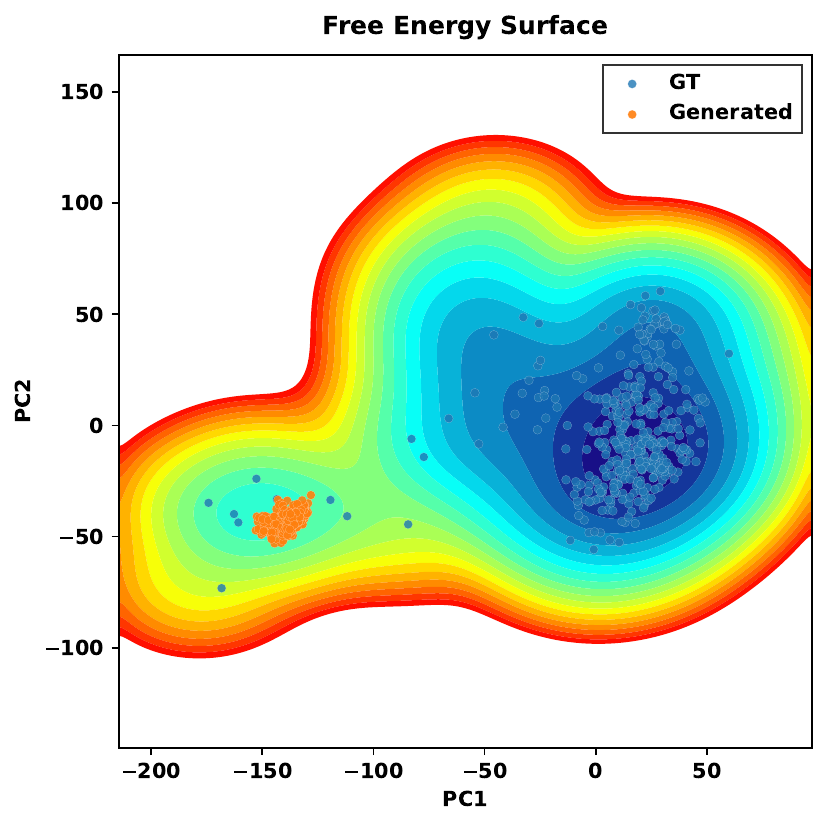}
    \includegraphics[width=0.245\textwidth]{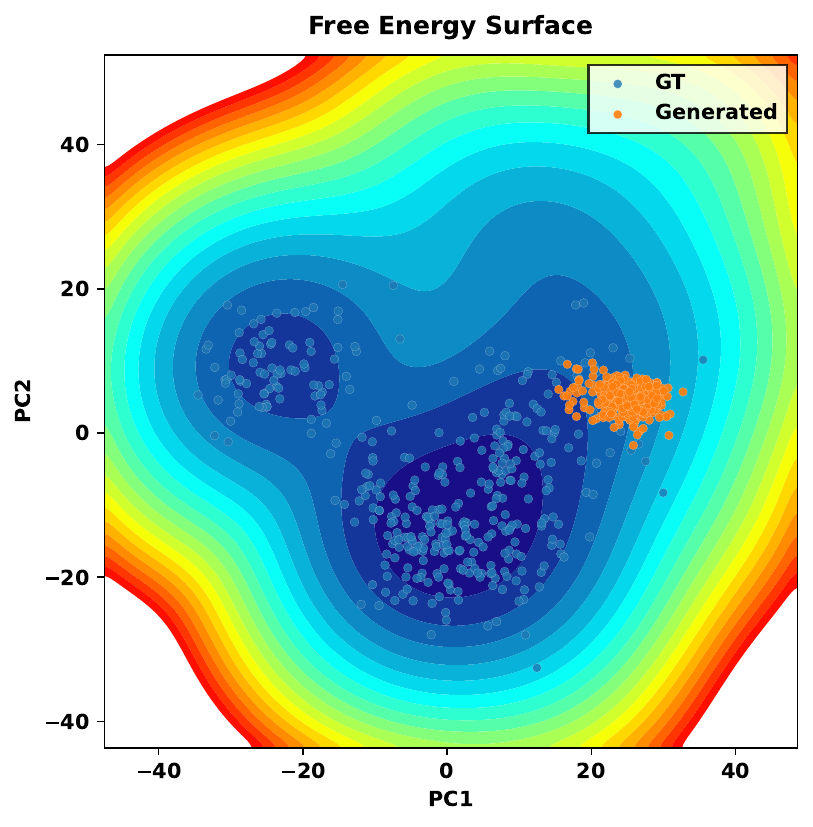}
    \includegraphics[width=0.245\textwidth]{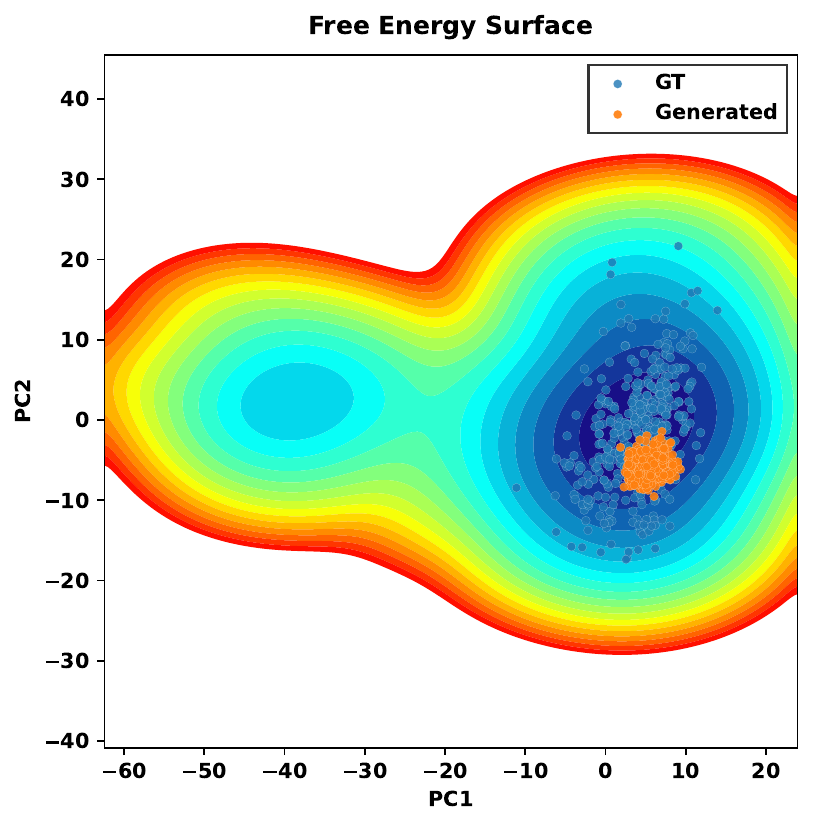}

    \includegraphics[width=0.245\textwidth]{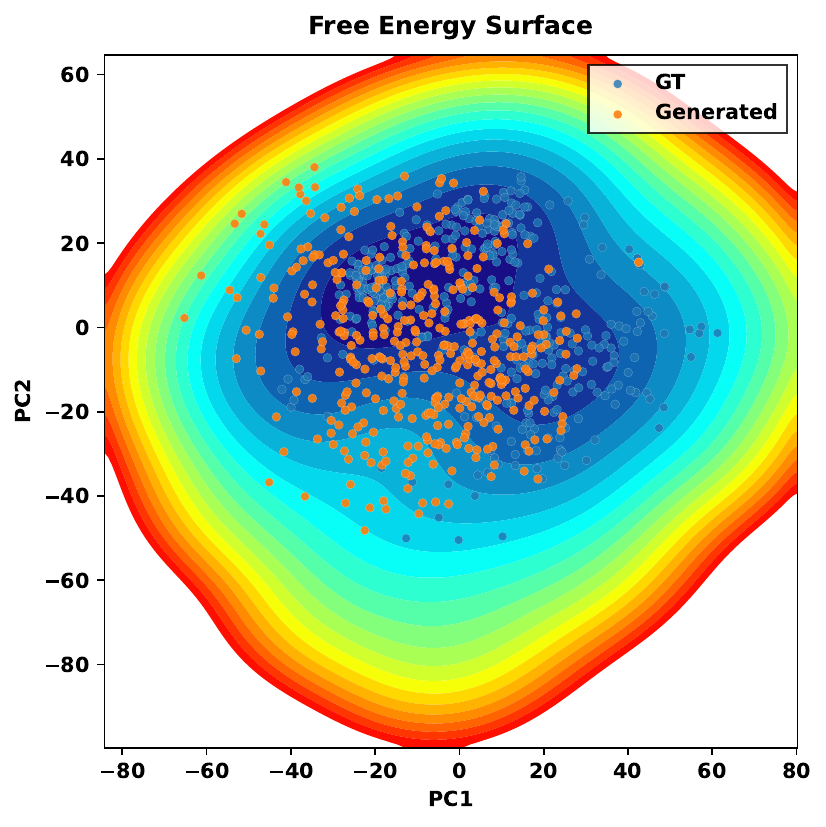}
    \includegraphics[width=0.245\textwidth]{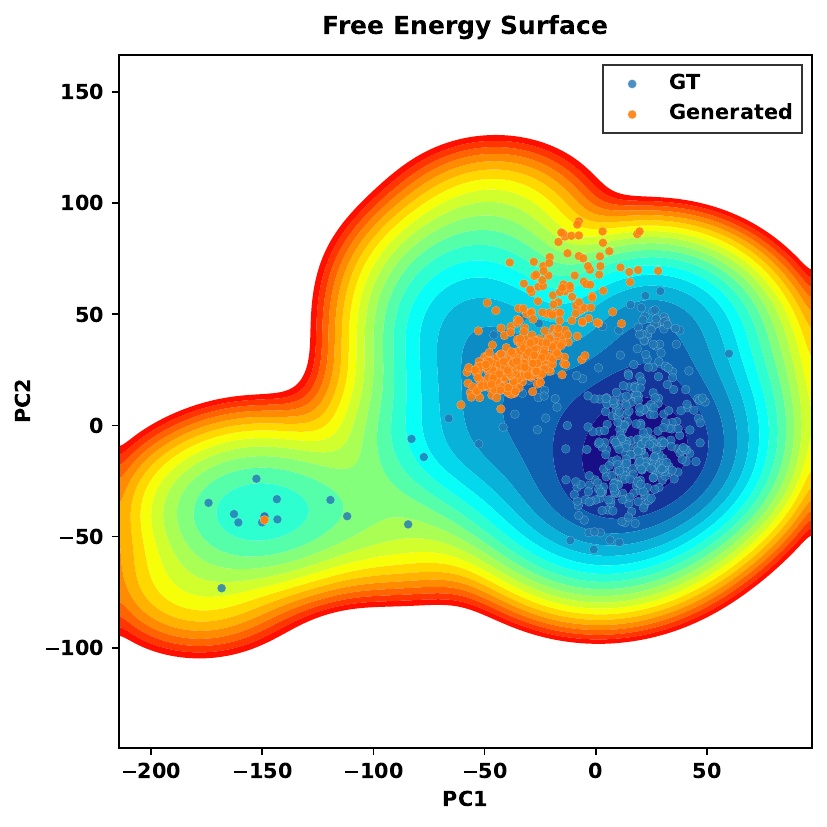}
    \includegraphics[width=0.245\textwidth]{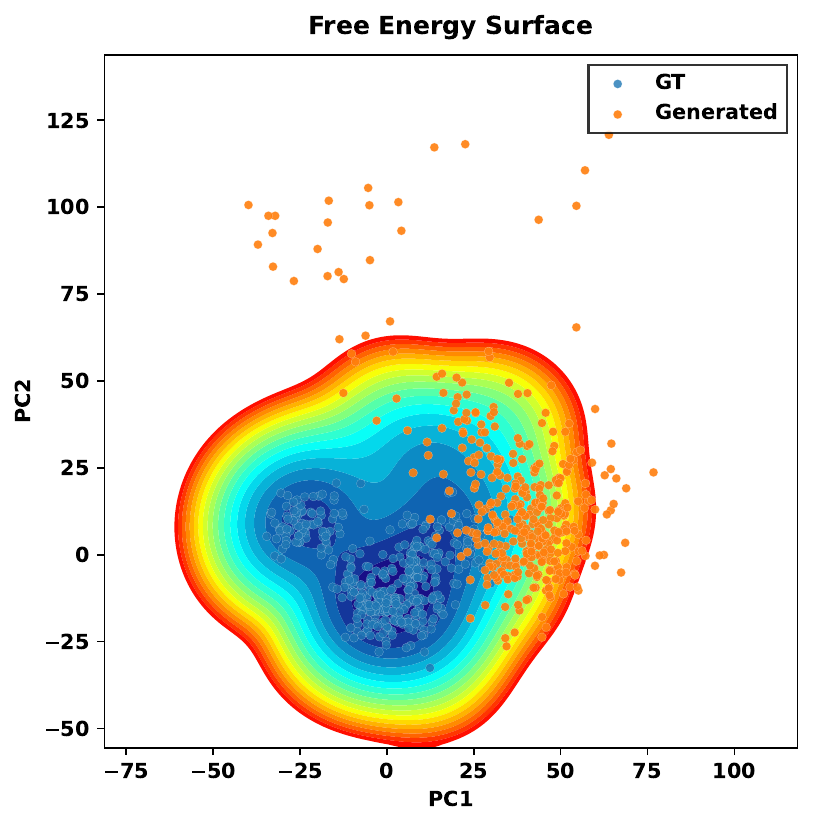}
    \includegraphics[width=0.245\textwidth]{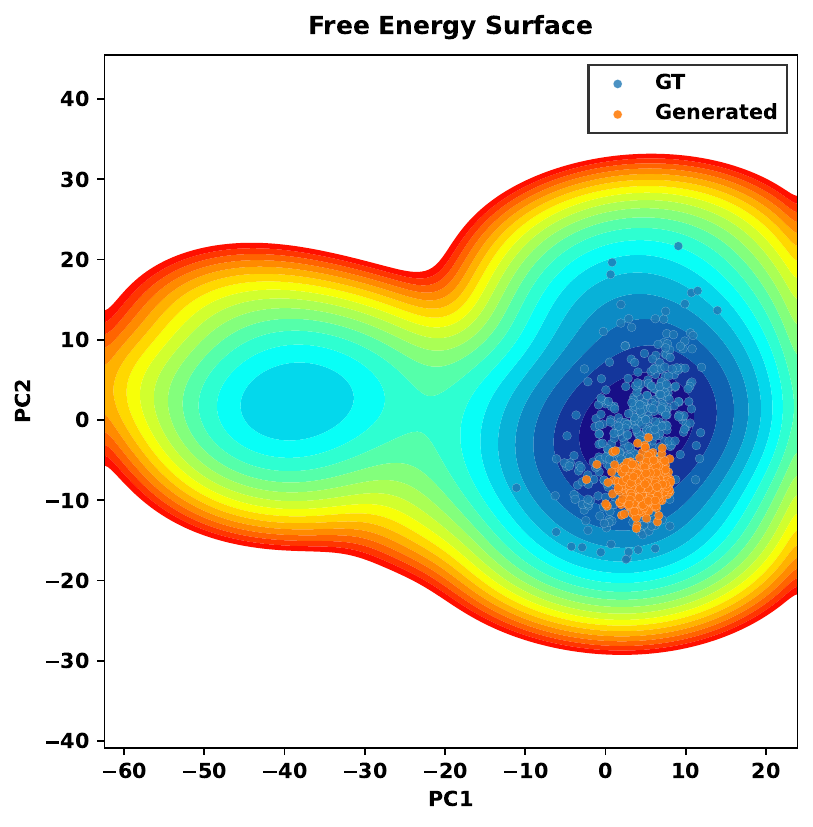}

    \includegraphics[width=0.245\textwidth]{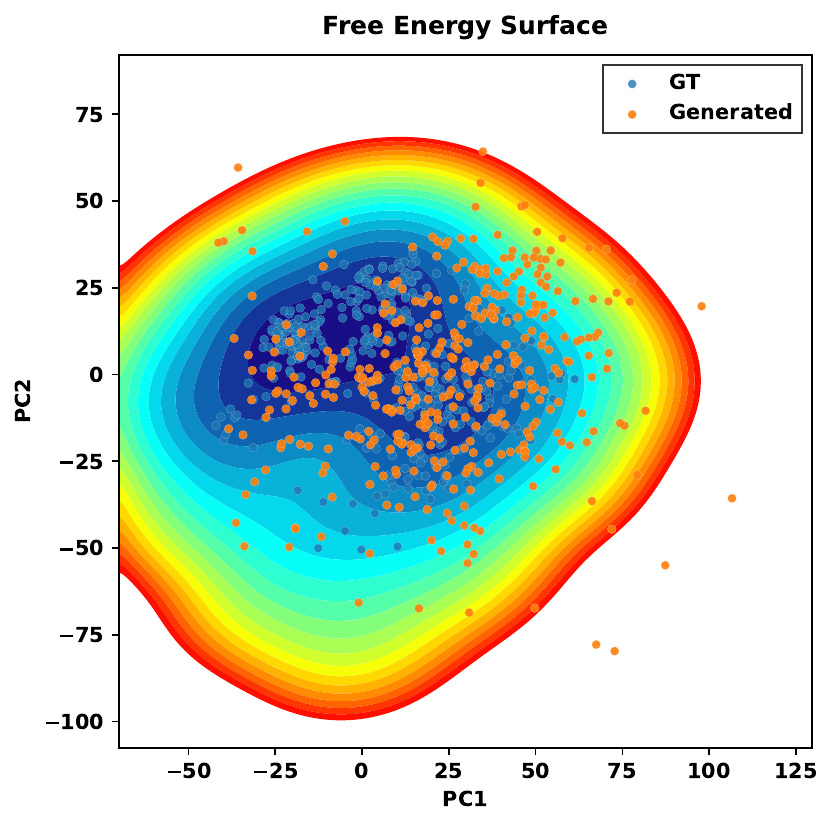}
    \includegraphics[width=0.245\textwidth]{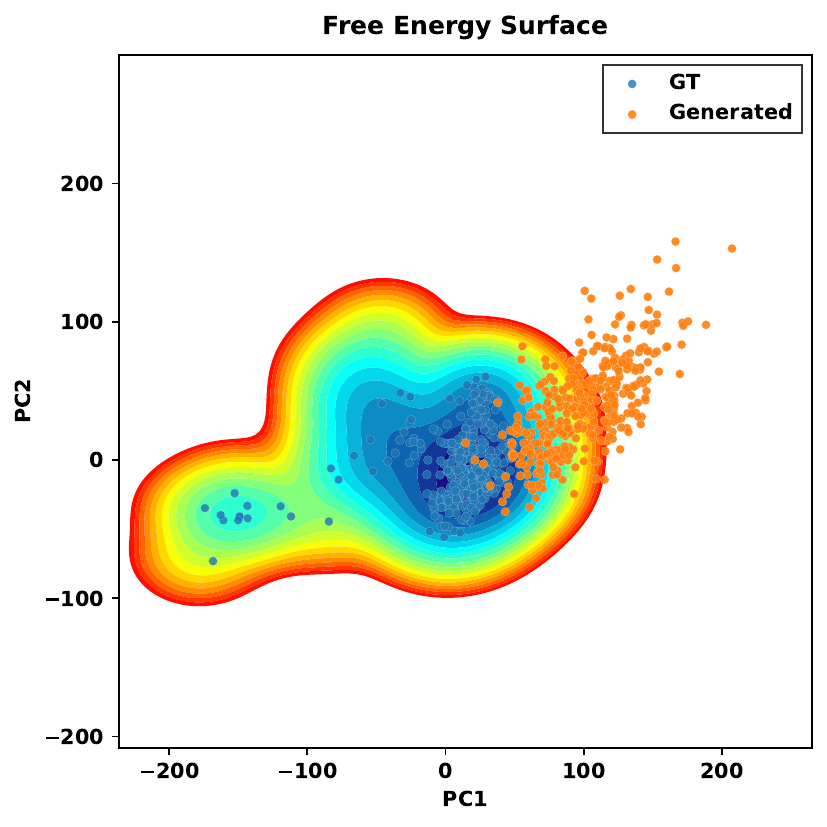}
    \includegraphics[width=0.245\textwidth]{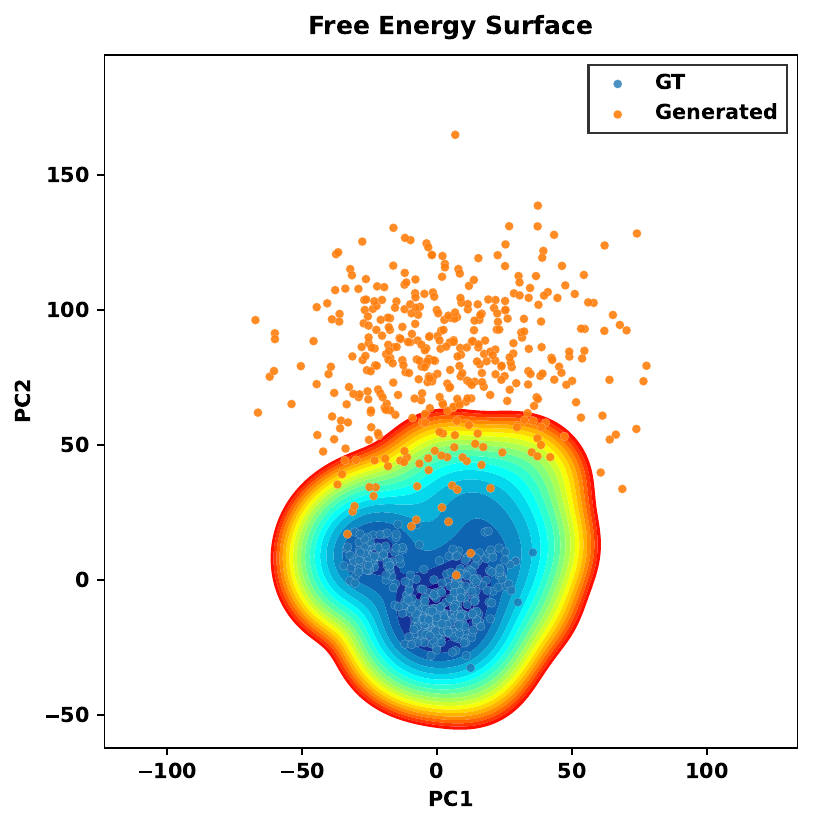}
    \includegraphics[width=0.245\textwidth]{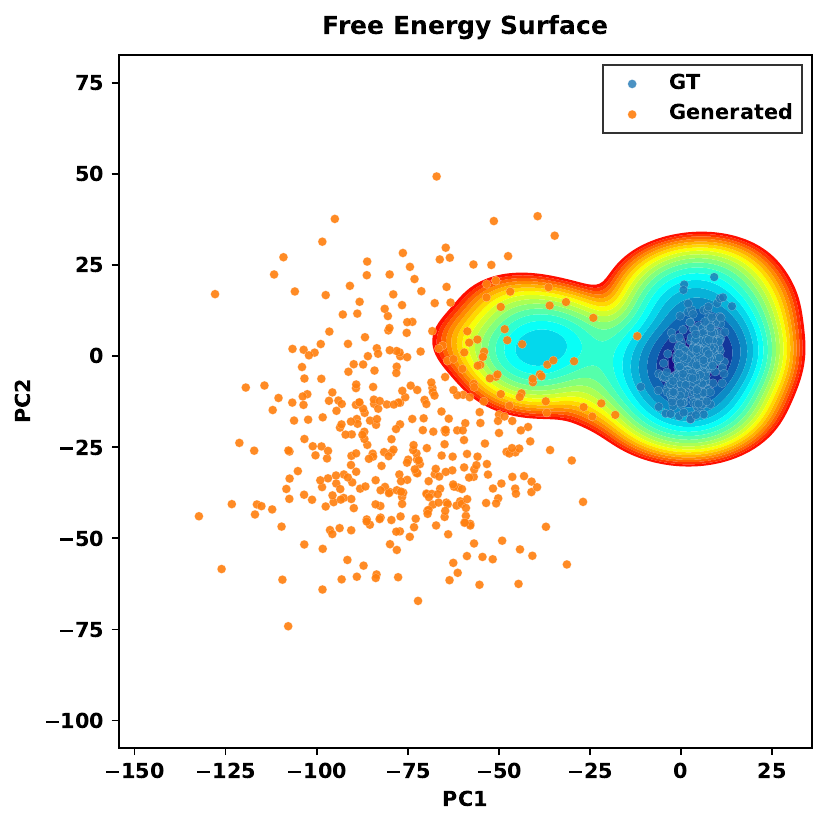}

    \caption{Comparison of FES for four randomly selected test proteins (from left to right: 3cj8B02, 3cx5E01, 3f6kA03, 4fomA03). Results are shown for TEMPO, MDGen, AlphaFlow, and BioEMU (from top to bottom).}
    \label{fig:appendix_fes}
\end{figure}

To complement the FES analysis, we randomly selected four additional proteins from our test set for detailed comparison. As shown in Figure~\ref{fig:appendix_fes}. Based on the FES analysis across four randomly selected test proteins, we observe several distinctive patterns in conformational sampling strategies: TEMPO demonstrates precise conformational sampling that closely aligns with the ground truth distributions. The generated conformations (orange points) are concentrated within physically meaningful energy basins, suggesting coherent trajectory generation. MDGen shows similar sampling quality, though with slightly more dispersed distributions in some cases.

In contrast, AlphaFlow exhibits broader but less focused sampling, often deviating from the main energy wells. BioEMU shows the most scattered sampling patterns, frequently generating physically implausible conformations that lie outside the main energy basins. This comparison highlights the advantage of our trajectory-aware approach over independent sampling methods - while methods like AlphaFlow and BioEMU may achieve wider conformational coverage, they often do so at the cost of physical realism.

These observations consistently demonstrate that TEMPO's multi-scale framework effectively balances conformational exploration with physical constraints, producing trajectories that maintain both continuity and thermodynamic plausibility. The results validate our design choice of incorporating temporal dependencies, which proves crucial for generating biologically meaningful protein dynamics.

\section{Per-Residue RMSF.}
\label{appendix:rmsf_analysis}
\begin{figure}[h]
    \centering
    \includegraphics[width=1\linewidth]{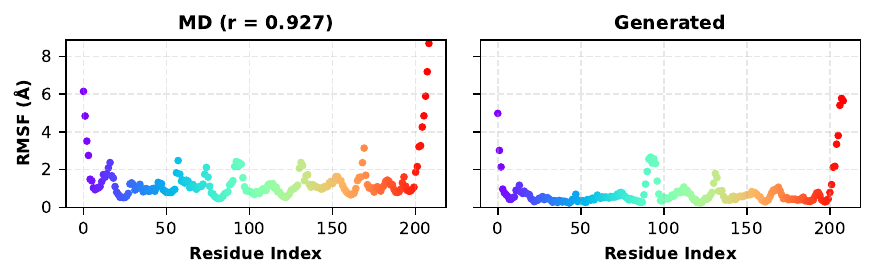}
    \caption{Comparison of Root Mean Square Fluctuation (RMSF) between MD simulation trajectory and generated trajectory for protein 4impA02. The RMSF values reflect the C$\alpha$ fluctuations of protein residues during the simulation. The Pearson (r) between the two RMSFs is $0.93$.}
    \label{fig:rmsf}
\end{figure}

\begin{figure}[htbp]
    \begin{minipage}{\textwidth}
        \includegraphics[width=\linewidth]{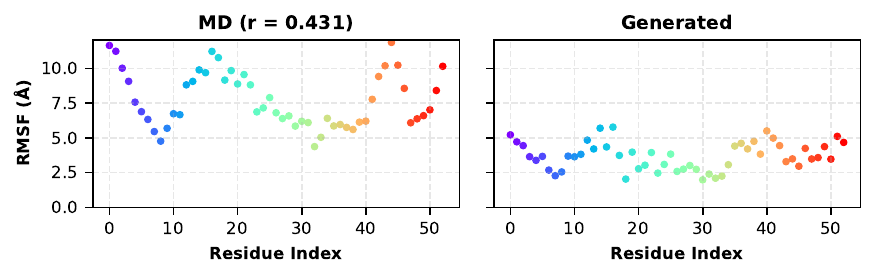}
    \end{minipage}
    
    \vspace{2mm}
    \begin{minipage}{\textwidth}
        \includegraphics[width=\linewidth]{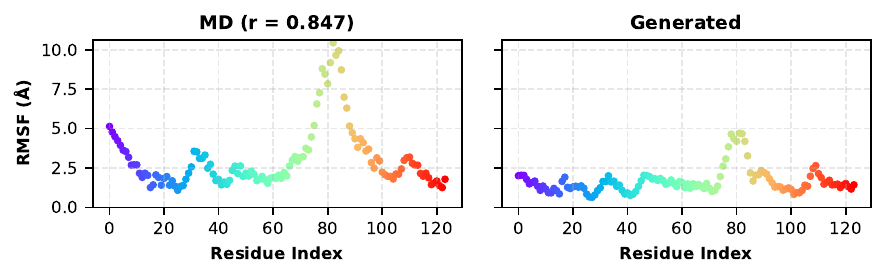}
    \end{minipage}
    
    \vspace{2mm}
    \begin{minipage}{\textwidth}
        \includegraphics[width=\linewidth]{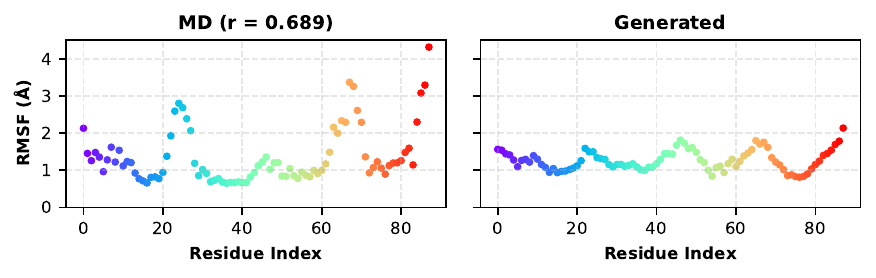}
    \end{minipage}
    
    \vspace{2mm}
    \begin{minipage}{\textwidth}
        \includegraphics[width=\linewidth]{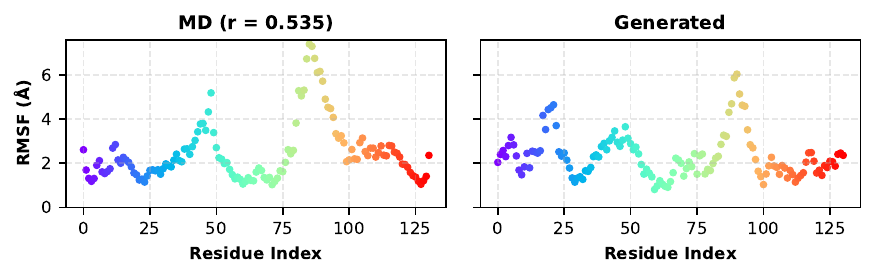}
    \end{minipage}

    \caption{RMSF profiles comparison between MD simulations and TEMPO generated trajectories for $4$ randomly selected test proteins. Each plot demonstrates the C$\alpha$ atomic fluctuations along the protein sequence (Protein from top to bottom: 1y4mA00, 3b8xA02,3bjdA01, 3gyxA02).}
    \label{app:rmsf}
\end{figure}

The RMSF analysis reveals our model's ability to capture local protein dynamics across different scales. Figure~\ref{fig:rmsf} illustrates a representative case study using protein 4impA02, where the generated trajectories closely mirror the MD simulation's C$\alpha$ fluctuation patterns. In this example, both profiles exhibit characteristic mobility signatures, with enhanced fluctuations at the N-terminus (residues 0-20) and C-terminus (residues 190-210), while maintaining relatively stable conformations (RMSF < 2\AA) in the central regions. Across our entire test set, the generated trajectories maintain a reasonable correlation (average Pearson r = $0.67$) with MD simulations in terms of RMSF profiles, suggesting that our model consistently reproduces biologically relevant flexibility patterns. This global performance indicates that the model has learned meaningful protein dynamics patterns rather than generating arbitrary motions. Additional case studies are provided in Figure~\ref {app:rmsf}.


\clearpage
\section*{NeurIPS Paper Checklist}

\begin{enumerate}

\item {\bf Claims}
    \item[] Question: Do the main claims made in the abstract and introduction accurately reflect the paper's contributions and scope?
    \item[] Answer: \answerYes{} 
    \item[] Justification: In Section~\ref{results}, our approach demonstrates distinct advantages across three critical dimensions of protein dynamics modeling and achieves SOTA on several metrics.
    \item[] Guidelines:
    \begin{itemize}
        \item The answer NA means that the abstract and introduction do not include the claims made in the paper.
        \item The abstract and/or introduction should clearly state the claims made, including the contributions made in the paper and important assumptions and limitations. A No or NA answer to this question will not be perceived well by the reviewers. 
        \item The claims made should match theoretical and experimental results, and reflect how much the results can be expected to generalize to other settings. 
        \item It is fine to include aspirational goals as motivation as long as it is clear that these goals are not attained by the paper. 
    \end{itemize}
\item {\bf Limitations}
    \item[] Question: Does the paper discuss the limitations of the work performed by the authors?
    \item[] Answer: \answerYes{} 
    \item[] Justification: In Appendix~\ref{limitaion}, we discuss the limitation of TEMPO.
    \item[] Guidelines:
    \begin{itemize}
        \item The answer NA means that the paper has no limitation while the answer No means that the paper has limitations, but those are not discussed in the paper. 
        \item The authors are encouraged to create a separate "Limitations" section in their paper.
        \item The paper should point out any strong assumptions and how robust the results are to violations of these assumptions (e.g., independence assumptions, noiseless settings, model well-specification, asymptotic approximations only holding locally). The authors should reflect on how these assumptions might be violated in practice and what the implications would be.
        \item The authors should reflect on the scope of the claims made, e.g., if the approach was only tested on a few datasets or with a few runs. In general, empirical results often depend on implicit assumptions, which should be articulated.
        \item The authors should reflect on the factors that influence the performance of the approach. For example, a facial recognition algorithm may perform poorly when image resolution is low or images are taken in low lighting. Or a speech-to-text system might not be used reliably to provide closed captions for online lectures because it fails to handle technical jargon.
        \item The authors should discuss the computational efficiency of the proposed algorithms and how they scale with dataset size.
        \item If applicable, the authors should discuss possible limitations of their approach to address problems of privacy and fairness.
        \item While the authors might fear that complete honesty about limitations might be used by reviewers as grounds for rejection, a worse outcome might be that reviewers discover limitations that aren't acknowledged in the paper. The authors should use their best judgment and recognize that individual actions in favor of transparency play an important role in developing norms that preserve the integrity of the community. Reviewers will be specifically instructed to not penalize honesty concerning limitations.
    \end{itemize}

\item {\bf Theory assumptions and proofs}
    \item[] Question: For each theoretical result, does the paper provide the full set of assumptions and a complete (and correct) proof?
    \item[] Answer: \answerNA{} 
    \item[] Justification: This paper does not include any theoretical result.
    \item[] Guidelines:
    \begin{itemize}
        \item The answer NA means that the paper does not include theoretical results. 
        \item All the theorems, formulas, and proofs in the paper should be numbered and cross-referenced.
        \item All assumptions should be clearly stated or referenced in the statement of any theorems.
        \item The proofs can either appear in the main paper or the supplemental material, but if they appear in the supplemental material, the authors are encouraged to provide a short proof sketch to provide intuition. 
        \item Inversely, any informal proof provided in the core of the paper should be complemented by formal proofs provided in appendix or supplemental material.
        \item Theorems and Lemmas that the proof relies upon should be properly referenced. 
    \end{itemize}

    \item {\bf Experimental result reproducibility}
    \item[] Question: Does the paper fully disclose all the information needed to reproduce the main experimental results of the paper to the extent that it affects the main claims and/or conclusions of the paper (regardless of whether the code and data are provided or not)?
    \item[] Answer: \answerYes{} 
    \item[] Justification: We present the implementation detail in Section~\ref{experiment} and also submit the code in the supplemental material.
    \item[] Guidelines:
    \begin{itemize}
        \item The answer NA means that the paper does not include experiments.
        \item If the paper includes experiments, a No answer to this question will not be perceived well by the reviewers: Making the paper reproducible is important, regardless of whether the code and data are provided or not.
        \item If the contribution is a dataset and/or model, the authors should describe the steps taken to make their results reproducible or verifiable. 
        \item Depending on the contribution, reproducibility can be accomplished in various ways. For example, if the contribution is a novel architecture, describing the architecture fully might suffice, or if the contribution is a specific model and empirical evaluation, it may be necessary to either make it possible for others to replicate the model with the same dataset, or provide access to the model. In general. releasing code and data is often one good way to accomplish this, but reproducibility can also be provided via detailed instructions for how to replicate the results, access to a hosted model (e.g., in the case of a large language model), releasing of a model checkpoint, or other means that are appropriate to the research performed.
        \item While NeurIPS does not require releasing code, the conference does require all submissions to provide some reasonable avenue for reproducibility, which may depend on the nature of the contribution. For example
        \begin{enumerate}
            \item If the contribution is primarily a new algorithm, the paper should make it clear how to reproduce that algorithm.
            \item If the contribution is primarily a new model architecture, the paper should describe the architecture clearly and fully.
            \item If the contribution is a new model (e.g., a large language model), then there should either be a way to access this model for reproducing the results or a way to reproduce the model (e.g., with an open-source dataset or instructions for how to construct the dataset).
            \item We recognize that reproducibility may be tricky in some cases, in which case authors are welcome to describe the particular way they provide for reproducibility. In the case of closed-source models, it may be that access to the model is limited in some way (e.g., to registered users), but it should be possible for other researchers to have some path to reproducing or verifying the results.
        \end{enumerate}
    \end{itemize}

\item {\bf Open access to data and code}
    \item[] Question: Does the paper provide open access to the data and code, with sufficient instructions to faithfully reproduce the main experimental results, as described in supplemental material?
    \item[] Answer: \answerYes{} 
    \item[] Justification: All the datasets are obtained from open-source libraries. The code is available in the supplemental material.
    \item[] Guidelines:
    \begin{itemize}
        \item The answer NA means that paper does not include experiments requiring code.
        \item Please see the NeurIPS code and data submission guidelines (\url{https://nips.cc/public/guides/CodeSubmissionPolicy}) for more details.
        \item While we encourage the release of code and data, we understand that this might not be possible, so “No” is an acceptable answer. Papers cannot be rejected simply for not including code, unless this is central to the contribution (e.g., for a new open-source benchmark).
        \item The instructions should contain the exact command and environment needed to run to reproduce the results. See the NeurIPS code and data submission guidelines (\url{https://nips.cc/public/guides/CodeSubmissionPolicy}) for more details.
        \item The authors should provide instructions on data access and preparation, including how to access the raw data, preprocessed data, intermediate data, and generated data, etc.
        \item The authors should provide scripts to reproduce all experimental results for the new proposed method and baselines. If only a subset of experiments are reproducible, they should state which ones are omitted from the script and why.
        \item At submission time, to preserve anonymity, the authors should release anonymized versions (if applicable).
        \item Providing as much information as possible in supplemental material (appended to the paper) is recommended, but including URLs to data and code is permitted.
    \end{itemize}

\item {\bf Experimental setting/details}
    \item[] Question: Does the paper specify all the training and test details (e.g., data splits, hyperparameters, how they were chosen, type of optimizer, etc.) necessary to understand the results?
    \item[] Answer: \answerYes{} 
    \item[] Justification: We state the experimental setup in Section~\ref{experiment}.
    \item[] Guidelines:
    \begin{itemize}
        \item The answer NA means that the paper does not include experiments.
        \item The experimental setting should be presented in the core of the paper to a level of detail that is necessary to appreciate the results and make sense of them.
        \item The full details can be provided either with the code, in appendix, or as supplemental material.
    \end{itemize}

\item {\bf Experiment statistical significance}
    \item[] Question: Does the paper report error bars suitably and correctly defined or other appropriate information about the statistical significance of the experiments?
    \item[] Answer: \answerYes{} 
    \item[] Justification: In Section~\ref{results}, we show the statistical significance of our results.
    \item[] Guidelines:
    \begin{itemize}
        \item The answer NA means that the paper does not include experiments.
        \item The authors should answer "Yes" if the results are accompanied by error bars, confidence intervals, or statistical significance tests, at least for the experiments that support the main claims of the paper.
        \item The factors of variability that the error bars are capturing should be clearly stated (for example, train/test split, initialization, random drawing of some parameter, or overall run with given experimental conditions).
        \item The method for calculating the error bars should be explained (closed form formula, call to a library function, bootstrap, etc.)
        \item The assumptions made should be given (e.g., Normally distributed errors).
        \item It should be clear whether the error bar is the standard deviation or the standard error of the mean.
        \item It is OK to report 1-sigma error bars, but one should state it. The authors should preferably report a 2-sigma error bar than state that they have a 96\% CI, if the hypothesis of Normality of errors is not verified.
        \item For asymmetric distributions, the authors should be careful not to show in tables or figures symmetric error bars that would yield results that are out of range (e.g. negative error rates).
        \item If error bars are reported in tables or plots, The authors should explain in the text how they were calculated and reference the corresponding figures or tables in the text.
    \end{itemize}

\item {\bf Experiments compute resources}
    \item[] Question: For each experiment, does the paper provide sufficient information on the computer resources (type of compute workers, memory, time of execution) needed to reproduce the experiments?
    \item[] Answer: \answerYes{} 
    \item[] Justification: In Section~\ref{results}, we report the memory and computational efficiency.
    \item[] Guidelines:
    \begin{itemize}
        \item The answer NA means that the paper does not include experiments.
        \item The paper should indicate the type of compute workers CPU or GPU, internal cluster, or cloud provider, including relevant memory and storage.
        \item The paper should provide the amount of compute required for each of the individual experimental runs as well as estimate the total compute. 
        \item The paper should disclose whether the full research project required more compute than the experiments reported in the paper (e.g., preliminary or failed experiments that didn't make it into the paper). 
    \end{itemize}
    
\item {\bf Code of ethics}
    \item[] Question: Does the research conducted in the paper conform, in every respect, with the NeurIPS Code of Ethics \url{https://neurips.cc/public/EthicsGuidelines}?
    \item[] Answer: \answerYes{} 
    \item[] Justification: We make sure to preserve anonymity.
    \item[] Guidelines:
    \begin{itemize}
        \item The answer NA means that the authors have not reviewed the NeurIPS Code of Ethics.
        \item If the authors answer No, they should explain the special circumstances that require a deviation from the Code of Ethics.
        \item The authors should make sure to preserve anonymity (e.g., if there is a special consideration due to laws or regulations in their jurisdiction).
    \end{itemize}

\item {\bf Broader impacts}
    \item[] Question: Does the paper discuss both potential positive societal impacts and negative societal impacts of the work performed?
    \item[] Answer: \answerNA{} 
    \item[] Justification: There is no societal impact on the work performed.
    \item[] Guidelines:
    \begin{itemize}
        \item The answer NA means that there is no societal impact of the work performed.
        \item If the authors answer NA or No, they should explain why their work has no societal impact or why the paper does not address societal impact.
        \item Examples of negative societal impacts include potential malicious or unintended uses (e.g., disinformation, generating fake profiles, surveillance), fairness considerations (e.g., deployment of technologies that could make decisions that unfairly impact specific groups), privacy considerations, and security considerations.
        \item The conference expects that many papers will be foundational research and not tied to particular applications, let alone deployments. However, if there is a direct path to any negative applications, the authors should point it out. For example, it is legitimate to point out that an improvement in the quality of generative models could be used to generate deepfakes for disinformation. On the other hand, it is not needed to point out that a generic algorithm for optimizing neural networks could enable people to train models that generate Deepfakes faster.
        \item The authors should consider possible harms that could arise when the technology is being used as intended and functioning correctly, harms that could arise when the technology is being used as intended but gives incorrect results, and harms following from (intentional or unintentional) misuse of the technology.
        \item If there are negative societal impacts, the authors could also discuss possible mitigation strategies (e.g., gated release of models, providing defenses in addition to attacks, mechanisms for monitoring misuse, mechanisms to monitor how a system learns from feedback over time, improving the efficiency and accessibility of ML).
    \end{itemize}
    
\item {\bf Safeguards}
    \item[] Question: Does the paper describe safeguards that have been put in place for responsible release of data or models that have a high risk for misuse (e.g., pretrained language models, image generators, or scraped datasets)?
    \item[] Answer: \answerNA{} 
    \item[] Justification: The paper poses no such risks.
    \item[] Guidelines:
    \begin{itemize}
        \item The answer NA means that the paper poses no such risks.
        \item Released models that have a high risk for misuse or dual-use should be released with necessary safeguards to allow for controlled use of the model, for example by requiring that users adhere to usage guidelines or restrictions to access the model or implementing safety filters. 
        \item Datasets that have been scraped from the Internet could pose safety risks. The authors should describe how they avoided releasing unsafe images.
        \item We recognize that providing effective safeguards is challenging, and many papers do not require this, but we encourage authors to take this into account and make a best faith effort.
    \end{itemize}

\item {\bf Licenses for existing assets}
    \item[] Question: Are the creators or original owners of assets (e.g., code, data, models), used in the paper, properly credited and are the license and terms of use explicitly mentioned and properly respected?
    \item[] Answer: \answerYes{} 
    \item[] Justification: Properly credited.
    \item[] Guidelines:
    \begin{itemize}
        \item The answer NA means that the paper does not use existing assets.
        \item The authors should cite the original paper that produced the code package or dataset.
        \item The authors should state which version of the asset is used and, if possible, include a URL.
        \item The name of the license (e.g., CC-BY 4.0) should be included for each asset.
        \item For scraped data from a particular source (e.g., website), the copyright and terms of service of that source should be provided.
        \item If assets are released, the license, copyright information, and terms of use in the package should be provided. For popular datasets, \url{paperswithcode.com/datasets} has curated licenses for some datasets. Their licensing guide can help determine the license of a dataset.
        \item For existing datasets that are re-packaged, both the original license and the license of the derived asset (if it has changed) should be provided.
        \item If this information is not available online, the authors are encouraged to reach out to the asset's creators.
    \end{itemize}

\item {\bf New assets}
    \item[] Question: Are new assets introduced in the paper well documented and is the documentation provided alongside the assets?
    \item[] Answer: \answerYes{} 
    \item[] Justification: We will release the code.
    \item[] Guidelines:
    \begin{itemize}
        \item The answer NA means that the paper does not release new assets.
        \item Researchers should communicate the details of the dataset/code/model as part of their submissions via structured templates. This includes details about training, license, limitations, etc. 
        \item The paper should discuss whether and how consent was obtained from people whose asset is used.
        \item At submission time, remember to anonymize your assets (if applicable). You can either create an anonymized URL or include an anonymized zip file.
    \end{itemize}

\item {\bf Crowdsourcing and research with human subjects}
    \item[] Question: For crowdsourcing experiments and research with human subjects, does the paper include the full text of instructions given to participants and screenshots, if applicable, as well as details about compensation (if any)? 
    \item[] Answer: \answerNA{} 
    \item[] Justification: The paper does not involve crowdsourcing nor research with human subjects.
    \item[] Guidelines:
    \begin{itemize}
        \item The answer NA means that the paper does not involve crowdsourcing nor research with human subjects.
        \item Including this information in the supplemental material is fine, but if the main contribution of the paper involves human subjects, then as much detail as possible should be included in the main paper. 
        \item According to the NeurIPS Code of Ethics, workers involved in data collection, curation, or other labor should be paid at least the minimum wage in the country of the data collector. 
    \end{itemize}

\item {\bf Institutional review board (IRB) approvals or equivalent for research with human subjects}
    \item[] Question: Does the paper describe potential risks incurred by study participants, whether such risks were disclosed to the subjects, and whether Institutional Review Board (IRB) approvals (or an equivalent approval/review based on the requirements of your country or institution) were obtained?
    \item[] Answer: \answerNA{} 
    \item[] Justification: The paper does not involve crowdsourcing nor research with human subjects.
    \item[] Guidelines:
    \begin{itemize}
        \item The answer NA means that the paper does not involve crowdsourcing nor research with human subjects.
        \item Depending on the country in which research is conducted, IRB approval (or equivalent) may be required for any human subjects research. If you obtained IRB approval, you should clearly state this in the paper. 
        \item We recognize that the procedures for this may vary significantly between institutions and locations, and we expect authors to adhere to the NeurIPS Code of Ethics and the guidelines for their institution. 
        \item For initial submissions, do not include any information that would break anonymity (if applicable), such as the institution conducting the review.
    \end{itemize}

\item {\bf Declaration of LLM usage}
    \item[] Question: Does the paper describe the usage of LLMs if it is an important, original, or non-standard component of the core methods in this research? Note that if the LLM is used only for writing, editing, or formatting purposes and does not impact the core methodology, scientific rigorousness, or originality of the research, declaration is not required.
    \item[] Answer: \answerNA{} 
    \item[] Justification: LLM is used only for writing.
    \item[] Guidelines:
    \begin{itemize}
        \item The answer NA means that the core method development in this research does not involve LLMs as any important, original, or non-standard components.
        \item Please refer to our LLM policy (\url{https://neurips.cc/Conferences/2025/LLM}) for what should or should not be described.
    \end{itemize}

\end{enumerate}

\end{document}